\documentclass{tlp} 

\makeatletter
\def\tabular{
  \def\@halignto{}\tabskip\tabcolsep \@plus 1fil\@ttabular
}

\newenvironment{myquote}
               {\list{}{\leftmargin0.3cm\rightmargin0cm
                        \topsep0.05cm \parsep0.03cm \itemsep0.03cm}%
                \item\relax}
               {\endlist}

\makeatother

\usepackage{latexsym}
\usepackage{url}
\usepackage{pstricks,pst-node}
\usepackage{dbicons}

\newtheorem{Example}{Example}
\newtheorem{Notation}{Notation}
\newtheorem{Definition}{Definition}
\newtheorem{Theorem}{Theorem}
\newtheorem{Proposition}[Theorem]{Proposition}
\newtheorem{Lemma}[Theorem]{Lemma}
\newtheorem{Corollary}[Theorem]{Corollary}
\newtheorem{Remark}{Remark}
\newenvironment{Proof}
  {\noindent\textbf{Proof}}{}

\makeatletter
\def\save@sf@q#1{{\ifhmode
  \edef\@SF{\spacefactor\the\spacefactor}\else
  \let\@SF\empty \fi \leavevmode #1\@SF}}
\def\@flq{\relax\ifmmode <\else \save@sf@q{\penalty\@M
    \raise .27ex\hbox{$\m@th\scriptscriptstyle <$}}\fi}
\def\flq{\protect\@flq}
\def\@frq{\relax\ifmmode >\else \save@sf@q{\penalty\@M
    \raise .27ex\hbox{$\m@th\scriptscriptstyle >$}}\fi}
\def\frq{\protect\@frq}

\def\htmlpage{\@ifnextchar[{\htmlpage@i}{\htmlpage@i[c]}}
\def\htmlpage@i[#1]#2#3{%
 \@ifnextchar[{\htmlpage@ii{#1}{#2}{#3}}%
              {\htmlpage@ii{#1}{#2}{#3}[fillstyle=none]}}
\def\htmlpage@ii#1#2#3[#4]#5{%
 \@ifnextchar[{\htmlpage@iii{#1}{#2}{#3}{#4}{#5}}%
              {\htmlpage@iii{#1}{#2}{#3}{#4}{#5}[#5]}}
\def\htmlpage@iii#1#2#3#4#5[#6]{%
 \psset{nodesep=0pt}%
 \rnode{#6}{%
 \begin{tabular}[1]{c}%
 \rnode{lu#6}{}\hspace{#2}\rnode{ru#6}{}\\[#3]
 \rnode{ll#6}{}\hspace{#2}\rnode{rl#6}{}%
 \end{tabular}}%
 \nodeconnections{%
 \ncline[linestyle=none]{lu#6}{ru#6}
   \ncput[npos=0.5]{\rnode{mu#6}{}}
   \ncput[npos=0.7]{\rnode{xx#6}{}}
 \ncline[linestyle=none]{ll#6}{rl#6}
   \ncput[npos=0.5]{\rnode{ml#6}{}}
 \ncline[linestyle=none]{ru#6}{rl#6}
   \ncput[npos=0.3]{\rnode{yy#6}{}}
 \ncbar[linestyle=none,#4,angleA=180,angleB=180,armB=0]{yy#6}{ll#6}
 \ncbar[linestyle=none,#4,angleA=-90,angleB=90,armB=0]{xx#6}{ll#6}
 \ncbar[#4,angleA=180,angleB=180,armB=0]{xx#6}{ll#6}
 \ncbar[#4,angleA=-90,angleB=-90,armB=0]{yy#6}{ll#6}
 \ncbar[angleA=-90,angleB=-90,armB=0]{xx#6}{yy#6}
 \ncline{xx#6}{yy#6}
 \ncline[linestyle=none]{mu#6}{ml#6}\ncput[npos=0.6]{#5}
 }}

\makeatletter

\newcommand{\smallvdots}{\strut\raisebox{-0.05cm}{\smash{\small\vdots}}}

\mathcode`+="002B  
\mathcode`-="007B

\begin{document}

\newcommand{\Florid}{\textsc{Florid}}
\newcommand{\FLORID}{\textsc{Florid}}
\newcommand{\FloX}{\textsc{FloX}}
\newcommand{\LoPiX}{\textsc{LoP}\textup{iX}}
\newcommand{\Oracle}{\textsc{Oracle}}
\newcommand{\TERRA}{\textsc{Terra}}
\newcommand{\Mondial}{\textsc{Mondial}}

\newcommand{\pl}{\mathcal}
\newcommand{\st}{\ensuremath{\ |\ }} 
\newcommand{\Res}[1]{{\rm Res}(#1)}
\newcommand{\Bdgs}[1]{{\rm Bdgs}(#1)}
\newcommand{\Var}{\mbox{\sf Var}}
\newcommand{\preds}{\mbox{\sf preds}}
\newcommand{\answers}[2]{\mbox{\sf answers$_{#1}$(\textsf{#2})}}
\newcommand{\free}{{\sf free}}
\newcommand{\join}{\bowtie}
\newcommand{\union}{\cup}
\newcommand{\cons}{\,\circ\,}
\newcommand{\atomize}{\textsf{atomize}}
\newcommand{\bel}{\beta}
\newcommand{\true}{\textsf{true}}
\newcommand{\false}{\textsf{false}}
\newcommand{\TX}{T\!\!X}

\newcommand{\AnnotatedResults}{\mbox{AnnotatedResults}}
\newcommand{\Bindings}{\mbox{Var\_Bindings}}
\newcommand{\BindingsLists}{\mbox{Var\_BindingsLists}}
\def\N{\ensuremath{\mathrm{I\! N}}} 

\newcommand{\DefIff}{~:\Leftrightarrow~}
\newcommand{\Iff}{~\Leftrightarrow~}
\newcommand{\Imp}{~\Rightarrow~}
\newcommand{\imp}{\rightarrow}
\renewcommand{\iff}{\leftrightarrow}

\newgray{lightgray}{0.75}
\newgray{verylightgray}{0.9}
\newgray{superlightgray}{0.97}

\newenvironment{expl}%
  {\begin{myquote}
   \sf\begin{tabular}[t]{l@{}l@{}l@{}l}}%
  {\end{tabular}\end{myquote}}


\newcommand{\isa}{\ensuremath{\;\textsf{isa}\;}}
\newcommand{\subcl}{\ensuremath{\;\textsf{subcl}\;}}

\newcommand{\anyd}{\mbox{$\leadsto$}}                      
\newcommand{\fd}{\ensuremath{\rightarrow}}                      
\newcommand{\bfd}{\mbox{$\bullet\!\!\!\fd$}}                
\newcommand{\bbfd}{\mbox{$\bullet\!\bullet\!\!\!\fd$}}                
\newcommand{\mvd}{\mbox{$\twoheadrightarrow$}}  
\newcommand{\bmvd}{\mbox{$\bullet\!\!\!\mvd$}}              
\newcommand{\Fd}{\mbox{$\Rightarrow$}}                      
\newcommand{\Mvd}{\mbox{$\Rightarrow\!\!\!\!\Rightarrow$}}  


\title[A Logic-Programming Style XML Data Manipulation Language]
{XPath-Logic and XPathLog: A Logic-Programming Style XML Data
  Manipulation Language}
\author[W. May]{WOLFGANG MAY \\
  Institut f\"ur Informatik, Universit\"at G\"ottingen\\
  \email{may@informatik.uni-goettingen.de}} \maketitle

\begin{abstract}
  We define XPathLog as a Datalog-style extension of XPath.  XPathLog
  provides a clear, declarative language for querying and manipulating
  XML whose perspectives are especially in XML data integration.
  
  In our characterization, the formal semantics is defined wrt.\ an
  \emph{edge-labeled graph-based} model which covers the XML data
  model.  We give a complete, logic-based characterization of XML data
  and the main language concept for XML, XPath.  XPath-Logic extends
  the XPath language with variable bindings and embeds it into
  first-order logic.  XPathLog is then the Horn fragment of
  XPath-Logic, providing a Datalog-style, rule-based language for
  querying and manipulating XML data.  The model-theoretic semantics
  of XPath-Logic serves as the base of XPathLog as a
  logic-programming language, whereas also an equivalent answer-set
  semantics for evaluating XPathLog queries is given.  In contrast to
  other approaches, the XPath syntax and semantics is also used for a
  declarative specification how the database should be \emph{updated}:
  when used in rule heads, XPath filters are interpreted as
  specifications of elements and properties which should be added to
  the database.
\end{abstract}

\noindent
\textbf{Keywords:} XML, XPath, Logic Programming, Information Integration.

\noindent
{\small Submitted: February 26, 2002; revised: September 16, 2002; 
  accepted: April 7, 2003.}

\section{Introduction}\label{sec-introduction}

Logic-based languages have proven useful in many areas since they
allow for small, declarative, and extendible programs.  For the
database area, Datalog has been investigated for querying and
rule-based data manipulation.  Extending the Datalog idea, more
complex logic-based frameworks like F-Logic
\cite{kifer-lausen-SIGMOD-89,kifer-lausen-wu-JACM-95}, or the
languages of the \textsc{Tsimmis} project
\cite{garcia-molina-papakonstantinou-quass-JIIS-97,abiteboul-quass-mchugh-widom-wiener-jodl-97}
have been successfully applied for knowledge representation and data
integration.  The experiences with a powerful language like F-Logic
were the motivation to have a similar ``native'' language for the XML
world that is much simpler than F-Logic, and that is based on the
standard XPath language.  As a result, we present XPathLog as an
XPath-based Datalog-style language for querying and manipulating XML
data.  By extending XPath with variable bindings and providing a
constructive semantics for XPath in rule reads, a declarative XML data
manipulation language is obtained.  Since both XPath and rule-based
programming by using variable bindings are well-known, intuitive
concepts, the ``effect'' of the language is easy to understand on an
intuitive basis.  Additionally, the well-known logic programming
semantics provide concise \emph{global} semantics of such programs
which coincide with the intuitive ideas.  Queries and rules for
manipulating and restructuring the internal XML database can be
expressed much easier than e.g.\ in XQuery \cite{XQuery-W3C} (where
update functionality is still in a prototypical state).

\paragraph{Semistructured Data and XML.}
XML has been designed and accepted as \emph{the} framework for
semi-structured data where it plays the same role as the relational
model for classical databases.  The XML data model applies both to
\emph{documents} and to \emph{databases}: The SGML language was
originally defined for \emph{documents} in the publishing area. On the
other hand, the interest in research on \emph{semistructured} data in
the 1990s\footnote{We list the approaches in the temporal order of
  their presentation.} (e.g., F-Logic
\cite{kifer-lausen-SIGMOD-89,kifer-lausen-wu-JACM-95}, GraphLog
\cite{consens-mendelzon-pods-90}, UnQL
\cite{buneman-davidson-hillebrandt-suciu-SIGMOD-96,buneman-fernandez-suciu-VLDBJ-2000},
\textsc{Tsimmis}
\cite{garcia-molina-papakonstantinou-quass-JIIS-97,abiteboul-quass-mchugh-widom-wiener-jodl-97}
with the \emph{OEM} data model and the \emph{MSL}, \emph{WSL}, and
\emph{Lorel} languages, \textsc{Strudel/StruQL}
\cite{fernandez-florescu-levy-suciu-SIGMODREC-97,fernandez-florescu-kang-SIGMOD-98},
and \textsl{YAT/YATL} \cite{cluet-delobel-simeon-smaga-SIGMOD-98}) was
motivated by the \emph{database} community, searching for a data model
for \emph{data integration} and a data format for \emph{electronic
  data interchange}. Here, also the combination of document-oriented
aspects with database aspects was an important motivation to go beyond
classical data models which then resulted in the design of XML.

The XML data model is a hierarchical model which defines an ordered
tree with attributes that can easily be interpreted as a document.
The natural relationships in documents are either (i) substructures,
or (ii) references to other parts of the document (where the term
\emph{reference} here means simply a cross-reference in a document).
The nested elements define a document structure whose leaves are the
text contents.  Elements (i.e., the structuring components) are
annotated by attributes which do \emph{not} belong to the document
contents.  Inside the tree, cross-references (\texttt{IDREF}
attributes) are allowed.

In contrast, for databases, a hierarchical structure is in general not
intuitive.  Here, several kinds of relationships have to be
represented, between substructures and pure references.  When using
XML for a database-like application, these relationships have to be
represented by reference attributes. On the other hand, order is
often not relevant in databases.

\paragraph{Mainstream XML Languages.}
Specialized languages have been defined for XML querying, e.g., XQL
\cite{XQL-W3C-99}, XML-QL
\cite{deutsch-fernandez-florescu-etal-WWW-99}, then XPath
\cite{XPath-W3C} developed from the experiences with XQL
and XSL Patterns (and XPointer) as an addressing language. XQuery
\cite{XQuery-W3C} extends XPath with SQL-like constructs. XSLT
\cite{XSLT-W3C} is an XPath-based language for transforming XML data.
A proposal for extending XQuery with update constructs (XUpdate) has
been published in \cite{tatarinov-ives-halevy-weld-SIGMOD-2001}; a more
detailed proposal is described in \cite{lehti-DA-01}.


\paragraph{Other Approaches to XML.}

XML-GL \cite{xml-gl-WWW8-1999,comai-damiani-fraternali-tois-01}
continued the idea of GraphLog for XML.  Elog
\cite{baumgartner-flesca-gottlob-LPAR-01} is based on Datalog and
classical first-order logic, flattening XML into predicates. It is
used as an internal language in Lixto
\cite{baumgartner-flesca-gottlob-VLDB-01}.
\cite{bry-schaffert-iclp-02} present the Xcerpt language which regards
XML trees as terms, similar to UnQL.

\subsection{Comparison of Design Concepts}\label{sec-design-concepts}

Amongst the existing languages for handling semistructured data and
XML, there are several facets for distinguishing them in terms of the
concepts they use.  A more detailed comparison with individual
languages can be found in Section~\ref{sec-related-and-conclusion}.

\paragraph{Data model.}
Semistructured data can be regarded as a general graph (OEM, UnQL,
Strudel, GraphLog, and F-Logic) or as a tree (YATL and XML). Moreover,
node-labeled graphs/trees (XML) or edge-labeled graphs (as in Strudel,
UnQL, GraphLog, and F-Logic) can be distinguished; for OEM both
representations can be found.  It is easy to represent a node-labeled
instance in a labeled model, whereas the other way requires node
replication. Also, ordered (e.g., XML) and unordered (e.g. in OEM,
F-Logic, and UnQL) tree/graph models are distinguished.

\paragraph{Access mechanism.} 
Generally, there are two approaches for selecting items in a
semistructured data tree or graph:

\begin{itemize}
\item by matching patterns and templates (GraphLog, MSL/WSL, UnQL,
  YATL, and later for XML-QL and XML-GL).  In UnQL and Xcerpt,
  (bi)simulation between semistructured data trees is used.  If a
  simulation of a match pattern with variables by the underlying
  database is found, the appropriate variable bindings are returned
  and used for generating an answer tree.
\item navigational access, like in object-oriented database languages
  (OQL), as done in Lorel, StruQL, and F-Logic, and later in XPath and
  also in our XPathLog approach.  UnQL provides both patterns and a
  navigational syntax.
\end{itemize}

\paragraph{Functionality.}
There are different approaches to either \emph{generating} an answer
by instantiating a generating pattern in the rule head according to
the variable bindings (UnQL, StruQL, and later XML-QL, XQuery and
XML-GL), or \emph{manipulating} the underlying structure by adding
information to the underlying database (GraphLog, F-Logic and
XUpdate).

Note that this distinction did not exist when considering classical
rule-based languages, e.g., Datalog for predicate logic. The facts
derived in the rule head were added to the database -- either
extensionally, or intensionally as view definitions -- without
directly interfering with the already stored facts. Other rules of the
program could easily use both the original data and the derived data.
For XML, a semantics where rules generate separate structures is easy
to define. In contrast, a semantics where the rule heads interfere
with the database contents has to take into account that the
evaluation of rules may violate the tree structure.

\paragraph{Underlying Framework.}
Some of the languages are based on a kind of model-theoretic
semantics: UnQL and Xcerpt directly operate on tree-term structures,
employing and defining mechanisms like tree matching, term unification
and (bi)si\-mula\-tion unification.  Elog is based on Datalog and
classical first-order logic, flattening XML data into predicates.
F-Logic defines F-Structures that extend classical first-order logic,
and then applies logic programming mechanisms to such models.  For the
other languages (Tsimmis, Strudel, XML-QL, XQuery, XUpdate), the
semantics is directly defined in terms of data structures.

\paragraph{Rule-based vs.\ Logic Programming.}

All languages discussed above are \emph{rule-based} and
\emph{declarative}, generating variable bindings by a
matching/selection part in the ``rule body'' and then using these
bindings in the ``rule head'' for generating output or updating the
database. This rule-based nature is more or less explicit: F-Logic,
MSL/WSL (Tsimmis), Elog, and Xcerpt use the \mbox{``:-''} Prolog
syntax, whereas UnQL, Lorel, StruQL, XML-QL and XQuery/XUpdate cover
their rule-like structure in an SQL-like clause syntax.  These clausal
languages allow for a straightforward extension with update constructs
(as it has been done for Lorel and proposed with XUpdate for XQuery).
GraphLog and XML-GL use a graphical representation.

The first, ``pure'' group separates strictly between the selection
part in the rule body and generation/update part in the rule head,
whereas UnQL, StruQL, XML-QL and XQuery allow for nested
selection-generation parts in the rule bodies.

The global semantics of these languages is influenced by their
functionality, distinguishing between query/transformation and
query/update languages: UnQL, Xcerpt, XML-QL, and XQuery
\emph{generate} (output) structures in their head which are not feeded
back into the input or internal database.

Only MSL/WSL, Elog, and F-Logic allow to for \emph{additions to the
  database} or \emph{view definitions} (depending whether bottom-up or
top-down semantics is considered), and to use the derived facts in the
selection/matching part of other rules.  StruQL, XML-QL, and XQuery
overcome this restriction by nesting selection-generation parts in the
rule bodies. The \textsf{traverse} construct of UnQL (applying a
subquery by structural recursion to arbitrary depth) also comes near
to local view definitions.  Note that these languages require regular
path expressions to compute the transitive closure of a binary
relation (see \cite{fernandez-florescu-levy-suciu-SIGMODREC-97}). We
consider the ability to compute a transitive closure as an important
feature for a language for handling semistructured data (especially,
for a ``logic-programming'' language, since that is what makes the
distinction between Datalog and the relational algebra/calculus).

The difference between \emph{rule-based transformation languages} and
\emph{logic programming} languages is mirrored by the fact that the
semantics of UnQL, Xcerpt, XML-QL, and XQuery is completely given by
he semantics of their rules (qualifying them as \emph{rule-based}
languages). In contrast, the \emph{global} semantics of F-Logic and
Elog also requires the notions of the $T_P$ operator and of minimal or
well-founded models (qualifying them as \emph{logic programming}
languages).  As a consequence, they require both a
\emph{model-theoretic semantics}, and an \emph{answer semantics} for
queries.

\paragraph{Design Principles for XPathLog.}
XPathLog follows a logic-based approach which has been motivated by
the experiences with F-Logic: XML instances are mapped to a semantical
structure for interpreting \emph{XPath-Logic} formulas.  XPath-Logic
is based on (i) first-order logic, and (ii) XPath expressions extended
with variable bindings.  The Horn fragment of XPath-Logic, called
\emph{XPathLog}, provides a declarative, Datalog-style
logic-programming language for manipulation of XML documents.

\noindent
Regarding the above design principles, XPathLog is positioned as
follows:

\begin{itemize}
\item XPathLog is completely XPath-based (i.e., navigational access).
  Matching and generation/update part are strictly distinguished.
  
  An extended XPath syntax is used for querying (rule bodies)
  \emph{and} generating/manipulating data (rule heads). The rule body
  serves for generating variable bindings which are then used in the
  head for \emph{extending} the current XML database, thereby defining
  an update semantics for XPath expressions.
  
\item XPathLog uses an \emph{edge-labeled graph} model, which is
  advantageous when defining several tree views of the internal
  database. The data model is partly ordered like in XML: subelements
  are ordered, attributes are unordered.
  
\item XPathLog is a logic-programming language according to the above
  characterization.  It works on an abstract semantical model which
  represents an \emph{XML database} supporting multiple overlapping
  XML trees.  These \emph{X-Structures} together with the logic,
  called \emph{XPath-Logic}, provide for a logical characterization of
  XML data.  XPathLog combines the intuitive ``local'' semantics of
  addressing XML data by XPath with the appeal of the ``global''
  semantics of logic programming.  As an update language, it is based
  on a bottom-up semantics.
  
\item In contrast to XML-QL, XQuery, and XSLT, the language does not
  use additional constructs whose semantics has to be defined
  separately: the only semantic prerequisite is the bottom-up
  evaluation strategy of Datalog or any other logic programming
  language.

\end{itemize}

\noindent
In this paper, we describe the data model, its logical foundation,
the internal semantics of queries, rules, and programs of XPathLog as
a true logic programming language for XML.  Some aspects have been
published in \cite{may-DBPL-01,may-behrends-FMLDO-01} and with the
\LoPiX\ prototype in \cite{may-vldb-demo-01}.  Here, we focus on the
theoretical issues of modeling XML and the semantics of a language for
queries and basic, elementary updates.  A full report on XPathLog can
be found in \cite{may-habil-01}.

A possible application area for XPathLog is e.g.\ the integration of
XML data from several sources as done in the case study
\cite{lopix-mondial-01}.  Here, the power of the combination of XPath
expressions with additional variable bindings allows for short and
concise declarative and flexible rules.  Both, queries and rules for
manipulating and restructuring the internal XML database can be
expressed much easier than e.g.\ in XQuery (where update functionality
is still in a prototypical state).

\paragraph*{Structure of the paper.}
Section~\ref{sec-xpathlogic} defines X-Structures as semantical
structures which represent XML documents and presents XPath-Logic.
The answer semantics of XPathLog as an XML query language is
investigated in Section~\ref{sec-xpathlog}.
Section~\ref{sec-xpathlog-rules} defines the semantics of rule heads
for generating and modifying XML data, and the semantics and
evaluation of XPathLog programs.  The implementation in the \LoPiX\ 
(\emph{\textbf{Lo}gic \textbf{P}rogramming \textbf{i}n \textbf{X}ML})
system and a case study are described in Section~\ref{sec-lopix}.  An
analysis, a general discussion of related work, and the conclusion can
be found in Section~\ref{sec-related-and-conclusion}.  Additional
proofs can be found in Appendix~\ref{sec-proofs}.

\section{XPath-Logic: The Model-Theoretic Framework}\label{sec-xpathlogic}

XPath-Logic and its Horn fragment, XPathLog, extend XPath
\cite{XPath-W3C} with Datalog-style variable bindings. XPath-Logic
provides the model-theoretic framework for defining a global,
logic-programming style semantics for XPathLog.

XPath \cite{XPath-W3C} is the common language for addressing node
sets in XML documents. It is based on navigation through the XML tree
by \emph{path expressions} of the form
\textsf{\textsl{root}/\textsl{axisStep}/.../\textsl{axisStep}} where
\textsf{\textsl{root}} specifies a starting point of the expression
(the root of a document, or a variable that is bound to a node in an
XML instance).  Every \textsf{\emph{axisStep}} is of the form
\textsf{\textsl{axis}::\textsl{nodetest}[\textsl{qualifier}]*}. The
\emph{axes} define navigation directions in an XML tree: Given an
element $e$, the \textsf{child} axis contains all its children and the
attribute axis contains all its attributes.  Analogously,
\textsf{parent}, \textsf{ancestor}, \textsf{descendant},
\textsf{preceding-sibling} and \textsf{following-sibling} axes are
defined.  They enumerate the respective nodes by traversing the
document tree starting from $e$.

First, along the chosen axis, all elements which satisfy the
\textsf{\textsl{nodetest}} (which specifies the nodetype or an
elementtype which nodes should be considered) are selected; the
resulting list is called the \emph{context}. Then, the given
\textsf{\textsl{qualifier(s)}} (also called \textsf{\textsl{filters}})
are applied to each of the nodes (as the \emph{context node}) for
finer selection.  Inside qualifiers, \emph{relative location paths}
are allowed that implicitly start at the context node.  Starting with
this (local) result set, the next step expression is applied (for
details, see \cite{XPath-W3C} or
\cite{XQuery-Formal-Semantics-W3C}).  The most frequently used axes
are abbreviated as \textsf{\textsl{path}/nodetest} for
\textsf{\textsl{path}/child::\textsl{nodetest}}, \ 
\textsf{\textsl{path}/@\textsl{nodetest}} for
\textsf{\textsl{path}/attribute::\textsl{nodetest}}, and
\textsf{\textsl{path}//\textsl{nodetest}} for
\textsf{\textsl{path}/descendant-or-self::*/child::\textsl{nodetest}}.


\begin{Example}[XML, XPath, Result Sets]\label{ex-xml-instance}
  Consider the of the \Mondial\ database \cite{Mondial-www} for
  illustrations; the DTD is given as follows:

{\sf
\begin{tabbing} 
==\===\=\kill
 \flq!ELEMENT mondial (country+, organization+, \ldots)\frq \\
 \flq!ELEMENT country (name, population, encompassed+, border*, city+, \ldots)\frq \\
\> \flq!ATTLIST country \ \= car\_code ID \#REQUIRED 
 \ capital IDREFS \#REQUIRED  \\
\>\> memberships IDREFS \#IMPLIED\frq\\
\flq!ELEMENT name (\#PCDATA)\frq \\
\flq!ELEMENT encompassed EMPTY\frq \\ 
\> \flq!ATTLIST encompassed \ \= continent CDATA \#REQUIRED \\
\>\> percentage CDATA \#REQUIRED\frq  \\
\flq!ELEMENT border EMPTY\frq \\ 
\> \flq!ATTLIST border \ \= country IDREF \#REQUIRED \ \
 length CDATA \#REQUIRED\frq \\
\flq!ELEMENT city (name, population*)\frq \ 
 \flq!ATTLIST city country IDREF \#REQUIRED\frq \\
\flq!ELEMENT population (\#PCDATA)\frq \ 
 \flq!ATTLIST population year CDATA \#IMPLIED\frq\\
\flq!ELEMENT organization (name, abbrev, members+)\frq\\
\> \flq!ATTLIST organization \ \=id ID \#REQUIRED 
                             \ \ headq IDREF \#IMPLIED\frq \\
\flq!ELEMENT abbrev (\#PCDATA)\frq \\
\flq!ELEMENT members EMPTY\frq \\
\> \flq!ATTLIST members \ \=type CDATA \#REQUIRED 
                     \ country IDREFS \#REQUIRED\frq 
\end{tabbing}}

\noindent
An excerpt of the instance is given below (and depicted as a graph can
be found in Figure~\ref{fig-x-structure-example} when X-Structures are
considered).

{\sf\begin{tabbing} 
==\===\=\kill
\flq country car\_code=``B'' capital=``cty-Brussels''
                 memberships=``org-eu org-nato \ldots''\frq \\
\> \flq name\frq Belgium\flq/name\frq \ \
   \flq population\frq 10170241\flq/population\frq\\
\> \flq encompassed continent=``Europe'' percentage=``100''/\frq\\
\> \flq border country=``NL'' length=``450''/\frq \ \
   \flq border country=``D'' length=``167''/\frq\\
\>\>\smallvdots\\
\> \flq city id=``cty-Brussels'' country=``B''\frq\ \
     \flq name\frq Brussels\flq/name\frq\\
\>\> \flq population year=``95''\frq 951580\flq/population\frq \\
\> \flq/city\frq\\
\> ~~\smallvdots \\
\flq/country\frq
\\[0.1cm]
\flq country car\_code=``D'' capital=``cty-Berlin''
                 memberships=``org-eu org-nato \ldots''\frq \\
\>\smallvdots\\
\flq/country\frq
\\[0.1cm]
\flq organization id=``org-eu'' headq=``cty-Brussels''\frq \\
==\=\kill
\>\flq name\frq European Union\flq/name\frq\ \ 
  \flq abbrev\frq EU\flq/abbrev\frq \\
\>\flq members \=type=``member'' 
           country=``GR F E A D I B L \ldots''/\frq \\
\>\flq members type=``membership applicant'' 
           country=``AL CZ \ldots''/\frq \\
\flq/organization\frq \\[0.1cm]
\flq organization id=``org-nato'' headq=``cty-Brussels'' \ldots\frq \\
\> \smallvdots \\
\flq/organization\frq  
\end{tabbing}}

\noindent
The XPath expression
\begin{expl}
  //country[name]/city[population/text()$>$100000 and @zipcode]/name/text()
\end{expl}
returns all names of cities such that the city belongs (i.e., is a
subelement) to a country where a \textsf{name} subelement exists, the
city's population is higher than 100000, and its zipcode is known.
\end{Example}

\noindent
XPath is only an addressing mechanism, not a full query language.
It provides the base for most XML query languages, which extend it
with their special constructs (e.g., functional style in XSLT, and
SQL/OQL style (e.g., joins) in XQuery). In the case of XPath-Logic and
XPathLog, the extension feature are Prolog/Datalog style variable
bindings, joins, and rules.

\begin{Remark}[Relationship to W3C Documents]
  We restrict the considerations to the core concepts of XPath as an
  addressing and navigation formalism for XML data, i.e., stepwise
  navigation along the axes and step qualifiers/filters.  For the
  XPath syntax and non-formal semantics, we always refer to the
  \emph{W3C XPath 2.0 Working Draft} \cite{XPath-W3C}. Note that the
  syntax and semantics of the core concepts of XPath is the same as in
  XQL \cite{XQL-W3C-99}, XPointer, early drafts of XPath, XPath 1.0
  \cite{XPath-W3C}, although both the presentation and the naming have
  been changed several times.
  
  A formal semantics of XPath has been given as a \emph{denotational}
  semantics in \cite{wadler-misc-99} that already covers these central
  notions of XPath. Later, it has been re-formulated first in the
  \emph{W3C XML Query Algebra} \cite{XML-Query-Algebra-W3C} and then
  in the \emph{W3C Query Formal Semantics}
  \cite{XQuery-Formal-Semantics-W3C} where a description in terms of
  type inference rules and value inference rules is given.  Note that
  since the early implementations (e.g., xt \cite{XT}), the actual
  semantics of XSL Patterns/XPath as its ``behavior'' has not changed.
  For comparing our approach with the formal semantics of XPath, we
  refer to \cite{wadler-misc-99} which gives a short and concise
  definition of the central concepts that is best suited as a
  reference.
\end{Remark}

\subsection{Syntax of XPath-Logic}\label{sec-xpathlog-syntax}

Inspired by the derivation of F-Logic from first-order logic as a
logic for dealing with structures containing complex objects,
XPath-Logic is defined for expressing properties of XML structures.
The main difference between XPath-Logic and first-order logic is that
XPath-Logic has an additional type of atomic formulas: \emph{reference
  expressions} which turn out to be a special kind of predicates with
a built-in semantics. The ``basic'' components of the language are
XPath-Logic \emph{PathExpressions} which are syntactically derived
from XPath's \emph{PathExpressions} by extending Path with
Prolog/Datalog style variable bindings.

\begin{Definition}[XPath-Logic: Syntax]\label{def-xpathlog-syntax-2}
  The set of basic formulas of an XPath-Logic language is defined 
  as follows:

  \begin{itemize}
  \item every language contains an infinite set $\Var$ of variables.  
  \item a specific XPath-Logic language is given by its
    \emph{signature} $\Sigma$ of element names, attribute
    names, function names, constant symbols, and predicate names.
  \item \emph{XPath-Logic reference expressions} over the above names
    extend the XPath \emph{path expressions}: The syntax of
    \emph{AxisSteps},
    \textsf{\textsl{axis}::\textsl{name}[\textsl{stepQualifier}]$^*$},
    may be extended to bind the selected nodes to variables by
    ``\texttt -$\frq$ Var'':
      \begin{verbatim}
Step ::= Axis "::" NodeTest StepQualifiers 
         | Axis "::" NodeTest StepQualifiers "->" Var StepQualifiers
         | Axis "::" Var StepQualifiers
         | Axis "::" Var StepQualifiers "->" Var StepQualifiers\end{verbatim}
    For an XPath-Logic reference expression, the \emph{underlying
      XPath expression} is obtained by removing the inserted variable
    binding constructs.
  \item An \emph{XPath-Logic predicate} is a predicate over reference
    expressions.
  \item \emph{terms} and \emph{atomic formulas} are defined
    analogously to first-order logic.
  \item XPath-Logic \emph{compound formulas} are built over predicates
    and reference expressions, using $\land$, $\lor$, $\neg$,
    $\exists$, and $\forall$.
  \item XPath-Logic allows to have formulas in step qualifiers.
\end{itemize}
\end{Definition}
Note that XPath-Logic does not use the explicit dereference operator
\textsf{$\Fd$} from XPath 2.0; instead implicit dereferencing of
attributes in paths is supported.\footnote{XPath-Logic has been
  designed before XPath 2.0 replaced XPath's \textsf{id(.)}  function
  by the dereferencing operator. Furthermore, we use a data model that
  directly incorporates references.}

\noindent
The goal of the paper is to introduce the Horn fragment from
XPath-Logic, called XPathLog as a Datalog-style XML query and
manipulation language. The following example gives some XPathLog
queries that review the basic XPath constructs, and illustrate the use
of the additional variable binding syntax.

\begin{Example}[XPathLog: Introductory Queries]
The following examples are evaluated against the \Mondial\ database.
\begin{description}
\item[Pure XPath expressions:] pure XPath expressions (i.e., without
  variables) are interpreted as \emph{existential queries} which
  return \textsf{true} if the result set is non-empty:
  \begin{expl}
    ?- //country[name/text() = ``Belgium'']//city/name/text(). \\
    true
  \end{expl}
  since the \textsf{country} element which has a \textsf{name}
  subelement with the text contents ``Belgium'' contains at least one
  \textsf{city} descendant with a \textsf{name} subelement with
  non-empty text contents.
\item[Output Result Set:] The query ``\textsf{?- \textsl{xpath}\fd
    N}'' for any XPath expression \textsf{\textsl{xpath}} binds $N$ to
  all nodes belonging to the result set of \textsf{\textsl{xpath}}:
  \begin{expl}
    ?- //country[name/text() = ``Belgium'']//city/name/text()\fd N. \\
    N/``Brussels''\\
    N/``Antwerp''\\
    \smallvdots
  \end{expl}
  respectively, for a result set consisting of elements, logical ids
  are returned:
  \begin{expl}
    ?- //country[name/text() = ``Belgium'']//city\fd C. \\
    C/\textsl{brussels} \\
    C/\textsl{antwerp} \\
    \smallvdots
  \end{expl}
\item[Additional Variables:] XPathLog allows to bind all nodes which
  are traversed by an expression:
  The following expression returns all tuples $(N_1,C,N_2)$ such that
  the city with name $N_2$ belongs to the country with name $N_1$ and
  car code $C$:
  \begin{expl}
  ?- //country[name/text()\fd N1 and @car\_code\fd C]%
        //city/name/text()\fd N2. \\
  N2/``Brussels''   C/``B''   N1/``Belgium'' \\  
  N2/``Antwerp''   C/``B''   N1/``Belgium''  \\
  \hspace*{0.5cm}\smallvdots\\
  N2/``Berlin''   C/``D''   N1/``Germany''  \\
  \hspace*{0.5cm}\smallvdots
  \end{expl}
  
\item[Dereferencing IDREF Attributes:] For every organization, give
  the name of city where the headquarter is located and all names and
  types of members: {\sf\begin{tabbing}
      ?- //organization\=[\= name/text()\fd N and abbrev/text()\fd A and \\
      \>\> @headq/name/text()\fd  SN]\\
      \>/members[@type\fd MT]/@country/name/text()\fd MN.
\end{tabbing}}
One element of the result set is e.g.,
\begin{expl}
  N/``\ldots'' ~~ A/``EU'' ~~ SN/``Brussels'' ~~ 
  MT/``member'' ~~ MN/``Belgium''   
\end{expl}

\item[Schema Querying:] The use of variables at name positions allows
  for schema querying, e.g., to give all names of subelements of
  elements of type \textsf{city}:
  \begin{expl}
    ?- //city/\fbox{SubElName}. \\
    SubElName/name \\
    SubElName/population \\
    \smallvdots
  \end{expl}

\item[Navigation Variables:] Search for all things that have the name
  ``Monaco''. More explicitly, give the element type of all elements
  that have a name subelement with the text contents ``Monaco'':
  \begin{expl}
  ?- //\fbox{Type}\fd X[name/text()\fd``Monaco''].\\
  ~~~ \begin{tabular}{ll}
      Type/country \ & X/\textsl{country-monaco} \\
      Type/city \ & X/\textsl{city-monaco} 
    \end{tabular}
  \end{expl}
\end{description}
\end{Example}

\noindent
Closed XPath-Logic formulas can e.g.\ be used for expressing integrity
constraints.

\begin{Example}[Integrity Constraints]%
\label{ex-xpathlogic-integrity-constraints}
There are some application-specific integrity constraints on the
\Mondial\ database:

\begin{description}
\item[Range restrictions:] The text contents of population elements and 
  the value of area attributes must be a non-negative number:
  \begin{expl}
    $\forall$ X: ((//population/text()\fd X or //@area\fd X) 
                    $\leadsto$ X $\ge$ 0).
  \end{expl}
  The sum of percentages of ethnic groups in a country is at most 100\%:
  \begin{expl}
    $\forall$ C: (//country\fd C $\leadsto$
                   sum\{N [C]; C/ethnicgroups/@percentage\fd N\} $\le$ 100).
  \end{expl}
\item[Bidirectional relationships:]
  The membership of countries in organizations is
  represented bidirectionally:
  \begin{expl}
  $\forall$ C,O: (&%
     //country\fd C[@memberships\fd O] $\iff$ \\
     & $\exists$ T: //organization\fd O/members[@type\fd T and @country\fd C]).
  \end{expl}
\item[Other conditions:] The \textsf{country} attribute of
  \textsf{border} subelements of \textsf{country} elements must
  reference a country which is encompassed by the same continent:
  \begin{expl}  
  $\forall$ C,C2: %
    \begin{tabular}[t]{@{}l}
      (//country\fd C/border[@country\fd C2] $\leadsto$ \\
      ~~(\begin{tabular}[t]{@{}l}
       //country\fd C2 and 
       $\exists$ Cont: (\begin{tabular}[t]{@{}l}
           C/encompassed/@continent\fd Cont and \\
           C2/encompassed/@continent\fd Cont))).
         \end{tabular}
       \end{tabular}
     \end{tabular}
   \end{expl}
\end{description}
\end{Example}

\subsection{XML Instances as Semantical Structures}\label{sec-x-structures}

Next, we need a basis for a model-theoretic semantics of XPath-Logic.
The \emph{information} that is carried by an XML instance is
\emph{abstractly} defined in the \emph{XML Information Set}
specification \cite{XML-Infoset-W3C}. It can be represented in
different ways -- e.g., as the \emph{human-readable} ASCII-based
notation, or by using the DOM \cite{DOM-W3C} that provides an abstract
datatype for \emph{implementations}. There are approaches that regard
XML trees as database items where the languages operate on (UnQL,
Xcerpt). In our approach, the atomic items are the edges of a
\emph{graph} (than can be an XML tree, but that can also represent
overlapping tree views on an internal graph-like database), called
\emph{XTreeGraph}.  In contrast to the DOM model and the XML Query
Data Model \cite{XML-Query-Data-Model-W3C} which use a node-labeled
tree (i.e., the element and attribute names are associated with the
nodes), the XTreeGraph is an \emph{edge-labeled} model.  Using an
edge-labeled model proves useful for data manipulation and integration
(see \cite{may-behrends-FMLDO-01}).  Recall that XML-QL
\cite{deutsch-fernandez-florescu-etal-WWW-99} also uses an
\emph{edge-labeled} graph which especially defines the same handling
of text contents as ours; influenced by the experiences with the
\textsc{Strudel/StruQL} \cite{fernandez-florescu-kang-SIGMOD-98}
project for data integration.

Formally, the XTreeGraph is represented by an \emph{X-Structure} (that
interprets a signature consisting of element and attribute names,
similar to a first order structure).  The advantage of that approach
is that it allows for \emph{manipulating} an internal database by
adding edges to the graph.  Thus, XPathLog is not only a query
language, but also a manipulation language.  Its rule heads do not
necessarily \emph{construct} new XML trees/terms, but can update the
X-Structure.  As a prerequisite for mapping XML instances to
X-Structures, some notation for handling lists is needed:

\begin{Notation}[Lists]
  Throughout this work, the following usual notation is used:
\begin{itemize}
\item For two sets $A$ and $B$, the set of mappings from $A$ to $B$
  is denoted by $B^A$.
\item A list over a domain $D$ is a mapping from $\N$ to $D$. Thus, 
  the set of lists over $D$ is denoted by $D^\N$.
\item the empty list is denoted by $\varepsilon$; a unary list 
  containing only the element $x$ is denoted by $(x)$;
  list concatenation as an operator is denoted by $\cons$.
\item $\textsf{set}
     ( expr_1(x_1,\ldots,x_n) \st expr_2(x_1,\ldots,x_n))$ \
  stands for 
  \[ \{expr_1(x_1,\ldots,x_n) \st expr_2(x_1,\ldots,x_n)\} \]
  (i.e., the set of all $expr_1(x_1,\ldots,x_n)$ such that the
  condition $expr_2(x_1,\ldots,x_n)$ holds).
  In the following, sets are sometimes used as lists exploiting 
  the fact that a set can be seen as a list by an arbitrary
  enumeration.
\item In a similar way, a list can be constructed by enumerating its
  elements.  For a list $\ell = (i_1,i_2,\ldots)$, \ 
  $\textsf{list}_{i\in \ell} ( expr_1(i) \st expr_2(i))$ \
  is the list of all $expr_1(i_j)$ where $expr_2(i_j)$ holds. \
  Similar to \textsf{list}, \ 
  $\textsf{concat}_{i\in I} ( expr_1(i) \st expr_2(i))$ \
  does the same if $expr_1(i)$ is already a list.
\item For a finite list $\ell = (x_1,\ldots,x_n)$, 
  $\textsf{reverse}(\ell) = (x_n,\ldots,x_1)$. 
\item For a list $\ell$, $\ell[i,j]$ denotes the sublist that consists
  of the $i$th to $j$th elements,
\item For a list $\ell$ of pairs i.e.,   
  $\ell = ((x_1,y_1), (x_2,y_2),\ldots)$,
  $\ell\downarrow_1$ denotes the projection of
  the list to the first component of the list elements, i.e.,
  $\ell\downarrow_1 := (x_1,x_2,\ldots)$.
\end{itemize}
\end{Notation}

\paragraph*{X-Structures.}
When representing XML instances as X-Structures, (i) their
elements/subelement structure, and (ii) the elements' attributes have
to be represented. The universe consists of the \emph{element nodes}
of the XML instance and the \emph{literals} used as attribute values
and text contents. \emph{Element nodes} have properties, defined by
(i) subelements (which are ordered) and (ii) attributes (which are
unordered). Multivalued attributes (\texttt{NMTOKENS} and
\texttt{IDREFS}) are silently split, and reference attributes are
silently resolved.  Additionally, X-Structures support named constants
and predicates as known from first-order logic.

\noindent
Each XML instance is represented as a structure with a universe $\pl
U$ over a signature $\Sigma=(\Sigma_N,\Sigma_F,\Sigma_C,\Sigma_P)$
which consists of
\begin{itemize}
\item $\Sigma_N$: element names and attribute names,
\item $\Sigma_F$: names of XML-built-in functions,
\item $\Sigma_C$: constant symbols, denoting elements in the XML
  instance (e.g., \textsf{germany}, interpreted as
  the element addressed by
  \textsf{/mondial/country[name=``Germany'']}).
\item $\Sigma_P$: predicates (with arity).
\item Additionally, a basic set of literal constants is assumed.
\end{itemize}

\noindent
An X-Structure contains only the basic facts about the XML tree, i.e.,
the child and attribute relationships (similar to the DOM).  Note that
our approach which associates the order with the children of elements,
differs from e.g., the DOM and XML-QL approaches where a \emph{global}
order of all elements is assumed.

\begin{Definition}[X-Structure]\label{def-X-structure}
  An X-Structure over a given signature $\Sigma$ is a tuple \
  $\pl X = (\pl V
            , \pl L
            , \pl N
            , \pl I%
            , \pl E
            , \pl A
)$ 
where the universe $\pl U$ consists of three sets $\pl V$, $\pl L$,
and $\pl N$: $\pl V$ is a set of nodes (from the graph point of view,
vertices), identified by internal names, $\pl L$ is a set of literals
(integers, floats, strings), $\pl N$ is the set of names (as e.g.,
occurring in node tests). Names may be further distinguished into $\pl
N_\pl E$, containing the element names (and a special element
\textsf{text()} for handling text children), and $\pl N_\pl A$
containing the attribute names.

\begin{itemize}
\item 
  $\pl I$ is a (partial) mapping, which interprets the signature: \
  $\pl I_\pl E : \ \Sigma_N \to \pl N_\pl E$ \ and \ 
  $\pl I_\pl A : \ \Sigma_N \to \pl N_\pl A$
  interpret the names in $\Sigma_N$ by element and attribute names.
  $\pl I_C : \Sigma_C \to \pl V $ \ interprets the constant symbols
  in $\Sigma_C$ by nodes in $\pl V$, and \\
  $\pl I_F : \ \pl V \times \Sigma_F \times 
       (\pl V \cup \pl L \cup \pl N)^*
        \to \pl V \cup \pl L \cup \pl N$ \
  represents the interpretation of built-in functions (as defined in
   \cite{XPath-XQuery-Functions-Operators-W3C}). Finally, \
  $\pl I_P\ : \ \ \Sigma_P \times (\pl V \cup \pl L \cup \pl N)^*
            \to \{\textsl{true},\textsl{false}\}$ \
  represents the interpretation of predicates.
\item $\pl E$ is a (partial) mapping \ $\pl E : \ \pl V \times \N
  \times \pl N_\pl E \to \pl V \cup \pl L$ \ (subelement relationship
  and text contents; from the graph point of view, an ordered set of
  edges).
\item $\pl A$ is a (partial) mapping \ $\pl A : \ \pl V \times \pl
  N_\pl A \to 2^\pl V \cup 2^\pl L$ \ (attribute values).  \emph{XML
    attribute nodes} do not belong to $\pl V$, but their literal
  values belong to $\pl L$. For reference attributes (\texttt{IDREF}),
  the ``results'' are not the \texttt{ID}-strings, but the target
  nodes in $\pl V$ themselves.
\end{itemize}
\end{Definition}
  
\noindent
Note that $\pl E$ and $\pl A$ are not direct interpretations of
$\Sigma$, but mappings that ``interprete'' $\pl N$.  $\Sigma$ is
mapped to $\pl N$ before being interpreted by $\pl I$, making
attribute and element names full citizens of the language (as, e.g.,
in F-Logic).

\begin{figure}[htbp]
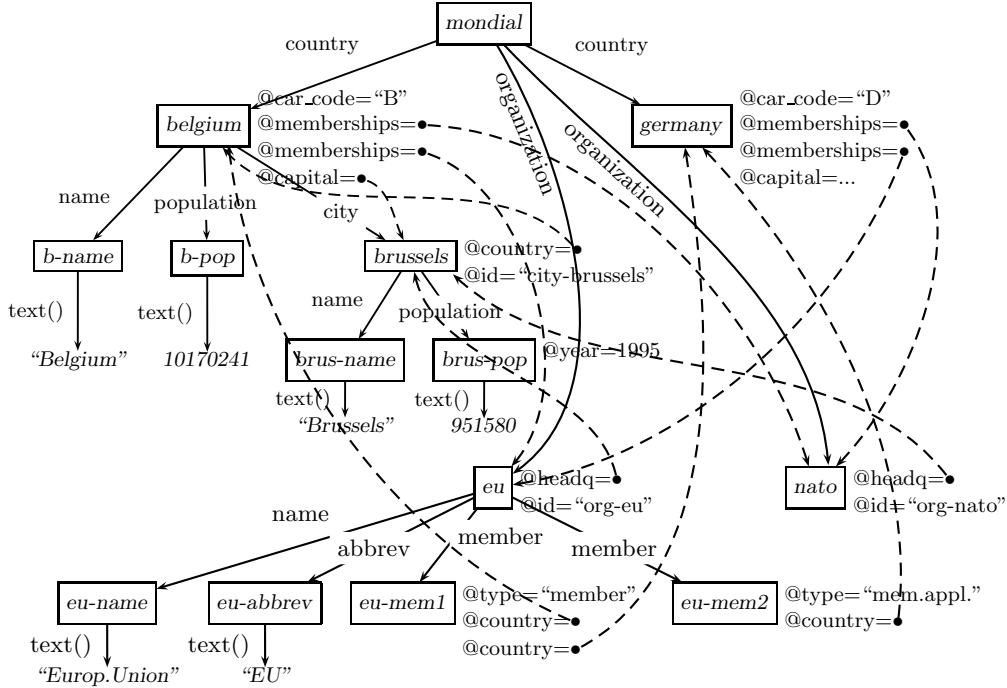

  \scalebox{0.9 0.9}{%
  \begin{tabular}{c}
   \begin{tabular}{ccccc}
     &&& \rnode{mondial}{\fbox{\textsl{\strut mondial}}}
      \\[1cm]
     & \rnode{belgium}{\fbox{\textsl{\strut belgium}}} 
           \rput[l]{0}(0,.5){\small @car\_code=``B''}%
           \rput[l]{0}(0,.1){\small @memberships=\rnode{mem-b-nato}{$\bullet$}}%
           \rput[l]{0}(0,-.3){\small @memberships=\rnode{mem-b-eu}{$\bullet$}}%
           \rput[l]{0}(0,-.7){\small @capital=\rnode{b-cap}{$\bullet$}}%
     &&& \hspace*{1cm}\rnode{germany}{\fbox{\textsl{\strut germany}}} 
           \rput[l]{0}(0,.5){\small @car\_code=``D''}%
           \rput[l]{0}(0,.1){\small @memberships=\rnode{mem-d-nato}{$\bullet$}}%
           \rput[l]{0}(0,-.3){\small @memberships=\rnode{mem-d-eu}{$\bullet$}}%
           \rput[l]{0}(0,-.7){\small @capital=...}%
     \\[1.5cm]
     \rnode{bname}{\fbox{\textsl{b-name}}}%
     & \rnode{bpop}{\fbox{\textsl{b-pop}}}%
     & \multicolumn{2}{c}{\rnode{brussels}{\fbox{\textsl{brussels}}}%
           \rput[l]{0}(0.1,.2){\small @country=\rnode{refB}{$\bullet$}}%
           \rput[l]{0}(0.1,-.2){\small @id=``city-brussels''}}
     \\[1cm]
     \rnode{bnamestring}{\textsl{``Belgium''}}
     &\rnode{bpopstring}{\textsl{10170241}}
     &\rnode{brusname}{\fbox{\textsl{\strut brus-name}}}
     &\rnode{bruspop}{\fbox{\textsl{\strut brus-pop}}}%
          \rput[l]{0}(0.1,.2){\small @year=1995} 
             \\[0.6cm]
     &
     &\rnode{brusstring}{\textsl{``Brussels''}}
     &\rnode{bruspopnum}{\textsl{951580}}\\~
   \end{tabular}\hfill~
  \nodeconnections{%
    \ncline{->}{mondial}{belgium}\nbput{\small country}
    \ncline{->}{mondial}{germany}\naput{\small country}
    \ncline{->}{belgium}{bname}\nbput[npos=0.7]{\small name}
    \ncline{->}{belgium}{bpop}\ncput*[npos=0.6]{\small population}
    \ncline{->}{bname}{bnamestring}\nbput{\small text()}
    \ncline{->}{bpop}{bpopstring}\nbput{\small text()}
    \ncline{->}{belgium}{brussels}\ncput*[npos=0.7]{\small city}
    \ncline{->}{brussels}{brusname}\nbput[npos=0.6]{\small name}
    \ncline{->}{brussels}{bruspop}\ncput*[npos=0.6]{\small population}
    \ncline{->}{brusname}{brusstring}\nbput{\small text()}
    \ncline{->}{bruspop}{bruspopnum}\nbput{\small text()}
    \ncarc[linestyle=dashed,angleB=40,arcangleA=-30,arcangleB=30]
            {->}{refB}{belgium}
    \ncarc[linestyle=dashed, angleA=0,arcangleA=40]{->}{b-cap}{brussels} }
    \\
   ~\hfill\begin{tabular}{cccc}
     && \rnode{eu}{\fbox{\textsl{\strut eu}}}
           \rput[l]{0}(0,.2){\small @headq=\rnode{eu-headq}{$\bullet$}}
           \rput[l]{0}(0,-.2){\small @id=``org-eu''}
     & \rnode{nato}{\fbox{\textsl{\strut nato}}}  
           \rput[l]{0}(0,.2){\small @headq=\rnode{nato-headq}{$\bullet$}}
           \rput[l]{0}(0,-.2){\small @id=``org-nato''}
      \\[1.2cm]
     \rnode{euname}{\fbox{\textsl{\strut eu-name}}}%
     & \rnode{euabbrev}{\fbox{\textsl{\strut eu-abbrev}}}%
     & \rnode{eumem1}{\fbox{\textsl{\strut eu-mem1}}}%
           \rput[l]{0}(0.1,.2){\small @type=``member''}
           \rput[l]{0}(0.1,-.2){\small @country=\rnode{mB}{$\bullet$}}
           \rput[l]{0}(0.1,-.6){\small @country=\rnode{mD}{$\bullet$}}
      \phantom{@country=MMMM}
     & \rnode{eumem2}{\fbox{\textsl{\strut eu-mem2}}}%
           \rput[l]{0}(0.1,.2){\small @type=``mem.appl.''} 
           \rput[l]{0}(0.1,-.2){\small @country=\rnode{mD2}{$\bullet$}}
      \phantom{@country=MMMM}
     \\[0.7cm]
     \rnode{eunamestring}{\textsl{``Europ.Union''}}
     & \rnode{euabbrevstring}{\textsl{``EU''}}
   \end{tabular}%
  \end{tabular}}%
  \nodeconnections{%
    \ncarc[angleA=-45,arcangleA=30,arcangleB=55]{->}{mondial}{eu}
              \nbput[labelsep=1pt,npos=0.2,nrot=:U]{\small organization}
    \ncarc[angleA=-40,arcangleB=30,offsetB=0.2cm]
         {->}{mondial}{nato}\nbput[labelsep=1pt,npos=0.3,nrot=:U]{\small organization}
    \ncline{->}{eu}{euname}\nbput{\small name}
    \ncline{->}{eu}{euabbrev}\ncput*[npos=0.6]{\small abbrev}
    \ncline{->}{eu}{eumem1}\naput*[labelsep=-2pt,npos=0.3]{\small member}
    \ncline{->}{eu}{eumem2}\ncput*[npos=0.6]{\small member}
    \ncline{->}{euname}{eunamestring}\nbput{\small text()}
    \ncline{->}{euabbrev}{euabbrevstring}\nbput{\small text()}
    \ncarc[linestyle=dashed,arcangle=30,offsetB=-0.3cm]{->}{mB}{belgium}
    \ncarc[linestyle=dashed,arcangleA=-50,arcangleB=-20]{->}{mD}{germany}
    \ncarc[linestyle=dashed,arcangleA=-30,arcangleB=-20]{->}{mD2}{germany}
    \ncarc[linestyle=dashed,arcangleA=40,arcangleB=30]{->}{mem-b-nato}{nato}
    \ncarc[linestyle=dashed,arcangleA=60,arcangleB=52]{->}{mem-b-eu}{eu}
    \ncarc[linestyle=dashed,arcangleA=50,arcangleB=30]{->}{mem-d-nato}{nato}
    \ncarc[linestyle=dashed,arcangleA=30,arcangleB=30]{->}{mem-d-eu}{eu}
    \ncarc[linestyle=dashed,arcangleA=-30,arcangleB=30]{->}{eu-headq}{brussels}
    \ncarc[linestyle=dashed,nodesepB=0.2cm,offsetB=-.2cm,arcangleA=-30,arcangleB=20]
              {->}{nato-headq}{brussels}}%
  \caption{Example X-Structure}
  \label{fig-x-structure-example}
\end{figure}

There is a canonical mapping from the set of XML instances to the set
of X-Structures.  The canonical X-Structure to an XML instance is a
single XML tree (cf.\ Figure~\ref{fig-x-structure-example}), covering
the DOM model.

\begin{Example}\label{ex-X-structure}
  Figure~\ref{fig-x-structure-example} shows the X-Structure of the
  running example given in Example~\ref{ex-xml-instance}.
\end{Example}

\noindent
The elements of $\pl V$ (representing the element nodes) do not carry
information in themselves, they are only of interest as anonymous
entities (similar to object ids) which have certain properties that
are given by $\pl E$, $\pl A$, and $\pl I$.  In the following,
mnemonic ids (e.g., \textsf{\emph{germany}}) are used for elements of
$\pl V$.  Also, $\Sigma_N$ is identified with $\pl N_\pl E$ and $\pl
N_\pl A$, omitting $\pl I_\pl E$ and $\pl I_\pl A$.

In full generality, an X-Structure can also contain subelement edges
and reference edges which are not conforming with the XML tree model,
but which are crucial for data integration: An element may be a
subelement of several other elements (as we will show in
Section~\ref{sec-left-hand-side} and Figure~\ref{fig-linking-2}), even
with different names of the subelement relationship (``overlapping
trees'' -- thus, the term \emph{XTreeGraph}
\cite{may-behrends-FMLDO-01} for the abstract data model). 

In the following, X-Structures serve for defining a semantics for
XPath-Logic, using the same terms as for XPath.

\begin{Definition}[Basic Result Sets: Axes]\label{def-result-sets-axes}
  For every node $x$ in an X-Structure $\pl X$ and every \emph{axis}
  $a$ as defined in XPath,
  \[ \pl A_{\pl X}(a,x) \in ((\pl V \cup \pl L) \times \pl N)^\N \]
  is the list of pairs $(value,name)$ generated by axis $a$ with $x$
  as context node (do not confuse $\pl A_{\pl X}$ with $\pl A$ which
  denotes the interpretation of attributes in $\pl X$).

\[\begin{array}{lll}
  \pl A_{\pl X}(\textsf{child},x) &:=& 
    \textsf{list}_{i\in \N}((y,name) \st \pl E(x,i,name) = y) \\
  \pl A_{\pl X}(\textsf{attribute},x) &:=&
    \textsf{list}((y,name) \st y \in \pl A(x,name))
    \quad \mbox{by some enumeration.}
  \end{array}
\]

\noindent
For the other axes, $\pl A_{\pl X}(a,x)$ is derived according to the
XPath specification:

\[\begin{array}{@{}lll}
   \pl A_{\pl X}(\textsf{parent},x) &:=& 
     \textsf{set}((p,\pl I(p,\textsf{name},())) \st  
       \mbox{ $x\in \pl A_{\pl X}(\textsf{child},p)\downarrow_1$}) \\
   \pl A_{\pl X}(\textsf{preceding-sibling},x) &:=& \\
   \multicolumn{3}{l}{\hspace*{1cm}
     \textsf{concat}_{p \in \pl A_{\pl X}({\sf parent},x)\downarrow_1}
       (\textsf{reverse}(\pl A_{\pl X}(\textsf{child},p)[1,i-1]) \st
              x =  \pl A_{\pl X}(\textsf{child},p)[i])}
   \\
   \pl A_{\pl X}(\textsf{following-sibling},x) &:=& \\
   \multicolumn{3}{l}{\hspace*{1cm}
     \textsf{concat}_{p \in \pl A_{\pl X}({\sf parent},x)\downarrow_1}
       (\pl A_{\pl X}(\textsf{child},p)[i+1,\textsf{last()}] \st 
              x =  \pl A_{\pl X}(\textsf{child},p)[i])} 
   \\
   \pl A_{\pl X}(\textsf{ancestor},x) &:=& 
     \textsf{concat}_{(p,n) \in \pl A_{\pl X}({\sf parent},x)}
       (((p,n))\ \cons\ \pl A_{\pl X}(\textsf{ancestor},p))  
   \\
   \pl A_{\pl X}(\textsf{descendant},x) &:=&
         \textsf{concat}_{(c,n) \in \pl A_{\pl X}({\sf child},x)}
                  ((c,n) \cons \pl A_{\pl X}(\textsf{descendant},c))
 \end{array}
\]
\end{Definition}

\begin{Remark}
  Recall that in the node-labeled XML/XPath data model, the semantics
  of expressions is always a list or a set of labeled nodes.  In
  contrast, $\pl A_{\pl X}$ does not return a list of (labeled) nodes,
  but a list of pairs $(node/literal, name)$, according to the
  edge-labeled data model underlying our approach.
\end{Remark}

\subsection{Semantics}\label{sec-xpathlogic-semantics}

The semantics of XPath-Logic is defined similar to that of first-order
logic by induction over the structure of expressions and formulas.
The main task here is to define the semantics of reference
expressions, handling navigation, order, and filtering.  A reference
expression simultaneously acts as a term (it has a result (list) and
can be compared to terms) and as a predicate (when used in a step
qualifier). 

The basic result lists are provided by $\pl A_{\pl X}(axis, v)$ for
every node $v$ of $\pl X$; recall that $\pl A_{\pl
  X}(\textsf{attribute},x)$ contains literals in case of non-reference
attributes, and element nodes in case of reference attributes.

\subsubsection{Semantics of Expressions}
\newcounter{mycounter}

As for first-order logic, a \emph{variable assignment} \ $\bel: \Var
\to \pl U$ \ maps variables to elements of the universe $\pl U$
(nodes, literals, and names) of the underlying X-Structure.  For a
variable assignment $\bel$, a variable $x$, and $d \in \pl U$, the
\emph{modified} variable assignment $\bel^d_x$ is identical with
$\bel$ except that it assigns $d$ to the variable $x$:
\[ \bel^d_x: \Var \to \pl U :
     \left\{
     \begin{array}{ll}
       y \mapsto \bel(y)  & \mbox{ if } y \neq x ~, \\
       x \mapsto d        & \mbox{ otherwise.}
     \end{array} \right. \]
For $\bel$ as above, and a variable $x$,
$\bel\setminus\{x\}$ denotes $\bel$ without the mapping for $x$.

Expressions are decomposed into their \emph{axis steps}. Every step
consists of choosing an axis, preselecting nodes by a \emph{node
  test}, and filtering the result by (i) ``normal'' predicates and
(ii) XPath \emph{context functions} (e.g., \textsf{position()} and
\textsf{last()}) which use the order of the intermediate result list
for selecting a certain element by its index.

\begin{Definition}[Semantics of XPath-Logic expressions]
\label{def-semantics-xpathlogic}
The semantics is defined by operators $\pl S$ and $\pl Q$ which are
derived from the formal semantics given in \cite{wadler-misc-99}.

\begin{itemize}
\item $\pl S_{\pl X}:$
   $\begin{array}[t]{l}
   \mbox{Reference\_Expressions} 
         \to (\pl V\cup\pl L\cup\pl N)^\N ~~, \mbox{ and}\\
   (\mbox{Axes} \times \mbox{Reference\_Expressions} \times \pl V
    \times \mbox{Var\_Assignments}) \to (\pl V\cup\pl L\cup\pl N)^\N
  \end{array}$ 
  \\
  evaluates reference expressions wrt.\ an axis, a context node, and a
  variable assignment and returns a result list.  In the second case,
  we use $any$ to denote that the actual value of the node does not
  matter, and we use $\pl S^{any}$ to denote that the actual value of
  $axis$ does not matter.
\item $\pl Q_{\pl X}$:
  $(\mbox{Predicate\_Expressions} \times \pl V
         \times\mbox{Var\_Assignments}) \to \mbox{Boolean}$ 
  \\
  evaluates step qualifiers wrt.\ a context node and a variable
  assignment.
\end{itemize}

\paragraph{Reference Expressions are evaluated by $\pl{S}$:}
\begin{enumerate}
\item For closed expressions, \
  $\pl S_{\pl X}({\it refExpr}) = 
        \pl S^{any}_{\pl X}({\it refExpr},any,\emptyset)$.
\item\label{def-semantics-xpathlogic-item-entry} 
  Reference expressions are translated into path expressions wrt.\
  a start node:
  \begin{itemize}
  \item rooted paths: \
  $\pl S^{any}_{\pl X}(/p,any,\bel) = \pl S^{any}_{\pl X}(p,root,\bel)$
  where $root$ is as follows:
  \begin{itemize}
  \item the unique root node if only one XML document is currently
    stored,
  \item the root node that has been used in the outer expression, if
    $/p$ occurs in an expression of the form $path[/p]$.
  \end{itemize}
\item rooted paths in other documents: \\
  \hspace*{1cm} 
    $\pl S^{any}_{\pl X}(\textsf{document(``http://\ldots'')}/p,any,\bel) = 
      \pl S^{any}_{\pl X}(p,root,\bel)$ \\
  where $root$ is the root node of the document
  stored at \textsf{http://\ldots}.
\item entry points specified by a constant $c$: \ $\pl S^{any}_{\pl
    X}(c/p,any,\bel) =
  \pl S^{any}_{\pl X}(p,\pl I_C(c),\bel)$ \\
  (this is mainly of interest when multiple documents are used and
  constants are associated with their roots or some nodes, see
  Section~\ref{sec-left-hand-side}).
\item  entry points specified by variables $v\in\Var$: \
  $\pl S^{any}_{\pl X}(v/p,any,\bel) = 
      \pl S^{any}_{\pl X}(p,\bel(v),\bel)$
\end{itemize}

\item Axis steps: \
 $\pl S^{any}_{\pl X}(axis::pattern,x,\bel) =
                 \pl S^{axis}_{\pl X}(pattern,x,\bel)$ \\
 where $pattern$ is of the form \textit{nodetest} \textit{remainder} 
 where \textit{remainder} is a sequence of step qualifiers and variable
 bindings. These are evaluated left to right, always applying the
 rightmost ``operation'' (step qualifier or variable) to the result of the
 left part:
\item Node test: \ 
  $\begin{array}[t]{lcl}
    \pl S^a_{\pl X}(name,x,\bel) &=& 
      \textsf{list}_{(v,n) \in \pl A_{\pl X}(a,x)} (v \st n = name)  \\
    \pl S^a_{\pl X}(\textsf{node()},x,\bel) &=& 
      \textsf{list}_{(v,n) \in \pl A_{\pl X}(a,x)} (v \st v \in \pl V)  \\
    \pl S^a_{\pl X}(\textsf{text()},x,\bel) &=& 
      \textsf{list}_{(v,n) \in \pl A_{\pl X}(a,x)} (v \st v \in \pl L)  \\
    \pl S^a_{\pl X}(N,x,\bel) &=& 
      \textsf{list}_{(v,n) \in \pl A_{\pl X}(a,x)} (v \st n = \bel(N)) 
    \end{array}$\smallskip

\item Step with variable binding:
 \[\begin{array}{lcl}
    \pl S^a_{\pl X}(pattern\fd V,x,\bel) &=& 
     \left\{
     \begin{array}{ll}
       (\bel(V)) & \mbox{if } (\bel(V)) \in \pl S^a_{\pl X}(pattern,x,\bel) \\
       \varepsilon      & \mbox{ otherwise.}
     \end{array} \right.
 \end{array}\]
 
\item \label{def-semantics-xpathlogic-item-filter}%
  Step qualifiers:
 \[ \pl S^a_{\pl X}(pattern[\textsl{stepQ}],x,\bel) =
 \textsf{list}_{y\in \pl S^a_{\pl X}(pattern,x,\bel)} 
     (y \st \pl Q_{\pl X}(\textsl{stepQ},y,
                          \bel^{k,n}_{\small\textsl{Pos},\textsl{Size}})) 
     \]
   where
   $L_1 := \pl S^a_{\pl X}(pattern,x,\bel)$ and $n := size(L_1)$, 
   and for every $y$, let $j$ the index of $y$ in $L_1$,
   $k := j$ if $a$ is a forward axis, and $k := n+1-j$ if $a$ is
   a backward axis
   (cf.\ \cite{wadler-misc-99}). $Pos$ and $Size$ are only used if the
   step qualifier contains a context function.
   
\item\label{def-semantics-xpathlogic-item-concat} 
Path: \ \
 $\pl S^a_{\pl X}(p_1/p_2,x,\bel) =
    \textsf{concat}_{y\in \pl S^a_{\pl X}(p_1,x,\bel)}
      (\pl S^{any}_{\pl X}(p_2,y,\bel))$

\setcounter{mycounter}{\value{enumi}}
\end{enumerate}

\paragraph{Step Qualifiers are evaluated by $\pl{Q}$:}
\begin{enumerate}\setcounter{enumi}{\value{mycounter}}
\item Reference expressions have an existential semantics in step qualifiers:
\[ \pl Q_{\pl X}({\it refExpr},y,\bel) \DefIff
       \pl S^{any}_{\pl X}({\it refExpr},y,\bel) \neq \emptyset
\]

\item \label{def-semantics-xpathlogic-item-predicates} Predicates
  (including comparison predicates): The semantics of predicates in
  XPath is \emph{element-oriented}:
  $p(\textsf{\textsl{refExpr},\textsl{term}})$ evaluates to
  \textsl{true} if at least one pair taken from the result sets of
  \textsf{\textsl{refExpr}} and \textsf{\textsl{term}} satisfies the
  predicate $p$ (either defined in $\pl I_P$, or a built-in predicate
  of XPath \cite{XPath-XQuery-Functions-Operators-W3C}):

 \[ \begin{array}{l}
    \pl Q_{\pl X}(pred(expr_1,\ldots,expr_n), y, \bel) \DefIff \\
    \hspace*{2cm}
      \begin{array}[t]{l}
      \mbox{ there are }
      x_1 \in \pl S^{any}_{\pl X}(expr_1,y,\bel),\ldots,
      x_n \in \pl S^{any}_{\pl X}(expr_n,y,\bel) \\
      \mbox{ such that }
       (x_1, \ldots, x_n) \in \pl I_P(pred)
    \end{array}
  \end{array}
    \] 

\item Boolean Connectives and Quantification are defined as usual.
\setcounter{mycounter}{\value{enumi}}
\end{enumerate}

\paragraph{Evaluation of Terms.}

\begin{enumerate}\setcounter{enumi}{\value{mycounter}}
\item Constants $c \in \Sigma_C$: \ $\pl S^a_{\pl X}(c, x,\bel) = \pl
  I_C(c)$. For literals, \ $\pl S^a_{\pl X}(lit, x,\bel) = lit$.
\item Variables: \ $\pl S^a_{\pl X}(var, x,\bel) = \bel(var)$.
\item \label{def-semantics-xpathlogic-item-functions}
 Functions and arithmetics are also defined element-wise: \\
 $\begin{array}[t]{l}
    \pl S^a_{\pl X}(f(expr_1,\ldots,expr_n),x,\bel) = \\
    \{ \pl I_F(\bel(x),f,x_1,\ldots,x_n) \st
       x_1 \in \pl S^{any}_{\pl X}(expr_1,x,\bel),\ldots,
       x_n \in \pl S^{any}_{\pl X}(expr_n,x,\bel)\}
    \end{array}$\smallskip
\item \label{def-semantics-xpathlogic-item-context-functions}
  Context-related functions use the extension of variable bindings
  by pseudo-variables $Size$ and $Pos$ in rule
  (\ref{def-semantics-xpathlogic-item-filter}):
 \[ \pl S^{any}_{\pl X}(\textsf{position()}, x, \bel) = \bel(\textsl{Pos}) 
    \qquad \textrm{and} \qquad 
    \pl S^{any}_{\pl X}(\textsf{last()}, x, \bel) = \bel(\textsl{Size})~.
  \]
\setcounter{mycounter}{\value{enumi}}
\end{enumerate}
\end{Definition}

\noindent
The following theorem states the equivalence of our semantics with
that given in \cite{wadler-misc-99}, which is in turn equivalent to
the one defined by the W3C for XPath in
\cite{XQuery-Formal-Semantics-W3C}.

\begin{Theorem}[Correctness of $\pl S$ and $\pl Q$ wrt.\ XPath]%
\label{theo-semantics-S-wadler}
For XPath reference expressions without splitting \texttt{NMTOKENS}
attributes, the semantics coincides with the one given in
\cite{wadler-misc-99} (which already covers all core constructs of
XPath as an addressing formalism): For every XPath expression $expr$,
\[ \pl S_{\pl X}(expr) = \pl S [[expr]](x)\]
  (for arbitrary $x$)
  where $\pl S[[expr]]$ is as defined in \cite{wadler-misc-99}.
  
  Note that $\pl S[[expr]]$ defines only a result \emph{set} that is
  implicitly ordered wrt.\ document order.  Our semantics coincides
  with the document order as long as no dereferencing is used.
\end{Theorem}

\noindent
The proof uses the following Lemma which contains the structural
induction (for proofs, see Appendix~\ref{sec-proofs}).

\begin{Lemma}[Correctness of $\pl S$ and $\pl Q$ wrt.\ XPath: 
  Structural Induction]\label{lemma-semantics-S-wadler}
  XPath-Logic reference expressions correspond to XPath as follows:
\begin{enumerate}
  \item For absolute expressions (i.e., $expr = /expr'$, and 
    no free variables): \\
   \hspace*{1cm}
   $\pl S^{any}_{\pl X}(expr,any,\emptyset) = \pl S [[expr]](x)$ \
     for arbitrary $x$.
  \item For expressions, for all $\bel$: \
   $\pl S^{any}_{\pl X}(expr,v,\bel) = \pl S [[expr]](v)$~.
  \item For step qualifiers, for all $\bel$: \
   $\pl Q_{\pl X}({\it stepQ},v,
                     \bel^{k,n}_{\small\textsl{Pos},\textsl{Size}}) \Iff
       \pl Q [[{\it stepQ}]](v, k, n)$~. 
  \item For arithmetic expressions and built-in functions, for all $\bel$:\\  
   \hspace*{1cm}
    $\pl S^{any}_{\pl X}
           (expr,v,\bel^{k,n}_{\small\textsl{Pos},\textsl{Size}}) 
        = \pl E [[expr]](v,k,n)$~.
\end{enumerate}
where $\pl Q[[expr]]$, $\pl S[[expr]]$, and $\pl E[[expr]]$ are as
defined in \cite{wadler-misc-99}.  Since XPath expressions are
variable-free, $\bel$ is empty except handling the pseudo variables
$Size$ and $Pos$ (which are often also empty).
\end{Lemma}

\noindent
The above behavior deviates from XPath for special kinds of
attributes: When navigating along reference attributes, the result is
not in document order, but in the same order as the referencing
elements were.  Additionally, \texttt{NMTOKENS} that are considered
as atomic in XPath, are split in XPath-Logic.

\subsubsection{Semantics of Formulas}

\begin{Definition}[Semantics of XPath-Logic Formulas]%
\label{def-semantics-xpathlog-formulas}
  Formulas are interpreted according to the usual first-order
  semantics
\[ \models\ \ \subseteq\ 
   (\mbox{X-Structures} \times \mbox{Var\_Assignments} 
         \times \mbox{Formulas})\]

\begin{enumerate}\setcounter{enumi}{\value{mycounter}}
\item Reference Expressions: The semantics of reference expressions
  corresponds to a \emph{predicate} in first-order logic, defining
  a purely existential semantics:
\[  (\pl X,\bel) \models {\it refExpr} \DefIff
     (\pl S_{\pl X}({\it refExpr},\bel)) \neq \emptyset\]
\item predicates and boolean connectives: same as in first-order
     logic.
\setcounter{mycounter}{\value{enumi}}
\end{enumerate}
\end{Definition}

\noindent
The above definitions associate a \emph{truth value semantics} with
XPath-Logic formulas. The $\models$ relationcan be used for expressing
integrity constraints on XML documents (see
Example~\ref{ex-xpathlogic-integrity-constraints}) and even sets of
documents, and for reasoning on X-Structures.  In contrast, when
defining XPathLog as a \emph{data manipulation language} in the next
section, a completely different formalization of the semantics is
given: there, as for Datalog queries, the \emph{answer substitutions}
for a formula containing free variables have to be computed.

\subsection{Annotated Literals}\label{sec-annotated-literals}

The XML data model distinguishes between elements and their text
contents.  Nevertheless, in several situations, elements containing
text contents are expected to act as numbers or strings:

\begin{Example}[Annotated Literals]
Consider again the \Mondial\ XML instance.  The XPath queries
\begin{expl}
  //country[population $>$ 5000000]/name/text() \qquad \textrm{and}\\
  //country[population/text() $>$ 5000000]/name/text()
\end{expl}
are equivalent and return ``Belgium'' in their result set.  In the
first query, the \emph{element} \ \textsf{\flq population\frq
  10170241\flq/population\frq} \ is implicitly casted into its literal
value.
\end{Example}

\noindent
What happens here is not evident in XML/DTD environments.  A
corresponding XML Schema instance would show that the
\emph{complexType} \textsf{population} is derived from a simple type
for integers. The idea here is that an element with a text contents
adds structure to a simple type by allowing subelements and
attributes.  Thus, text elements with attributes behave as
\emph{annotated literals}:

\begin{itemize}
\item comparisons, arithmetics, and (optionally) output use the
  literal value,
\item navigation expressions use the element node, and
\item in variable bindings, the variable is bound to the element, but
  it acts as described above when the variable is used e.g.\ in a
  comparison.
\end{itemize}

\section{XPathLog: The Horn Fragment of XPath-Logic}\label{sec-xpathlog}

Similar to the case of Datalog which is the function-free Horn
fragment of first-order predicate logic, XPathLog is a logic
programming language based on XPath-Logic.  The evaluation of a query
\ \textsf{?- L$_1$, \ldots, L$_n$} \ results in a set of variable
bindings (of the free variables of the query) to elements of the
universe.  The semantics of XPathLog programs --i.e., the semantics of
the evaluation of a set of XPathLog rules as a logic program-- is then
defined in Section~\ref{sec-xpathlog-rules} by combining the answer
semantics with the model-theoretic semantics defined in the preceding
section.

\begin{Definition}[XPathLog]\label{def-xpathlog-atoms}
Atoms are the basic components of XPathLog rules:
\begin{itemize}
\item an \emph{XPathLog atom} is either an XPath-Logic \emph{reference
    expression} which does not contain quantifiers or disjunction in
  step qualifiers, or a predicate expression over such expressions.
\item an XPathLog atom is \emph{definite} if it uses only the child,
  sibling, and attribute axes and the atom does not contain negation,
  disjunction, function applications, and context functions.  These
  atoms are allowed in rule heads (see
  Section~\ref{sec-left-hand-side}). The excluded features would cause
  ambiguities what update is intended, e.g., ``insert $x$ as a
  descendant'' does not specify where the element should actually be
  inserted.
\end{itemize}

\noindent
Similar to Datalog, an \emph{XPathLog literal} is an atom or a negated
atom and an \emph{XPathLog query} is a list \textsf{?- L$_1$, \ldots,
  L$_n$} of literals (in general, containing free variables). An
\emph{XPathLog rule} is a formula of the form \ $A_1,\ldots,A_k \gets
L_1, \ldots, L_n$ \ where $L_i$ are literals and $A_i$ are definite
atoms.  $L_1, \ldots, L_n$ is the \emph{body} of the rule, evaluated
as a conjunction.  $A_1,\ldots,A_k$ is the \emph{head} of the rule,
which may contain free variables that must also occur free in the
body.  In contrast to usual logic programming, we allow for lists of
atoms in the rule head which are interpreted as conjunctions.
\end{Definition}


\subsection{Queries in XPathLog}\label{sec-queries-xpathlog}

The semantics $\pl{SB}$ of XPathLog queries associates a result set
and a set of \emph{answer substitutions} with every XPathLog query by
extending the above definition of $\pl S$.  The semantics provides the
formal base for the implementation of an \emph{algebraic evaluation}
of XPathLog queries in \LoPiX\ (cf.\ Section~\ref{sec-lopix}).

\subsubsection{Answers Data Model}

Whereas in Datalog, the answer to a query \textsf{?- L$_1$, \ldots,
  L$_n$} is a set of variable bindings, the semantics of XPath-Logic
reference expressions is defined wrt.\ an X-Structure $\pl X$ as an
\emph{annotated result list}, i.e., the semantics of an expression is
\begin{itemize}
\item[(i)] a result list (corresponding to the result list of the
  underlying XPath expression -- i.e., without the additional variable
  bindings), and
\item[(ii)] with every element of the result list, a list of variable
  bindings is associated.
\end{itemize}

\noindent
The result list (i) is the same as defined by $\pl S$ in
Definition~\ref{def-semantics-xpathlogic}, equivalent to the one
defined for XPath expressions in \cite{wadler-misc-99} and by the W3C
for XPath \cite{XQuery-Formal-Semantics-W3C}.  Whereas from the XPath
point of view for ``addressing'' nodes, only the result list is
relevant, XPathLog queries are mapped to a set of variable bindings
based on the associated bindings lists.

\begin{Example}[Semantics]\label{ex-semantics-1}
  First, the semantics is illustrated by an example. Let
  $\pl X$ be the XML structure given in Example~\ref{ex-X-structure},
  and
  \begin{expl}
   $expr$ := //organization\fd O[%
       \begin{tabular}[t]{@{}l@{}l}
         member/@country[@car\_code\fd C and name/text()\fd N]]\\
         \multicolumn{2}{@{}l}{/abbrev/text()\fd A.}
       \end{tabular}
  \end{expl}

\noindent
The underlying XPath expression is 
\begin{expl}
   //organization[member/@country[@car\_code and name/text()]]%
      /abbrev/text()~.
\end{expl}
with the result list \textsf{(``UN'',``EU'',\ldots)}.  With each of
the results, a list of bindings for the variables \textsf{O, C, N},
and \textsf{A} is associated, yielding the annotated result list
\begin{expl}
  \begin{tabular}{r@{}c@{}ll}
  $\pl{SB}_{\pl{X}}(expr)$ = 
     list((&``UN'', \{ & (O/\textsl{un}, A/``UN'', C/``AL'', N/``Albania''), \\
         &           &(O/\textsl{un}, A/``UN'', C/``GR'', N/``Greece''), \\
         &           & \multicolumn{1}{c}{\smallvdots}&\}),\\
         (&``EU'', \{ & (O/\textsl{eu}, A/``EU'', C/``D'', N/``Germany''), \\
         &           & (O/\textsl{eu}, A/``EU'', C/``F'', N/``France''), \\
         &           & \multicolumn{1}{c}{\smallvdots} & \}),\\
         & \smallvdots    &             & ~~~)
       \end{tabular}
\end{expl}
\end{Example}

\begin{Definition}[Semantics]
  The domain of \emph{sets of} variable bindings for $V_1,\ldots,V_n$
  (i.e., the domain of the second component of our semantics -- i.e.,
  the possible answer sets for a query whose free variables are
  $V_1,\ldots,V_n$) is
\[ \Bindings_{V_1,\ldots,V_n} \ := \ 
   (2^{((\pl V \cup \pl L\cup \pl N)^n)})^{\{V_1,\ldots,V_n\}}~. \] 
   Thus, in the general case for a general set $\Var$ of variables
   where $n$ is unknown,
\[ \Bindings \ := \ 
   \bigcup_{n\in \N_0} (2^{((\pl V \cup \pl L\cup \pl N)^n)})^{(\Var^n)} \]
   is the set of sets of variable assignments. For an empty set of
   variables, $\{\textsl{true}\}$ is the only element in $\Bindings$;
   in contrast, $\emptyset$ means that there is no variable binding
   which satisfies a given requirement.  We use $\bel$ for denoting an
   individual variable binding, and $\xi \in \Bindings$ for denoting a
   set of variable bindings.

\noindent
\[ \AnnotatedResults \ := \ 
   ((\pl V \cup \pl L) \times \Bindings)
\]
is the set of annotated results (i.e., an annotated result is a pair
$(v,\xi)$ where $v$ is a node or a literal and $\xi$ is a set of
variable bindings (for the set of variables occurring free in a
certain formula)).
\end{Definition}

\begin{Definition}[Operators on Annotated Result Lists]
  
  From an annotated result list $\theta$, the result list is obtained
  as $\Res{\theta}$:
  \[ \begin{array}{ll}
     \mbox{Res}\ : &
          \AnnotatedResults^\N \to (\pl V \cup \pl L \cup \pl N)^\N \\
       & ((x_1,\xi_1), \ldots, (x_n,\xi_n)) \mapsto (x_1,\ldots,x_n)
     \end{array}
  \]
  For an annotated result list $\theta$ and a given $x
  \in\Res{\theta}$ contained in the result list, the set of variable
  bindings associated with $x$ is obtained by $\Bdgs{\theta,x}$:
  \[ \begin{array}{ll}
     \mbox{Bdgs}:&
          \AnnotatedResults^\N \times (\pl V \cup \pl L \cup \pl N)
          \to \Bindings \\
       & \hspace*{-0.4cm}
      (((x_1,\xi_1),\ldots,(x,\xi),\ldots,(x_n,\xi_n)),x) \mapsto \xi \ \
        \mbox{(let $\Bdgs{\theta,x} = \emptyset$ if $x \notin \Res{\theta}$)}
     \end{array}
  \]
\end{Definition}

\noindent
Note that the joins ($\join$) used in this section are always purely
relational joins that are applied to the bindings component.

\begin{Example}[Semantics (Cont'd)]
  Continuing Example~\ref{ex-semantics-1},
  $\Res{\pl{SB}_{\pl{X}}(expr)} = 
     (\mbox{\textsf{``UN'',``EU'', \ldots}})$ is
  the result list of the underlying XPath expression, and
\begin{expl}
  \begin{tabular}{r@{}ll}
  $\Bdgs{\pl{SB}_{\pl{X}}(expr),\textsf{``EU''}}$ = 
         \{ &(O/\textsl{eu}, A/``EU'', C/``D'', N/``Germany''), \\
            &(O/\textsl{eu}, A/``EU'', C/``F'', N/``France''), \\
            & \multicolumn{1}{c}{\smallvdots} & \}
       \end{tabular}
\end{expl}
yields the variable bindings that are associated with the result value
\textsf{``EU''}.
\end{Example}

\subsubsection{Safety}

The semantics definition evaluates formulas and expressions wrt.\ a
given set of variable bindings which e.g., results from evaluating
other subexpressions of the same query.  This approach allows for a
more efficient evaluation of joins (\emph{sideways information passing
  strategy}), and is especially needed for evaluating \emph{negated}
expressions (by defining negation as a relational ``minus'' operator).
Negated expressions which contain free variables are intended to
\emph{exclude} some bindings from a given set of potential results.
Thus, for variables occurring in the scope of a negation, the input
answer set to the negation must already provide potential bindings.
This leads to a \emph{safety} requirement similar to Datalog.

\begin{Definition}[Safe Queries]\label{def-safe-query} 
  First, safety of variables is decided for each individual ocurrence.
  A variable occurrence $V$ is \emph{safe} wrt.\ the query if at least
  one of the following holds:
\begin{itemize}
\item if the occurrence is in a literal $L$, and it is not inside the
  scope of a negation and not in a comparison predicate other than
  equality (e.g., $X<3$ is unsafe).
\item if the occurrence is in a literal $L_i$ inside a step qualifier
  $\textsl{\textsf{pattern}}%
  [L_1 \textsf{ and } \ldots \textsf{ and } L_n]$ and $V$ has a safe
  occurrence in \textsl{\textsf{pattern}} or in some $L_j$ such that
  $j<i$.
\item if the occurrence is in a literal $L_i$ of the query $?- L_1
  \textsf{ and } \ldots \textsf{ and } L_n$, and $V$ has a safe
  occurrence in some $L_j$ such that $j<i$.
\end{itemize}

\noindent
A query \textsf{?- L$_1$, \ldots, L$_n$} is \emph{safe} if all
variable occurrences in the query are safe.
\end{Definition}


\subsubsection{Semantics of Expressions}
\setcounter{mycounter}{0}

In the following, the semantics of safe queries is defined.  The basic
(non-annotated) result lists are again provided by $\pl
A_{\pl{X}}(axis, v)$ for every node $v$ of $\pl{X}$.

\begin{Definition}[Answer Semantics of XPath-Logic Expressions]
\label{def-semantics-xpathlog}
The semantics is defined by operators $\pl{SB}$ and $\pl{QB}$ derived
from $\pl S$ and $\pl Q$ as defined in
Definition~\ref{def-semantics-xpathlogic}; the $\pl B$ stands for the
extension with variable bindings:
\begin{itemize}
\item $\pl{SB}_{\pl{X}}  :  
   \begin{array}[t]{@{}l}
   (\mbox{Reference\_Expressions}) \to \AnnotatedResults^\N~~\mbox{, and} \\
   (\mbox{Axes} \times \pl V \times \mbox{Reference\_Expressions} 
       \times \Bindings) 
     \to \AnnotatedResults^\N 
   \end{array}$\smallskip\\
   evaluates reference expressions wrt.\ an axis, an (optional)
   context node and a given set of variable bindings and returns an
   annotated result list.
\item $\pl{QB}_{\pl{X}} \ :
   \begin{array}[t]{l}
    (\mbox{Predicate\_Expressions} \times \pl V \times \Bindings) \to \Bindings
  \end{array}$
  \smallskip\\
  evaluates step qualifiers wrt.\ a context node to sets
  of variable bindings.
\end{itemize}

\noindent
Expressions are evaluated by $\pl{SB}$:

\begin{enumerate}
\item If no input bindings are given, \
  $\pl{SB}_{\pl{X}}({\it refExpr}) = 
        \pl{SB}^{any}_{\pl{X}}({\it refExpr},any,\emptyset)$
\item Reference expressions are translated into path expressions wrt.\
  a start node:
  \begin{itemize}
  \item entry points: rooted path: \ 
    $\pl{SB}^{any}_{\pl{X}}(/p,any,Bdgs) = \pl{SB}^{any}_{\pl{X}}(p,root,Bdgs)$ \\
    where $root$ is the current root as in
    Definition~\ref{def-semantics-xpathlogic}
    (\ref{def-semantics-xpathlogic-item-entry}) for the same case.

  \item entry points: constants $c \in \Sigma_C$: \
   $\pl{SB}^{any}_{\pl{X}}(c/p,any,Bdgs) = \pl{SB}^{any}_{\pl{X}}(p,c,Bdgs)$

  \item rooted paths in other documents: \\
  \hspace*{1cm} 
    $\pl{SB}^{any}_{\pl X}(\textsf{document(``http://\ldots'')}/p,any,Bdgs) = 
      \pl S^{any}_{\pl X}(p,root,\bel)$ \\
  where $root$ is the root node of the document stored at 
  \textsf{http://\ldots}.
 
  \item  entry points: variables $V\in\Var$:
   \[ \pl{SB}^{any}_{\pl{X}}(V/p,any,Bdgs) = 
    \textsf{concat}_{x \in \pl V_{\rm active}}(\pl{SB}^{any}_{\pl{X}}
                     (p,x,Bdgs \join \{V/x\})) 
    \]
    where $\pl V_{\rm active}$ is the set of element nodes in the
    current database.
    
    Remark: Here, the input bindings are used for optimization: if
    every $\bel\in Bdgs$ provides already bindings for the variable
    $V$, the \emph{sideways information passing strategy} 
    directly effects the join $\{V/x\} \join Bdgs$, restricting the
    possible values for $V$ which in fact results in 

    \[\begin{array}{l}
      \pl{SB}^{any}_{\pl{X}}(V/p,any,Bdgs) = 
      \textsf{concat}_{\bel \in Bdgs, x = \bel(V)} 
        (\pl{SB}^{any}_{\pl{X}}(p,x,Bdgs \join \{V/x\}))
      \end{array}
    \]

    Thus, the propagation of bindings is not only necessary for handling
    negation but also provides a relevant optimization for positive
    literals.
    
    Note that in the recursive call 
    $\pl{SB}^{any}_{\pl{X}}(p,x,Bdgs \join \{V/x\})$, the propagated 
    bindings are already augmented with the bindings for $V$.
  \end{itemize}
\item\label{def-semantics-xpathlog-item-locationstep}%
Axis step: \
 $\pl{SB}^{any}_{\pl{X}}(axis::pattern,x,Bdgs) =
                           \pl{SB}^{axis}_{\pl{X}}(pattern,x,Bdgs)$ \\
 where $pattern$ is of the form \textit{nodetest} \textit{remainder} 
 where \textit{remainder} is a sequence of step qualifiers and variable
 bindings. These are evaluated from left to right, always applying the
 rightmost ``operation'' (qualifier or variable) to the result of the
 left part:
\item\label{def-semantics-xpathlog-item-nodetest}%
 Node test:
 \[\begin{array}{lcl}
    \pl{SB}^a_{\pl{X}}(name,x,Bdgs) &=& 
      \textsf{list}_{(v,n) \in \pl A_{\pl{X}}(a,x),\ n = name} 
            (v,\{true\} \join Bdgs) 
      \\
    \pl{SB}^a_{\pl{X}}(\textsf{node()},x,Bdgs) &=& 
      \textsf{list}_{(v,n) \in \pl A_{\pl{X}}(a,x),\ v \in \pl V} 
            (v,\{true\} \join Bdgs) 
      \\
    \pl{SB}^a_{\pl{X}}(\textsf{text()},x,Bdgs) &=& 
      \textsf{list}_{(v,n) \in \pl A_{\pl{X}}(a,x),\ v \in \pl L} 
            (v,\{true\} \join Bdgs) 
      \\
    \pl{SB}^a_{\pl{X}}(N,x,Bdgs) &=& 
      \textsf{list}_{(v,n) \in \pl A_{\pl{X}}(a,x)} (v,\{N/v\} \join Bdgs) 
   \end{array}
  \]

\item Step with variable binding: 
 \[\pl{SB}^a_{\pl{X}}(pattern\fd V,x,Bdgs) = 
         \textsf{list}_{(y,\xi) \in \pl{SB}^a_{\pl{X}}(pattern,x,Bdgs)}
            (y, \xi \join \{V/y\})
  \]

\item \label{def-semantics-xpathlog-item-filter}%
 Step qualifiers:
\[ \begin{array}{l}
    \pl{SB}^a_{\pl{X}}(pattern[{\it stepQ}],x,Bdgs) =  \\
    \hspace*{0.5cm}
     \textsf{list}_{(y,\xi)\in \pl{SB}^a_{\pl{X}}(pattern,x,Bdgs),\
             \pl{QB}_{\pl{X}}({\it stepQ},y,\xi') \neq\emptyset}
         (y,\ \pl{QB}_{\pl{X}}({\it stepQ},y,\xi')\setminus\{Pos,Size\})
       \end{array}
       \]
   If the step qualifier does not contain \emph{context functions},
   then $\xi' := \xi$, otherwise let
   $L:=\pl{SB}^a_{\pl{X}}(pattern,x,Bdgs)$,
   and then for every $(y,\xi)$ in $L$, $\xi'$ is obtained as follows,
   extending $\xi$ with bindings of the pseudo variables $Size$ and 
   $Pos$:
   \begin{itemize}
   \item start with $\xi'=\emptyset$,
   \item for every $\bel\in\xi$, the list \ $L' =
     \textsf{list}_{(y,\xi)\in L {\ \rm s.t.\ } \bel\in\xi}(y)$ \ 
     contains all nodes which are selected \emph{for the variable
       assignment $\beta$}.
   \item let $size := size(L')$, and for every $y$, let $j$ the
     index of $x_1$ in $L'$, $pos := j$ if $a$ is a forward axis, and
     $pos := size+1-j$ if $a$ is a backward axis.
   \item add $\bel^{Size,Pos}_{size,pos}$ to $\xi'$.
   \end{itemize}

\item Path: \
 $\pl{SB}^a_{\pl{X}}(p_1/p_2,x,Bdgs) =
    \textsf{concat}_{(y,\xi)\in \pl{SB}^{any}_{\pl{X}}(p_1,x,Bdgs)}
      \pl{SB}^a_{\pl{X}}(p_2,y,\xi)$

\setcounter{mycounter}{\value{enumi}}
\end{enumerate}

\paragraph{Step Qualifiers are evaluated by $\pl{QB}$:}
\begin{enumerate}\setcounter{enumi}{\value{mycounter}}
\item Reference expressions (existential semantics) in step qualifiers:
\[ \pl{QB}_{\pl{X}}({\it refExpr}, x, Bdgs) =
       \bigcup_{(y,\xi)\in \pl{SB}^{any}_{\pl{X}}({\it refExpr},x,Bdgs)}
       \xi
\]

\item The built-in equality predicate ``$=$'' is not only a comparison
  if both sides are bound, but also serves as an assignment if the
  left-hand side is a variable $V\in\Var$ which is \emph{not} bound in
  $Bdgs$:
 \[\pl{QB}_{\pl{X}}(V=expr, x, Bdgs) =
       \bigcup_{(y,\xi) \in \pl{SB}^{any}_{\pl{X}}(expr,x,Bdgs)}
        \xi \join \{V/y\} \]
 All other built-in comparisons require all variables to be bound:
 \[ \pl{QB}_{\pl{X}}(expr_1\ \textsf{\textsl{op}}\ expr_2,x,Bdgs) = 
       \bigcup_{(x_i,\xi_i) \in \pl{SB}_{\pl{X}}(expr_i,x,Bdgs)~,~
                x_1\ {\sf\sl op}\ x_2
               }
       \hspace*{-1.2cm}
        \xi_1 \join\xi_2~.
  \]
\item Predicates except built-in comparisons:
  \[\pl{QB}_{\pl{X}}(pred(expr_1,\ldots,expr_n), x, Bdgs) =
       \bigcup_{\begin{array}{c}
         \scriptstyle
          (x_i,\xi_i) \in \pl{SB}_{\pl{X}}(expr_i,x,Bdgs)\\[-0.4ex]
         \scriptstyle
          (x_1, \ldots, x_n) \in \pl I(pred)
       \end{array}
       }
       \hspace*{-1.3cm}
        \xi_1 \join \ldots \join \xi_n
      \] 

\item Negated expressions which do \emph{not contain any free variable}:
 \[ \pl{QB}_{\pl{X}}(\mbox{\sf not } A, x, Bdgs) = 
    \left\{\begin{array}{ll}
       Bdgs & \mbox{if $\pl{QB}_{\pl{X}}(A,x,\emptyset) = \emptyset$,} \\
       \emptyset & 
          \mbox{otherwise, i.e., if 
                   $\pl{QB}_{\pl{X}}(A,x,\emptyset) = \{true\}.$} 
     \end{array}\right.
   \]
   
 \item \label{def-semantics-xpathlog-item-negation}
   For negated expressions which contain free variables,
   negation is interpreted as the ``minus'' operator (as known e.g.,
   from the relational algebra) wrt.\ the given input bindings.  Thus,
   all variables which occur free in $A$ must be safe, i.e., every
   input variable binding has to provide a value for them.
   
   For two variable bindings $\bel_1,\bel_2$, we write $\bel_1 \le
   \bel_2$ if all variable bindings in $\bel_1$ occur also in
   $\bel_2$.  Intuitively, in this case, if $\bel_1$ is ``abandoned'',
   $\bel_2$ should also be abandoned.
  
 \[ \begin{array}{l}
    \pl{QB}_{\pl{X}}(\mbox{\sf not } expr, x,Bdgs) = \\
    \hspace*{1cm}
        Bdgs \;-\; \{ \bel\in Bdgs \st 
               \mbox{ there is a $\bel' \in \pl{QB}_{\pl{X}}(expr,x,Bdgs)$
                      s.t.\ $\bel\le\bel'$}\}
    \end{array}
   \]

\item\label{def-semantics-xpathlog-item-log-conjunction} 
 Conjunction:
 \[ \begin{array}{l}
    \pl{QB}_{\pl{X}}(expr_1 \mbox{\sf\ and } expr_2, x, Bdgs) = \\
    \hspace*{1cm}
       \pl{QB}_{\pl{X}}(expr_1,x,Bdgs) \join 
       \pl{QB}_{\pl{X}}(expr_2,x,\pl{QB}_{\pl{X}}(expr_1,x,Bdgs))
     \end{array}
     \]
     Here, in case of negated conjuncts in the step qualifier, the
     safety of variables has to be considered. The above definition
     assumes that by a left-to-right evaluation of conjuncts, the
     evaluation is safe.  \setcounter{mycounter}{\value{enumi}}
\end{enumerate}

\paragraph{Evaluation of Terms}

\begin{enumerate}\setcounter{enumi}{\value{mycounter}}
\item Constants: for literals,
  $\pl{SB}^{any}_{\pl{X}}(lit, x, Bdgs) = (lit, Bdgs)$.
  For constants $c \in \Sigma_C$, 
 $\pl{SB}^{any}_{\pl{X}}(c, x, Bdgs) = (\pl I_C(c), Bdgs)$.
\item\label{def-semantics-xpathlog-item-vars}
  Variables: the variable occurrence must be safe, then: \
 $\pl{SB}^{any}_{\pl{X}}(var, x, Bdgs) = 
      \textsf{list}_{\bel \in Bdgs}(\bel(var), \bel)$.

\item Function terms and arithmetics:
 \[ \begin{array}{l}
    \pl{SB}^{any}_{\pl{X}}(f(arg_1,\ldots,arg_n)),x,Bdgs) =  \smallskip\\
    \hspace*{1cm}
       \textsf{list}_{
          (x_i,\xi_i) \in \pl{SB}^{any}_{\pl{X}}(arg_1,x,Bdgs),\ldots, 
        }
        (f(x_1, \ldots, x_n), \xi_1 \join \ldots \join \xi_n)
   \end{array}
   \]
  where $f(x_1, \ldots, x_n)$ results from the built-in evaluation of $f$.
\item Context-related functions use the extension of variable bindings
  by pseudo-variables $Size$ and $Pos$ in rule
  (\ref{def-semantics-xpathlog-item-filter}):
 \[ \begin{array}{l}
    \pl{SB}^{any}_{\pl{X}}(\textsf{position()}, x, Bdgs) = 
        \textsf{list}_{\bel \in Bdgs} 
           (\bel(Pos), \{\bel' \in Bdgs\st \bel(Pos)=\bel'(Pos)\}) \\
    \pl{SB}^{any}_{\pl{X}}(\textsf{last()}, x, Bdgs) = 
        \textsf{list}_{\bel \in Bdgs} 
           (\bel(Pos), \{\bel' \in Bdgs\st \bel(Size)=\bel'(Size)\})
  \end{array}
  \]
\end{enumerate}
\end{Definition}

\noindent
The above semantics is an algebraic characterization of the logical
semantics of XPath-Logic expressions which has been defined in
Section~\ref{def-semantics-xpathlogic}:

\begin{Theorem}[Correctness of $\pl{SB}$ and $\pl{QB}$]%
\label{thm-correctness-SB-QB}
  For every (in general, containing free variables) XPathLog
  expression $expr$,
\[ \Res{\pl{SB}_{\pl{X}}(expr)} = 
     \bigcup_{\beta \in (\pl V\cup\pl L\cup\pl N)^{\free(expr)}}
       \pl S_{\pl{X}} (expr,\bel)~. \]
More detailed, for all $x \in \pl V\cup\pl L\cup\pl N$,
\[ (x \in  \Res{\pl{SB}_{\pl{X}}(expr)} \mbox{ and }
    \bel \in \Bdgs{\pl{SB}_{\pl{X}}(expr),x})
   \Iff
   x \in \pl S_{\pl{X}}(expr,\bel)~.
\]
\end{Theorem}

 \noindent
Again, the theorem uses a lemma which encapsulates the structural induction.

\begin{Lemma}[Correctness of $\pl{SB}$ and $\pl{QB}$: Structural Induction]%
\label{lemma-correctness-SB-QB}
The correctness of the answers semantics of XPathLog expressions
mirrors the generation of answer sets by the evaluation: The input set
$Bdgs$ may contain bindings for the free variables of an expression.
If for some variable $var$, no binding is given, the result extends
$Bdgs$ with bindings of $var$.  If bindings are given for $var$, this
specifies a constraint on the answers to be returned (expressed by
joins).

  \begin{itemize}
  \item For every absolute expression $expr$, (i.e., $expr = /expr'$)
    and every set $Bdgs$ of variable bindings,
    \[\begin{array}{l}
     (x \in \Res{\pl{SB}_{\pl{X}}(expr,Bdgs)} \mbox{ and }
      \bel \in \Bdgs{\pl{SB}_{\pl{X}}(expr,Bdgs),x})
      \Iff \\
    \hspace*{0.5cm}
      (x \in \pl S_{\pl{X}}(expr,\bel) 
       \mbox{ and $\bel$ completes some $\bel'\in Bdgs$ with $\free(expr)$})~.
     \end{array}
     \]
  \item For every expression $expr$, node $v$, and every 
  set $Bdgs$ of variable bindings,
    \[\begin{array}{l}
     (x \in \Res{\pl{SB}_{\pl{X}}(expr,v,Bdgs)} \mbox{ and }
      \bel \in \Bdgs{\pl{SB}_{\pl{X}}(expr,v,Bdgs),x})
      \Iff \\
    \hspace*{0.5cm}
      (x \in \pl S_{\pl{X}}(expr,v,\bel) 
       \mbox{ and $\bel$ completes some $\bel'\in Bdgs$ with $\free(expr)$})~. 
     \end{array}
     \]
  \item for every step qualifier {\it stepQ}, node $v$, and every 
  set $Bdgs$ of variable bindings,
  \[\begin{array}{l}
     \bel \in \pl{QB}_{\pl{X}}({\it stepQ},v,Bdgs) \Iff \\
    \hspace*{.5cm}
     \pl Q_{\pl{X}}({\it stepQ},v,\bel)
       \mbox{ and $\bel$ completes some 
              $\bel'\in Bdgs$ with $\free({\it stepQ})$})~. 
     \end{array}
     \]
   \end{itemize}
 \end{Lemma}

\noindent
The proof can be found in Appendix~\ref{sec-proofs}.

\subsubsection{Semantics of Queries}

According to Definition~\ref{def-xpathlog-atoms}, XPathLog queries are
conjunctions of XPathLog literals.  In the following, the evaluation
of safe queries is defined. The definition of safety guarantees that a
left-to-right evaluation of the body is well-defined (i.e., all
variable evaluations
in Definition~\ref{def-semantics-xpathlog}%
(\ref{def-semantics-xpathlog-item-vars}) are safe).
Definition~\ref{def-semantics-xpathlog}%
(\ref{def-semantics-xpathlog-item-log-conjunction}) already applied
left-to-right propagation when evaluating step qualifiers.

\begin{Definition}
The evaluation $\pl{QB}$ is extended to atoms by
\[\pl{QB}_{\pl{X}} \ :
   \begin{array}[t]{l}
    (\mbox{Atoms} \times \Bindings) \to \Bindings\\ 
    (A,Bdgs) \mapsto \pl{QB}_{\pl{X}}(A,root,Bdgs)
  \end{array}
\]

\noindent
For safe queries \textsf{?- \textsl{atom}} consisting of only one
atom, $\pl{QB}$ yields the answer bindings:
  \[  (\pl{X},\bel) \models {\it atom} \DefIff
      \bel \in \pl{QB}_{\pl{X}}({\it atom},\emptyset))~.\]
\end{Definition}

\begin{Definition}[Evaluation of Negated Literals]
  The evaluation of negated literals $L$ is defined wrt.\ a set of
  input bindings which must cover the free variables in $L$, similar
  to negation in step qualifiers in
  Def.~\ref{def-semantics-xpathlog}%
   (\ref{def-semantics-xpathlog-item-negation}):
  \[\begin{array}{l}
    \pl{QB}_{\pl{X}}(\textsf{not}\ A,Bdgs) := \\
    \hspace*{1cm}
        Bdgs \;-\; \{ \bel\in Bdgs \st 
               \mbox{ there is a $\bel' \in \pl{QB}_{\pl{X}}(A,Bdgs)$
                      s.t.\ $\bel\le\bel'$}\}~.
    \end{array}
   \]
 \end{Definition}
 
\begin{Definition}[Evaluation of Queries]
The evaluation of a safe query ~\textsf{?- $L_1$, \ldots, $L_n$}~
is defined similar to the evaluation of conjunctive step qualifiers in 
Definition~\ref{def-semantics-xpathlog}%
(\ref{def-semantics-xpathlog-item-log-conjunction}):
 \[ \begin{array}{l}
   \pl{QB}_{\pl{X}} \ :
    \mbox{Conj\_Literals} \to \Bindings\\ 
    \pl{QB}_{\pl{X}}(L_1 \land \ldots \land L_i) := 
    \begin{array}[t]{l}
       \pl{QB}_{\pl{X}}(L_1 \land \ldots \land L_{i-1}) \join \\
       \pl{QB}_{\pl{X}}(L_i,
               (\pl{QB}_{\pl{X}}(L_1 \land \ldots 
                                   \land L_{i-1}))|_{\free(L_i)})
     \end{array}
   \end{array}
   \]
Given an X-Structure $\pl{X}$, the answer to a query \
\textsf{?- L$_1$, \ldots, L$_n$} \ is the set \\
\hspace*{0.5cm}
 $\answers{\pl{X}}{L$_1$, \ldots, L$_n$} \ := \
      \pl{QB}_{\pl{X}}(\mbox{\textsf{L$_1{} \land \ldots\ \land {}$L$_n$}})$
\ \
 of variable bindings.
\end{Definition}

\begin{Theorem}[Correctness: Evaluation of Queries]%
\label{theo-semantics-xpathlog-S}
For all safe XPathLog queries $Q$, \ \
$\bel \in  \pl{QB}_\pl{X}(Q) 
                \Iff (\pl{X},\bel) \models Q$~.
\end{Theorem}

\noindent
Note that the semantics of formulas is not based on a Herbrand
structure consisting of ground atoms (as ``usual'' Herbrand semantics
are), but directly on the interpretations $\pl A_{\pl{X}}$ of the axes
in the X-Structure, and on an interpretation of predicate symbols that
can be represented as a finite set of tuples over $\pl V \cup \pl L
\cup \pl N$.

\section{XPathLog Programs}\label{sec-xpathlog-rules}

In logic programming, rules are used for a declarative specification:
if the body of a clause evaluates to \textsl{true} for some assignment
of its variables, the truth of the head atom for the same variable
assignment can be inferred.  Depending on the intention, this
semantics can be used for (top-down) checking if something is
derivable from a given set of facts, or (bottom-up) extending a given
set of facts by additional, derived knowledge. In this work, we mainly
investigate the bottom-up strategy, regarding XPathLog as an
\emph{update language for XML databases}: the evaluation of the body
wrt.\ a given structure yields variable bindings which are propagated
to the rule head where facts are added to the model.

Positive XPathLog programs (i.e., the rules contain only positive
literals; also step qualifiers may only contain positive expressions)
are evaluated bottom-up by a $T_P$-like operator over the X-Structure,
providing a minimal model semantics.  The formal definition of a $T_P$
operator will be given in Definition~\ref{def-semantics-xpathlog-once}
for XPathLog programs after explaining the semantics of insertions and
updates.


\subsection{Atomization}\label{sec-atomization}

In this section, an alternative semantics of conjunctions of
\emph{definite XPathLog atoms} is defined which provides the base for
the \emph{constructive} semantics of reference expressions in
\emph{rule heads}.  The semantics is defined by resolving reference
expressions syntactically into their constituting atomic steps in the
same way as in F-Logic (cf.\ \cite{frohn-lausen-uphoff-VLDB-94}).
A similar strategy for resolving expressions into atomic steps is
followed by several approaches which store XML data in relational
databases
\cite{deutsch-fernandez-suciu-SIGMOD-2000,shanmugasundaram-gang-tufte-etal-VLDB-99,florescu-kossman-TR-INRIA-99},
by flattening the XML instance to one or more universal relations.

\begin{Definition}[Atomization of Formulas]\label{def-atomize}
The function \ \ 
$\atomize: \textrm{XPathLogAtoms} \to 2^{\rm XPathLogAtoms}$ \ \
resolves a definite XPathLog atom into atoms of the form
\textsf{\textsl{node}[\textsl{axis}::\textsl{nodetest}\fd\textsl{result}]}
and predicates over variables and constants.  It will be used in
Definition~\ref{def-semantics-xpathlog-once} for specifying the
semantics of rule heads.  \atomize\ is defined by structural induction
corresponding to the induction steps when defining $\pl S_{\pl X}$.
In the following, $path$ stands for a \emph{path expression} (or a
variable), and $name$ for a name (or a variable).

\begin{itemize}
\item the entry case:  \
  $\atomize(/remainder) := \atomize(\textsf{root}/remainder)$
 \item Paths are resolved into steps and step qualifiers are isolated (since
   context functions are not allowed in definite atoms, it
   can be assumed that there is at most one step qualifier, optionally
   preceded by a variable assignment):
  \[\begin{array}{l}
    \atomize(path/axis::nodetest\fd var[\textit{stepQualifier}]~/remainder) := \\
    \hspace*{1.5cm}
      \atomize(path[axis::nodetest\fd  var]) \cup {}\\
    \hspace*{1.5cm}
      \atomize(var[\textit{stepQualifier}]) \cup 
      \atomize(var/remainder) ~,\\
    \atomize(path/axis::nodetest[\textit{stepQualifier}]~/remainder) := \\
    \hspace*{1.5cm}
      \atomize(path[axis::nodetest\fd  \_X]) \cup {}\\
    \hspace*{1.5cm}
      \atomize(\_X[\textit{stepQualifier}]) \cup
      \atomize(\_X/remainder) ~\\
    \hspace*{.5cm}
   \mbox{where $\_X$ is a new don't care variable.}
   \end{array}
   \]
\item Conjunctions in step qualifiers are separated:
  \[\begin{array}{l}
    \atomize(var[pred_1 \mbox{\sf and } \ldots \mbox{\sf and } pred_n]) := \\
    \hspace*{1.5cm}
     \atomize(var[pred_1]) \cup\ldots\cup\atomize(var[pred_n])
   \end{array}\]
\item Predicates in step qualifiers:
  \[ \atomize(var[pred(expr_1,\ldots,expr_n)]) := 
     \begin{array}[t]{@{}l}
     \atomize(equality(var,expr_1,\_X_1)) \cup \ldots \\
     \atomize(equality(var,expr_n,\_X_n)) \cup {}\\
     \{pred(\_X_1,\ldots,\_X_n)\}
   \end{array}
   \]
   where $equality(var,expr,X)$ is defined as follows (if $expr_i$
   is a constant, it is not replaced by a variable):
\begin{itemize}
\item $equality(var,expr,X) = ``expr \fd  X$'' if $expr$ is of the 
      form $//remainder$,
\item $equality(var,expr,X) = ``var/expr \fd  X$'' if $expr$ 
      is of the form $axis::nodetest~remainder$.
\end{itemize}
\item Predicate atoms are handled in the same way. Note that here all
  arguments are absolute expressions (rooted, or starting at a
  constant, or at a variable).
\end{itemize}
\end{Definition}

\begin{Example}[Atomization]
\begin{expl}
  ?- //organization\fd O[\begin{tabular}[t]{@{}l}
       name/text()\fd ON and  \\
       @headq = members/@country[name/text()\fd CN]/@capital].
     \end{tabular}
\end{expl}
is atomized into
\begin{expl}
  ?- \begin{tabular}[t]{l}
  root[descendant::organization\fd O], \quad
  O[name\fd \_ON], \_ON[text()\fd ON], \\
  O[@headq\fd \_S],
  O[members\fd \_M], \_M[@country\fd\_C], \_C[@country\fd\_Cap],  \\
  \_S = \_Cap, 
  \_C[child::name\fd \_CN], \_CN[text()\fd CN]. 
\end{tabular}
\end{expl}
\end{Example}

\begin{Theorem}[Correctness of \atomize]\label{theorem-nearly-equivalent}
  The above semantics is equivalent to the one presented in
  Definition~\ref{def-semantics-xpathlog} for all definite XPathLog
  atoms $A$ and every X-Structure $\pl{X}$, i.e.
  \[ \answers{\pl{X}}{$A$} = 
                \answers{\pl{X}}{\textsf{atomize($A$)}} \]
\end{Theorem}

\noindent
Again, the theorem uses a lemma which encapsulates the structural
induction, using the logical semantics for showing the correctness of
\atomize.

\begin{Lemma}[Correctness of \atomize: Structural Induction]
\label{lem-correctness-atomize}
  For every X-Structure $\pl{X}$ and every definite
  XPath-Logic atom $A$,
\begin{itemize}
\item for every variable assignment $\bel$ of $\free(A)$ such that
  $(\pl{X},\bel) \models A$, there exists a variable assignment
  $\bel' \supseteq \bel$ of $\free(\atomize(A))$ such that
  $(\pl{X},\bel') \models \atomize(A)$, and
\item for every variable assignment $\bel'$ of
  $\free(\atomize(A))$ such that
  $(\pl{X},\bel') \models \atomize(A)$, \
   $(\pl{X},\bel'|_{\free(A)}) \models A$.
\end{itemize}
\end{Lemma}

\noindent
The proof can be found in Appendix~\ref{sec-proofs}.

\subsection{Left Hand Side}\label{sec-left-hand-side}

Using \emph{logical expressions} for specifying an update is perhaps
the most important difference to approaches like XSLT, XML-QL, or
XQuery where the structure to be \emph{generated} is always specified
by XML patterns, or to the update proposal for XML described in
\cite{tatarinov-ives-halevy-weld-SIGMOD-2001}.  In contrast, in
XPathLog, existing nodes are communicated via variables to the head,
where they are \emph{modified} when appearing at host position of
atoms.  The semantics of the left hand side of XPathLog rules -- which
is a list of \emph{definite} XPathLog atoms -- is now investigated
based on the atomization of expressions. When used in the head, the
\textsf{``/''} operator and the \textsf{``[\ldots]''} construct
specify which properties should be added (thus, \textsf{``[\ldots]''}
does not act as a step qualifier, but as a \emph{constructor}).  When
using the child or attribute axis for updates, the host of the
expression gives the element to be updated or extended; when a sibling
axis is used, effectively the parent of the host is extended with a
new subelement. 

Note that the (pure) XPathLog language does \emph{not} allow to delete
or replace existing elements or attributes\footnote{suitable
  extensions, e.g., of the form
  \textsf{delete(\textsl{elem},\textsl{prop},\textsl{val})} can be
  defined.  Such extensions which would turn XPathLog into a
  rule-based imperative language are not investigated in this work.}
-- modifications are always monotonic in the sense that existing
``things'' remain.

\paragraph{Generation or Extension of Attributes.}
A ground-instantiated atom of the form $n[@a{\fd}v]$ specifies that
the attribute $@a$ of the node $n$ should be set or extended with $v$.
If $v$ is not a literal value but a node, a reference to $v$ is
stored.

\begin{Example}[Adding Attributes]
  We add the data code to Switzerland, and make it a member of the
  European Union:
\begin{expl}
 C[@datacode\fd ``ch''], C[@memberships\fd O] :- \\
\hspace*{0.5cm}
  //country\fd C[@car\_code=``CH''], //organization\fd O[abbrev/text()\fd ``EU''].
\end{expl}
results in
{\sf
\begin{tabbing}
==\===\==\=\kill    
\>\flq country \= \fbox{datacode=``ch''} car\_code=``CH'' 
     industry=``machinery chemicals watches''\\
\>\> memberships=``org-efta org-un \fbox{org-eu} \ldots''\frq \qquad
     \ldots\ \ \
  \flq/country\frq
\end{tabbing}
}
\end{Example}

\paragraph{Creation of Elements.}

Elements can be created as \emph{free} elements by atoms of the form
\textsf{/\textsl{name}[...]} (meaning ``some element of type
\textsf{\textsl{name}}'' -- this is interpreted to create an element
which is not a subelement of any other element), or as subelements.

\begin{figure}[htbp]
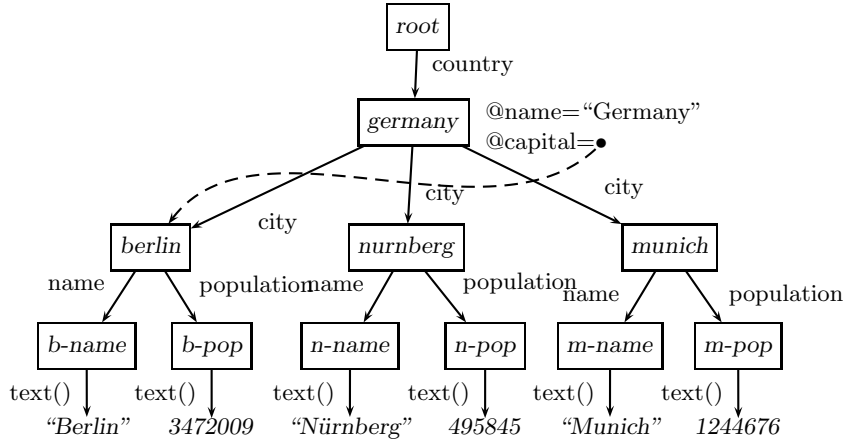

  \begin{center}
   \rnode{root}{\fbox{\textsl{\strut root}}}
   \\[0.6cm]
   \rnode{germany}{\fbox{\textsl{\strut germany}}} 
        \rput[l]{0}(0.1,.2){\small @name=``Germany''}%
        \rput[l]{0}(0.1,-.2){\small @capital=\rnode{toberlin}{$\bullet$}}%
   \\[1cm]
   \begin{tabular}{cccccc}
   \multicolumn{2}{c}{\rnode{bln}{\fbox{\textsl{\strut berlin}}}}
   & \multicolumn{2}{c}{\rnode{nrnbg}{\fbox{\textsl{\strut nurnberg}}}}
   & \multicolumn{2}{c}{\rnode{mch}{\fbox{\textsl{\strut munich}}}}
   \\[0.8cm]
   \rnode{bname}{\fbox{\textsl{\strut b-name}}}
   & \rnode{bpop}{\fbox{\textsl{\strut b-pop}}}%
   & \rnode{nname}{\fbox{\textsl{\strut n-name}}}
   & \rnode{npop}{\fbox{\textsl{\strut n-pop}}}%
   & \rnode{mname}{\fbox{\textsl{\strut m-name}}}
   & \rnode{mpop}{\fbox{\textsl{\strut m-pop}}}\\[0.7cm]
   \rnode{bstring}{\textsl{``Berlin''}}
   & \rnode{bpopnum}{\textsl{3472009}}
   & \rnode{nstring}{\textsl{``N\"urnberg''}}
   & \rnode{npopnum}{\textsl{495845}}
   & \rnode{mstring}{\textsl{``Munich''}}
   & \rnode{mpopnum}{\textsl{1244676}}
   \end{tabular}
  \nodeconnections{%
    \ncline{->}{root}{germany}\naput[npos=0.3]{\small country}
    \ncline{->}{germany}{bln}\naput[npos=0.6]{\small city}
    \ncline{->}{germany}{nrnbg}\naput[npos=0.6]{\small city}
    \ncline{->}{germany}{mch}\naput[npos=0.9]{\small city}
    \ncline{->}{bln}{bname}\nbput{\small name}
    \ncline{->}{bln}{bpop}\naput[npos=0.6]{\small population}
    \ncline{->}{nrnbg}{nname}\nbput{\small name}
    \ncline{->}{nrnbg}{npop}\naput[npos=0.6]{\small population}
    \ncline{->}{mch}{mname}\nbput[npos=0.7]{\small name}
    \ncline{->}{mch}{mpop}\naput[npos=0.8]{\small population}
    \ncline{->}{bname}{bstring}\nbput{\small text()}
    \ncline{->}{bpop}{bpopnum}\nbput{\small text()}
    \ncline{->}{nname}{nstring}\nbput{\small text()}
    \ncline{->}{npop}{npopnum}\nbput{\small text()}
    \ncline{->}{mname}{mstring}\nbput{\small text()}
    \ncline{->}{mpop}{mpopnum}\nbput{\small text()}
    \ncarc[linestyle=dashed,arcangleA=30,arcangleB=-40]{->}{toberlin}{bln}}
    \caption{Linking -- before}
    \label{fig-linking-1}
  \end{center}
\end{figure}

\begin{figure}[htbp]
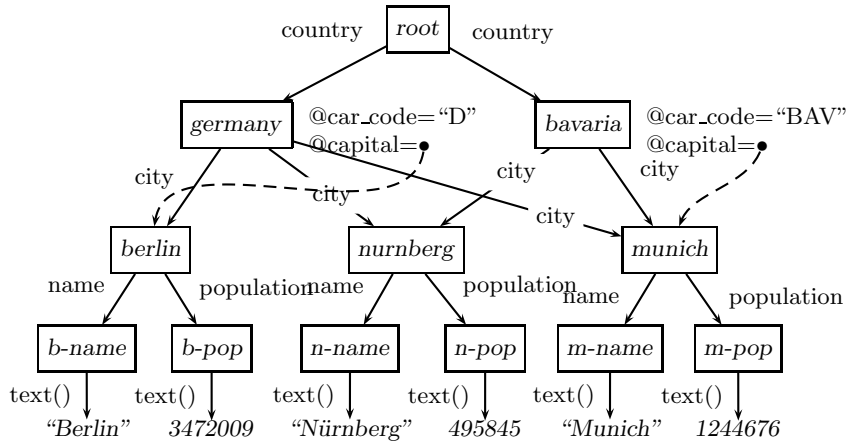

  \begin{center}
   \rnode{root}{\fbox{\textsl{\strut root}}}
   \\[0.6cm]
   \begin{tabular}{cc}
   \rnode{germany}{\fbox{\textsl{\strut germany}}} 
        \rput[l]{0}(0.1,.2){\small @car\_code=``D''}%
        \rput[l]{0}(0.1,-.2){\small @capital=\rnode{toberlin}{$\bullet$}}%
   \hspace*{3cm}
   \rnode{bavaria}{\fbox{\textsl{\strut bavaria}}} 
        \rput[l]{0}(0.1,.2){\small @car\_code=``BAV''}%
        \rput[l]{0}(0.1,-.2){\small @capital=\rnode{tomunich}{$\bullet$}}%
   \end{tabular}
   \\[1cm]
   \begin{tabular}{cccccc}
   \multicolumn{2}{c}{\rnode{bln}{\fbox{\textsl{\strut berlin}}}}
   & \multicolumn{2}{c}{\rnode{nrnbg}{\fbox{\textsl{\strut nurnberg}}}}
   & \multicolumn{2}{c}{\rnode{mch}{\fbox{\textsl{\strut munich}}}}
   \\[0.8cm]
   \rnode{bname}{\fbox{\textsl{\strut b-name}}}
   & \rnode{bpop}{\fbox{\textsl{\strut b-pop}}}%
   & \rnode{nname}{\fbox{\textsl{\strut n-name}}}
   & \rnode{npop}{\fbox{\textsl{\strut n-pop}}}%
   & \rnode{mname}{\fbox{\textsl{\strut m-name}}}
   & \rnode{mpop}{\fbox{\textsl{\strut m-pop}}}\\[0.7cm]
   \rnode{bstring}{\textsl{``Berlin''}}
   & \rnode{bpopnum}{\textsl{3472009}}
   & \rnode{nstring}{\textsl{``N\"urnberg''}}
   & \rnode{npopnum}{\textsl{495845}}
   & \rnode{mstring}{\textsl{``Munich''}}
   & \rnode{mpopnum}{\textsl{1244676}}
   \end{tabular}
  \nodeconnections{%
    \ncline{->}{root}{germany}\nbput[npos=0.3]{\small country}
    \ncline{->}{root}{bavaria}\naput[npos=0.3]{\small country}
    \ncline{->}{germany}{bln}\nbput[npos=0.6]{\small city}
    \ncline{->}{germany}{nrnbg}\ncput*[npos=0.6]{\small city}
    \ncline{->}{germany}{mch}\ncput*[npos=0.8]{\small city}
    \ncline{->}{bavaria}{nrnbg}\ncput*[npos=0.3]{\small city}
    \ncline{->}{bavaria}{mch}\naput[npos=0.5]{\small city}
    \ncline{->}{bln}{bname}\nbput{\small name}
    \ncline{->}{bln}{bpop}\naput[npos=0.6]{\small population}
    \ncline{->}{nrnbg}{nname}\nbput{\small name}
    \ncline{->}{nrnbg}{npop}\naput[npos=0.6]{\small population}
    \ncline{->}{mch}{mname}\nbput[npos=0.7]{\small name}
    \ncline{->}{mch}{mpop}\naput[npos=0.8]{\small population}
    \ncline{->}{bname}{bstring}\nbput{\small text()}
    \ncline{->}{bpop}{bpopnum}\nbput{\small text()}
    \ncline{->}{nname}{nstring}\nbput{\small text()}
    \ncline{->}{npop}{npopnum}\nbput{\small text()}
    \ncline{->}{mname}{mstring}\nbput{\small text()}
    \ncline{->}{mpop}{mpopnum}\nbput{\small text()}
    \ncarc[linestyle=dashed,arcangleA=60,arcangleB=-60]{->}{toberlin}{bln}
    \ncarc[linestyle=dashed,arcangleA=20,arcangleB=-20]{->}{tomunich}{mch}}
    \caption{Linking -- after}
    \label{fig-linking-2}
  \end{center}
\end{figure}

\begin{Example}[Creating Elements]\label{Ex-free-element}
  We create a new (free) country element with some properties
  (cf.\ Figures~\ref{fig-linking-1} and~\ref{fig-linking-2}):
  \begin{expl}
    /country[@car\_code\fd ``BAV'' and @capital\fd X 
              and city\fd X and city\fd Y] :- \\
    \multicolumn{2}{l}{\hspace*{0.5cm}      
       //city\fd X[name/text()=``Munich''], \ \
       //city\fd Y[name/text()=``Nurnberg''].}
  \end{expl}
  The two city elements are \emph{linked} as subelements.  This
  operation has no equivalent in the ``classical'' XML model: these
  elements are now children of \emph{two} country elements.  Thus,
  changing the elements effects both trees. \emph{Linking} is a
  crucial feature for efficient restructuring and integration of data
  (cf. \cite{may-behrends-FMLDO-01}).
\end{Example}

\paragraph{Insertion of Subelements.}
Existing elements can be assigned as subelements to other
elements: A ground instantiated atom $n[\textsf{child}::s\fd m]$ makes
$m$ a subelement of type $s$ of $n$.  In this case, $m$ is
\emph{linked} as $n/s$ at the end of $n$'s children list.

\begin{Example}[Inserting Subelements]\label{Ex-free-element-2}
The following two rules are equivalent to the above ones:
  \begin{expl}
    /country[@car\_code\fd ``BAV'']. \\
    C[@capital\fd X and city\fd X and city\fd Y] :-
            //country\fd C[@car\_code=``BAV''], \\
    \hspace*{0.5cm}      
       //city\fd X[name/text()\fd ``Munich''], \ \
       //city\fd Y[name/text()\fd ``Nurnberg'']. \\
    \hspace*{0.5cm}      
  \end{expl}
  Here, the first rule creates a free element, whereas the second rule
  uses the variable binding of $C$ to this element for inserting
  subelements and attributes.
\end{Example}

\noindent
In the above case, the position of the new subelement is not
specified. If the atom is of the form
$h[\textsf{child}(i){::}s{\fd}v]$ or
$h[\textsf{following/preceding-sibling}(j){::}s{\fd}v]$, this means
that the new element to be inserted should be made the $i$th
subelement of $h$ or $j$th following/preceding sibling of $h$,
respectively.

\paragraph{Generation of Elements by Path Expressions.}

Additionally, subelements can be created by \emph{reference
  expressions} in the rule head which create nested elements that
satisfy the given reference expression. The atomization introduces
local variables that occur \emph{only in the head} of the rule.  Here,
we follow the semantics of PathLog \cite{frohn-lausen-uphoff-VLDB-94}
which is implemented in \Florid\ \cite{ludaescher-himmeroeder-IS-98}
for object creation.  After the atomization, the resulting atoms are
processed in an order such that the local variables are bound to the
nodes/objects which are generated.

\begin{Example}[Inserting Text Children]
Bavaria gets a text subelement \textsf{name}:

\begin{expl}
  C/name[text()\fd ``Bavaria''] :- //country\fd C[@car\_code=``BAV'']. \\
\end{expl}

\noindent
Here, the atomized version of the rule is
\begin{expl}
  C[name\fd \_N], \_N[text()\fd ``Bavaria''] :- \\
  \hspace*{1cm}
  root[descendant::country\fd C], C[@car\_code=``BAV''].
\end{expl}
The body produces the variable binding \ \textsf{C/\textsl{bavaria}}.
When the head is evaluated, first, the fact
\textsf{\textsl{bavaria}[child::name\fd $x_1$]} is inserted, adding an
(empty) \textsf{name} subelement $x_1$ to \textsf{\textsl{bavaria}}
and binding the local variable $\_N$ to $x_1$.  Then, the second atom
is evaluated, generating the text contents to $x_1$.
\end{Example}

\paragraph{Once-for-each-Binding.}

In contrast to classical logic programming where it does not matter if
a fact is ``inserted'' into the database several times (e.g., once in
every $T_P$ round), here subelements must be created exactly once for
each instantiation of a rule.  We define a revised $T_P$-operator in
Definition~\ref{def-semantics-xpathlog-once}.

\paragraph{Using Navigation Variables for Restructuring.}
For data restructuring and integration, the intuitiveness and
declarativeness of a language gains much from variables ranging not
only over data, but also over schema concepts (as, e.g., in SchemaSQL
\cite{lakshmanan-sadri-subramanian-VLDB-96}).  Such features have
already been used for HTML-based Web data integration with F-Logic
\cite{ludaescher-himmeroeder-IS-98}.

Extending the XPath wildcard concept, XPathLog also allows to have
variables at name position. Thus, it allows for schema querying, and
also for generating new structures dependent on the data contents of
the original one.

\begin{Example}[Restructuring, Name Variables]%
\label{ex-integration-restructuring}

  Consider a data source which provides data about waters
  according to the DTD

\begin{expl} 
\flq!ELEMENT terra (water+, \ldots)\frq \\

\flq!ELEMENT water (...)\frq \quad
 \flq!ATTLIST water name CDATA \#REQUIRED \ldots\frq
\end{expl}

\noindent
which contains, e.g., the following elements:
\begin{expl}
\flq water type=``river'' name=``Mississippi''\frq\ ... \flq/water\frq \\
\flq water type=``sea'' name=``North Sea''\frq\ ... \flq/water\frq ~.
\end{expl}

\noindent
This tree should be converted into the target DTD
\begin{expl} 
\flq!ELEMENT geo ((river$|$lake$|$sea)*)\frq \\
 \flq!ELEMENT river (\ldots)\frq\qquad 
 \flq!ATTLIST river name CDATA \#REQUIRED \ldots\frq \\
  \hspace*{0.5cm} \emph{(analogously for lakes and seas)}
\end{expl}

\noindent
The first rule,  \
\textsf{\upshape
         result/T[@name\fd N] :- //water[@type\fd T and @name\fd N]} \\
creates 
\textsf{\upshape\flq river name=``Mississippi''/\frq} \ and \
\textsf{\upshape\flq sea name=``North Sea''/\frq}~.

\noindent
Attributes and contents are then transformed by separate rules which
copy properties by using variables at element name and attribute name
position:
\begin{expl}
  X[@A\fd V] & :- //water[@type\fd T and @name\fd N and @A\fd V], 
                //T\fd X[@name\fd N]. \\
  X[S\fd V] & :- //water[@type\fd T and @name\fd N and S\fd V], 
                //T\fd X[@name\fd N].
\end{expl}
\end{Example}

\subsection{Global Semantics of Positive XPathLog Programs}\label{sec-programs}

\noindent
An XPathLog program is a declarative specification how to manipulate
an XML database, starting with one or more input documents.  The
semantics of XPathLog programs is defined by bottom-up evaluation
based on a $T_P$ operator.  Thus, the semantics coincides with the
usual understanding of a stepwise process.

For implementing the once-for-each-binding approach, the $T_P$
operator has to be extended with bookkeeping about the instances of
inserted rule heads. Additionally, the insertion of subelements adds
some nonmonotonicity: adding an atom \textsf{n[child(i)::e\fd v]} adds
a new subelement at the $i$th position, making the original $i$th
child/sibling the $i+1$st etc.  In case of multiple extensions to the
same element, the positions are determined wrt.\ the original
structure.

\begin{Definition}[Extension of X-structures]%
\label{def-extension-DOM-HS}
Given an X-Structure $\pl{X}$ and a set $\pl I$ of ground-instantiated
atoms as obtained from \atomize\ to be inserted, the new
X-Structure \ $\pl{X}'= \pl{X} \prec \pl I$ \ is obtained as follows:
  \begin{itemize}
  \item initialize \ 
    $\begin{array}[t]{@{}l}
      \pl A_{\pl{X}'}(\textsf{child},x) := 
        \pl A_{\pl{X}}(\textsf{child},x),~
      \pl A_{\pl{X}'}(\textsf{attribute},x) := 
        \pl A_{\pl{X}}(\textsf{attribute},x)~, \\
     \preds(\pl{X}') :=  
         \preds(\pl{X}) \cup 
         \{ p \st \mbox{$p\in \pl I$ is a predicate atom}\}
      \end{array}$
  
    \noindent
    for all node identifiers $x$.
  \item for all elements of $\pl A_{\pl{X}}(\textsf{child},x)$, let \
    $\alpha(\pl A_{\pl{X}}(\textsf{child},x)[i]) := 
      \pl A_{\pl{X}'}(\textsf{child},x)[i]$ \\
     ($\alpha$ maps the indexing from the old list to the new one).
  \item for all atoms $x[\textsf{child}(i)::e\fd y] \in \pl I$,
     insert $(y,e)$ into $\pl A_{\pl{X}'}(\textsf{child},x)$ 
     immediately after \
     $\alpha(\pl A_{\pl{X}}(\textsf{child},x)[i])$.
  \item for all atoms $x[\textsf{child}::e\fd y] \in \pl I$,
     append $(y,e)$ at the end of $\pl A_{\pl{X}'}(\textsf{child},x)$.
  \item analogously for sibling axes.
  \item for all atoms $x[@a\fd y] \in \pl I$,
     append $(y,a)$ to $\pl A_{\pl{X}'}(\textsf{attribute},x)$.
  \end{itemize}
\end{Definition}

\begin{Proposition}[Extension of X-Structures]
  The extension operation is correct: 
  $\pl{X} \prec \pl I \models \pl I$,
  i.e., when querying the inserted atoms, the query evaluates to true.
\end{Proposition}

\noindent
With the correctness of \atomize, the insertion of
rule heads performs correctly:

\begin{Corollary}[Correctness of Insertions]
  For inserting the ground-instantiated head of a rule, it is correct
  to insert the atomized head: For all ground XPathLog atoms A, \
  $\pl X \prec \atomize(A) \models A$~.
\end{Corollary}

\begin{Definition}[$\TX_P$-Operator for XPath-Logic Programs]%
\label{def-semantics-xpathlog-once}
The $\TX$-operator works on pairs $(\pl{X}, Dic)$ where 
$\pl{X}$ is an X-Structure, and $Dic$ is a dictionary
which associates to every rule a set $\xi$ of bindings which have
been instantiated in the current iteration:

\noindent
\begin{eqnarray*}
  (\pl{X},Dic) + (\{(r_1,\bel_1),\ldots,(r_n,\bel_n)\}) &:=&
    (\begin{array}[t]{@{}l}
     \pl{X} \prec \{\bel_i(\atomize(head(r_i))) \st 1\le i \le n\}, \\
      Dic.insert(\{(r_1,\bel_1),\ldots,(r_n,\bel_n)\})) ~,
    \end{array} \\
  (\pl{X},Dic)\downarrow_1 &:=& \pl{X}~.
\end{eqnarray*}

\noindent
For an XPathLog program $P$ and an X-Structure $\pl{X}$,
\[\begin{array}{@{}l@{\,}l@{\;}l}  
  \TX_P(\pl{X}, B) & := & (\pl{X},B)  
      + \{ (r,\bel) \mid 
      \begin{array}[t]{@{}l}
      r = (h \leftarrow b) \in P  
      \mbox{ and } \pl{X} \models \bel(b)
      \mbox{, and } (r,\bel) \notin B\}~,
      \end{array} \hidewidth\\
  \TX^0_P(\pl{X}) &:=& (\pl{X},\emptyset)~, \smallskip\\
  \TX^{i+1}_P(\pl{X}) &:=& \TX_P(\TX^i_P(\pl{X}))~, \smallskip\\
  \TX_P^\omega(\pl{X}) &:=& \left\{
      \begin{array}{l}
        (\lim_{i\to\infty} \TX^i_P(\pl{X}))\downarrow_1 
              ~\mbox{ if 
            $\TX^0_P(\pl{X}),\TX^1_P(\pl{X}),\ldots$} \mbox{ converges,} \\
        \bot ~ \mbox{ otherwise.}
      \end{array} \right.
    \end{array}\]
\end{Definition}
  
\begin{Remark}
  Note that for pure Datalog programs $P$ (i.e., only predicates over
  first-order terms), the evaluation wrt.\ $\TX_P$ does not change the
  semantics, i.e., \ $\TX_P^\omega(\pl{X}) = T_P^\omega(\pl{X})$~.
\end{Remark}

\begin{Proposition}[Properties of the $\TX_P$ operator]
The $\TX_P$ operator extends the well-known $T_P$ operator. For all
positive XPathLog programs $P$, the following holds:
\begin{itemize}
\item without considering context functions, the $\TX_P$ operator is
  monotonous (which guarantees that a minimal fixpoint
  $\TX_P^\omega(\pl{X})$ exists),
\item $\TX_P^\omega(\pl{X}) \models P$,
\item $\TX_P$ is order-preserving: for all XPathLog reference
  expressions $expr$ which do not use negation or context functions,
  $\pl S_{\pl{X}}(expr)$ is a sublist of 
  $\pl S_{\TX_P(\pl{X})}(expr)$,
\item for all atoms $A$ that do not contain aggregations or function
  applications, if $A$ holds in $\pl X$, then it also holds after
  application of $\TX_P$:  \mbox{$\pl{X} \models A ~\Rightarrow~
  \TX_P(\pl{X}) \models A$.}
\end{itemize}
\end{Proposition}

\begin{Proof}
  Both properties follow immediately from the definition. The child
  and attribute axes are extended solely by appending and inserting
  new ``facts''.
\end{Proof}

\subsection{Semantics of General XPathLog Programs}

For logic programs which use negation (or similar nonmonotonic
features, such as aggregation), there is no \emph{minimal model
  semantics}. Instead, their semantics is defined wrt.\ perfect
models, well-founded models, or stable models.  For practical use -
especially when considering bottom-up evaluation -- the notion of
\emph{perfect models} and \emph{stratification}
\cite{przymusinski-INCOLL-minker-88} provides a solution to the
problems raised by negation and other nonmonotonic features.
Stratification expresses the intuitive notion of process which
executes as a sequence of steps.


Note that already not all Datalog programs are stratifiable.  For
logics over complex structures such as e.g., F-Logic, a reasonable
notion of stratification can be defined based on the names occurring
at property position -- as long as variables are not allowed at the
property position.  With variables allowed at property position, it
has been showed for F-Logic in \cite{frohn-PHD-98} that programs are
in general not stratifiable.  Since (i) even without variables at
property position, there are many programs which are not syntactically
stratifiable, and (ii) variables at the property position prove to be
very useful for data integration (cf.\ 
Example~\ref{ex-integration-restructuring}), syntax-based
stratification is not suitable for our approach.  Since the intention
of XPathLog programs is in general to implement a stepwise process by
bottom-up evaluation, often there is a natural, \emph{user-defined}
stratification.  User defined stratification is supported in the
\LoPiX\ system \cite{LoPiX-www} (cf.\ Section~\ref{sec-lopix}).  The
semantics is computed in the same way as for positive programs by
iterating the $\TX_P$ operator for each stratum.

\paragraph{Language Extensions.}

In addition to the pure language as described above, XPathLog supports
several extensions. A detailed description of, e.g., aggregation (as
known, e.g., from SQL) and a class hierarchy and signatures (taken
from F-Logic), and data-driven Web access can be found in
\cite{may-habil-01}.

\section{Implementation and Application}\label{sec-lopix}
\subsection{Implementation: The \LoPiX\ System}

\begin{figure}[htbp]
  \begin{center}
\setlength{\unitlength}{0.1mm}%
\begin{picture}(1300,1070)

\put(-60,0){\framebox(780,1070){}}
\put(160,160){\database[t]{3cm}{0.8cm}%
[fillstyle=solid,fillcolor=lightgray]{\begin{tabular}{c}Object\\Manager
                                      \end{tabular}}[OM]}

\thicklines
\put(-30,220){\rput{90}(0,0){Storage}}
\put(0,15){\framebox(680,465){}}
\put(10,330){\framebox(420,140)
     {\rnode{OMAccess}{\rnode{OMAccessLeft}{\strut}\bf OM Access}}}
\put(440,330){\framebox(230,140){\rnode{WebAccess}
  {\bf \begin{tabular}{c}WebAccess\\ ~~\\ ~~\end{tabular}}}}
\thinlines
\put(450,340){\framebox(100,80){\rnode{DTDParser}
  {\begin{tabular}{c}DTD\\ Parser\end{tabular}}}}
\put(560,340){\framebox(100,80){\rnode{XMLParser}
  {\begin{tabular}{c}XML\\ Parser\end{tabular}}}}
\thicklines
\put(-30,630){\rput{90}(0,0){Evaluation}}
\put(0,510){\framebox(680,230){}}
\put(10,520){\framebox(340,100){\rnode{AlgEval}
  {\bf \begin{tabular}{c}Algebraic\\ Evaluation ($\pl S$)\end{tabular}}}}
\put(360,520){\framebox(310,100){\rnode{AlgIns}
  {\bf \begin{tabular}{c}Algebraic\\ Insertion\end{tabular}}}}
\put(10,630){\framebox(660,100){\rnode{LogEval}
  {\bf \begin{tabular}{c}Logic Evaluation\\ (Bottom-up) $\TX_P$\end{tabular}}}}
\thicklines
\put(-30,840){\rput{90}(0,0){Execution}}
\put(0,790){\rnode{SysComm}{\framebox(230,90)
  {\begin{tabular}{c}System\\Commands\end{tabular}}}}
\put(240,790){\rnode{XPathLogParser}{\framebox(440,90)
  {\begin{tabular}{c}\bf XPathLog Parser\\ 
                     (Programs and Queries)\end{tabular}}}}

\put(0,940){\framebox(680,110){\rnode{UI}{\bf User Interface~~~~~~~~~~~~~~}}}
\thinlines
\put(400,950){\framebox(270,90){\rnode{PP}
  {\begin{tabular}{c}Pretty Printer\\ Bindings/XML\end{tabular}}}}
\put(1000,225){\rnode{Internet}
  {\framebox(50,340){\rput{90}(0,0){Internet}}}}
\put(1150,475){\htmlpage{1.2cm}{0.9cm}[fillstyle=solid,fillcolor=lightgray]
    {\begin{tabular}{c}XML\\[-0.1cm]url$_1$\end{tabular}}[XMLDoc]}
\put(1150,305){\htmlpage{1.2cm}{0.9cm}[fillstyle=solid,fillcolor=lightgray]
    {\begin{tabular}{c}DTD\\[-0.1cm]url$_2$\end{tabular}}[DTDDoc]}
\put(910,940){\rnode{OutputScreen}
  {\psframebox{\psframebox[framearc=.3]{%
      \begin{tabular}{c}
       ~\\[-0.2cm]interactive\\Output\\[-0.2cm]~\\
      \end{tabular}}}}}
\put(940,730){\htmlpage{1.2cm}{0.9cm}[fillstyle=solid,fillcolor=lightgray]
    {\begin{tabular}{c}XML\\[-0.1cm]output\end{tabular}}[OutputFile]}
\put(820,80){\begin{tabular}{ll}
    \rnode{l1}{\strut}\qquad~~\rnode{r1}{\strut}
    \ncline[doubleline=true,linecolor=blue,linestyle=dashed]{->}{l1}{r1}
    & Internet in- and output \\
    \rnode{l2}{\strut}\qquad~~\rnode{r2}{\strut}
    \ncline[doubleline=true,linecolor=blue]{->}{l2}{r2}
    & inserts to internal storage\\
    \rnode{l3}{\strut}\qquad~~\rnode{r3}{\strut}
    \ncline[linecolor=red]{->}{l3}{r3}
    & internal information flow\\
    \rnode{l4}{\strut}\qquad~~\rnode{r4}{\strut}
    \ncline[linecolor=magenta,linestyle=dashed]{->}{l4}{r4}
    & querying internal storage
  \end{tabular}}
\end{picture}
\nodeconnections{%
\ncline[linecolor=magenta,linestyle=dashed,offset=.1cm,nodesepA=.1cm]{<->}{OMAccess}{OM}
\ncline[doubleline=true,nodesepA=.2cm,linecolor=blue]{->}{WebAccess}{OM}
\ncline[doubleline=true,linecolor=blue,linestyle=dashed]{->}{Internet}{WebAccess}
  \SpecialCoor
  \pcline[doubleline=true]{<->}([angle=180]XMLDoc)%
  ([angle=180,nodesep=.9cm]XMLDoc)
  \pcline[doubleline=true]{<->}([angle=180,nodesep=-1pt]DTDDoc)%
  ([angle=180,nodesep=.9cm]DTDDoc)
  \NormalCoor
\ncline[doubleline=true,nodesep=-1pt]{<->}{Internet}{OutputFile}
\ncline[doubleline=true,linecolor=blue,nodesepB=.1cm]{->}{AlgIns}{OMAccess}
\ncline[linecolor=red,offsetA=.1cm, offsetB=-.15cm,nodesepB=.2cm]
   {<->}{AlgEval}{OMAccess}
\ncline[linecolor=red]{->}{AlgIns}{WebAccess}
\ncline[doubleline=true,linecolor=blue,nodesepA=.1cm,offset=-.1cm]
   {->}{OMAccess}{OM}
\ncline[linecolor=red]{->}{LogEval}{AlgIns}
\ncline[linecolor=red]{<->}{LogEval}{AlgEval}
\ncline[linecolor=red,offsetA=.2cm,nodesepA=-0.1cm]{->}{SysComm}{PP}
\ncarc[arcangleA=-35,arcangleB=-45,linecolor=red]{->}{LogEval}{PP}
\ncline[linecolor=red]{->}{XPathLogParser}{LogEval}
\ncline[linecolor=red,nodesepA=0.1cm,nodesepB=-.2cm,offsetB=-.3cm]
   {->}{UI}{SysComm}
\ncline[linecolor=red,nodesepA=0.1cm]{->}{UI}{XPathLogParser}
\ncline[linecolor=red,nodesepA=0.1cm,offsetA=.05cm,nodesepB=0.1cm]
   {<->}{SysComm}{OMAccessLeft}
\ncline[linecolor=blue,linestyle=dashed]{->}{PP}{OutputScreen}
\ncline[doubleline=true,linecolor=blue,linestyle=dashed]
        {->}{PP}{OutputFile}
}
    \caption{Architecture of the \LoPiX\ System}
    \label{fig-lopix-architecture}
  \end{center}
\end{figure}
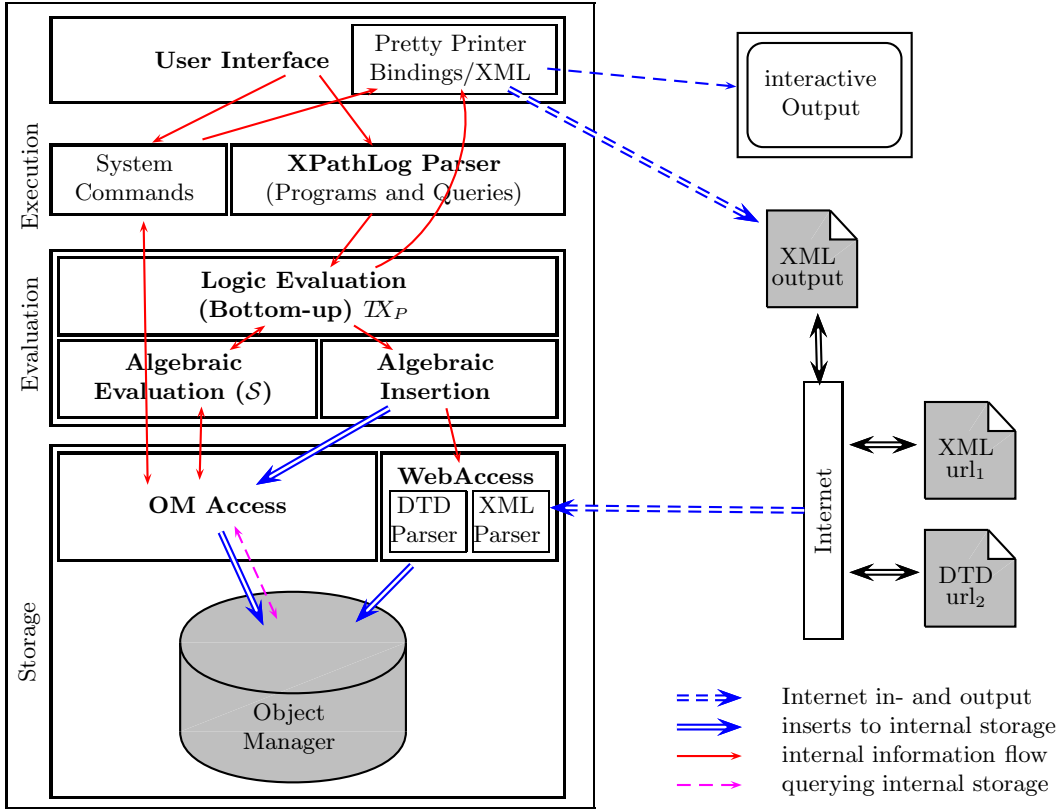

XPathLog has been implemented in the \LoPiX\ system \cite{LoPiX-www}
which extends the pure XPathLog language with a Web-aware environment
and additional functionality for data integration.  \LoPiX\ has been
developed using major components from the \Florid\ system
\cite{FLORID,ludaescher-himmeroeder-IS-98}, an implementation (in C++)
of F-Logic.  Due to the similarities between the F-Logic data model
and the XML data model in general, and XPathLogic's
multi-overlapping-tree model in particular, the \Florid\ modules
provided a solid base for an XPathLog implementation.  Especially the
functionality of the complete module for the evaluation of a deductive
language over a data model with complex objects could be reused.  The
system architecture of \LoPiX\ is depicted in
Figure~\ref{fig-lopix-architecture}.

\paragraph{Storage.}
The (extensional) database, is stored in the \emph{ObjectManager}.
Here, two variants have been developed: the first one uses a
proprietary integrated, frame-based model (from \Florid) that is
equipped with indexes for optimized access, whereas the second one is
based on a standard DOM implementation.

The \emph{ObjectManagerAccess} encapsulates the storage by
implementing the abstract XTreeGraph data model based on the contents
of the \emph{ObjectManager}.  This abstraction level also adds
intensional properties including derived axes, transitivity of class
hierarchy, downwards closure of signatures, inheritance, object
fusion, synonyms, built-in functionality for data conversion, string
handling including matching regular expressions, arithmetics, and
aggregation operators.

The \emph{WebAccess} functionality is closely intertwined with the
\emph{OMAccess} module: XML sources are mapped to trees in the
internal database.  Additionally, a method for mapping a DTD to
XPathLog signature atoms is provided.

\paragraph{Evaluation.}
The central \emph{Evaluation} module (\emph{LogicEvaluation,
  AlgebraicEvaluation}, and \emph{AlgebraicInsert}) is taken nearly
unchanged from \Florid\ and provides in fact a \emph{generic}
implementation of a deductive language over a data model with complex
objects.  \emph{LogicEvaluation} implements a seminaive bottom-up
evaluation of rules. \emph{AlgebraicEvaluation} translates rule bodies
and heads into the underlying object algebra and evaluates the
generated algebraic expressions using the query interface of
\emph{OMAccess}.  The object algebra implements the semantics of
XPathLog queries described in Section~\ref{sec-queries-xpathlog},
generating sets of tuples of variable bindings. \emph{AlgebraicInsert}
instantiates the rule heads with the generated variable bindings and
adds the corresponding facts into the database using again the
\emph{OMAccess} interface, implementing the $\TX_P$-semantics defined
in Section~\ref{sec-xpathlog-rules}.  The evaluation of algebraic
expressions does not materialize any intermediate result, but is
purely based on nested iterators.

\paragraph{Execution and UserInterface.}
The execution module provides the infrastructure for the system,
consisting of a \emph{Parser} (lex/yacc-based) and a
\emph{SystemCommands} module that implements (partially) non-logical
commands for controlling the evaluation process.  The
\emph{UserInterface} module allows to use \LoPiX\ from the command
shell by invoking system commands and stating interactive queries.
The \emph{PrettyPrinter} outputs answers in the variable bindings
format known from Datalog; additionally, the result of queries that
bind only a single variable can be output as a result set in XML ASCII
representation.  Additionally, \emph{result views}, i.e., the
projections of trees rooted in a given node to a given signature can be
exported.

\subsection{Case-Study: \Mondial}
\label{sec-mondial-case-study}

XPathLog/LoPiX has successfully been applied in the \Mondial\ case
study \cite{Mondial-www,lopix-mondial-01}. There, the practicability
of the approach for data integration is illustrated by integrating a
geographical database from the XML representations of its sources
(which have been created by \Florid\ wrappers in
\cite{may-MONDIAL-report-99}).

\begin{description}
\item[The CIA World Factbook:] The CIA World Factbook Country Listing
  (\textsf{cia:},
  \url{http://www.odci.gov/cia/publications/pubs.html}) provides
  political, economic, and social and some geographical information
  about the countries.  A separate part of the CIA World Factbook
  provides information about political and economical organizations
  (\textsf{orgs:}).  Here, the data sources overlap by the membership
  relation: with every organization, the member countries are stored
  in \textsf{orgs} by name (using the same names as in the
  \textsf{cia} part).

\item[Global Statistics: Cities and Provinces:]
    
  The \emph{Global Statistics} data (\textsf{gs:},
  \url{http://www.stats.demon.nl}) provides information (grouped by
  countries) about administrative divisions (area and population,
  sometimes capital) and main cities (population with year, and
  province). Whereas the country names are the same as in CIA, the
  names of cities, that are e.g.\ capitals of countries or where the
  headquarter of a political organization is located, may differ.
\end{description}

\noindent
The case-study showed that XPathLog allows for an effective, and
elegant programming of the integration process. The nature of an
XPathLog program as a list of rules allows for grouping rules which
together handle a certain task. The programs are modular which also
allows for adapting them to potential changes in the source structure
and ontology.

For data integration in general, not only ``simple'' updates are
desired, but also specialized operations on tree fragments.  The
result is constructed using subtrees, elements, and literals of the
input sources by the integration operations that extend the basic
XPathLog in \LoPiX.  These operations heavily depend on the use of the
XTreeGraph data model \cite{may-behrends-FMLDO-01}.

\paragraph{Fusing Elements and Subtrees.}
\emph{Fusing} elements that represent the same real-world entity from
different data sources into a unified element is an important task in
information integration. The result is still an element of \emph{both}
source trees, and collects the attributes and subelements of both
original elements.

\begin{Example}[Object Fusion]\label{ex-integra-1}
  Consider two data sources as shown below and in
  Figure~\ref{fig-element-fusion}(a).  Both describe countries, where
  \textsf{cia} contains information about name, area, population, and
  capital, and \textsf{gs} contains information about cities.
{\small\sf
\begin{tabbing} 
==\===\=\kill
\flq!ELEMENT cia (country+)\frq \\[0.05cm]
\flq!ELEMENT country (border*)\frq \\
 \hspace*{0.5cm}
 \flq!ATTLIST country \ 
       \= name CDATA \#REQUIRED \ \ \ \ \ \= car\_code ID \#REQUIRED \\
       \> area CDATA \#IMPLIED  \> population CDATA \#IMPLIED \\
       \>capital CDATA \#REQUIRED\frq  
\\[0.05cm]
\flq!ELEMENT border (\#PCDATA)\frq \quad\
\flq!ATTLIST border country IDREF \#REQUIRED\frq \\[0.1cm]

\flq!ELEMENT gs (country+)\frq \\[0.05cm]

\flq!ELEMENT country (city+)\frq\ \
 \flq!ATTLIST country \ 
       \= name CDATA \#REQUIRED\frq  \\[0.05cm]

\flq!ELEMENT city EMPTY\frq \\
\hspace*{0.5cm}
\flq!ATTLIST city name CDATA \#REQUIRED\ pop CDATA \#REQUIRED\frq
\end{tabbing}
}

\paragraph{Excerpts of the instances:}\ \\
\begin{minipage}[t]{.48\textwidth}
{\small\sf
 \begin{tabbing} 
=\=\kill
\flq cia\frq \\
\> \flq country \=
    car\_code='D' \ capital='Berlin' \\
    \>\>name='Germany' \ area='356910' \\
    \>\>population='83536115'\frq\\
\> \hspace*{0.3cm}
  \flq border country='F'\frq 451\flq /border\frq\ \\
\> \hspace*{0.3cm}
  \flq border country='A'\frq 784\flq /border\frq\ \\
\> \hspace*{0.6cm} : \\
\>\flq /country\frq  \\
\> \hspace*{0.3cm} : \\
\flq /cia\frq
\end{tabbing}
}\end{minipage}
\begin{minipage}[t]{.48\textwidth}
{\small\sf
 \begin{tabbing} 
=\=\kill
\flq gs\frq\ \\
\> \flq country  name='Germany'\frq\\
\> \hspace*{0.2cm}
  \flq city name=``Berlin'' pop=``3472009''/\frq \\
\> \hspace*{0.2cm}
  \flq city name=``Hamburg'' pop=``1705872''/\frq \\
\> \hspace*{0.4cm} : \\
\>\flq /country\frq  \\
\> \hspace*{0.2cm} : \\
\flq /gs\frq
\end{tabbing}
}\end{minipage}
\smallskip

\noindent
An obvious and typical integration step is to unify the
\textsf{countries} in the \textsf{cia} tree with the
\textsf{countries} in the \textsf{gs} tree.  In XPathLog, this is done
by the rule
\begin{expl}
  C1 = C2 :- cia/cia:country\fd C1[@cia:name\fd N],\
             gs/gs:country\fd C2[@gs:name\fd N].
\end{expl}
The example is continued below,
Figure~\ref{fig-element-fusion}(b) depicts the final result.
\end{Example}

\paragraph{Synonyms.}
Names are also subject of operations, e.g., the integrated database
uses a unified terminology that differs from the source terminologies.
Instead of generating new relationships between nodes, target
terminology is introduced by synonyms for already existing
relationships.

\begin{Example}[Integration: Synonyms]
  Especially, synonyms are an efficient means for taking a whole
  property from a source tree (and namespace) to the result tree:
  Consider the situation obtained in Example~\ref{ex-integra-1} where
  the following synonyms are defined:

  \begin{expl}
   cia:name = name.     &  gs:city = city.     & ~~~~~ gs:text() = text(). \\
   cia:area = area.             &  gs:name = name. \\
   cia:population = population.~~~~~ &  gs:pop = population.
  \end{expl}
\end{Example}

\paragraph{Adding Links.}
The integrated database often contains additional links (by subelement or
reference attribute relationships) between elements that originally
belong to different sources.

\begin{Example}[Integration: Additional Links]
  The integration is completed by linking the country subtrees to a
  result tree and adding the \textsf{capital} reference attributes,
  here, using \textsf{\textsl{germany}[@cia:capital=``Berlin'']} and
  \textsf{\textsl{berlin}[name=``Berlin'']}. The resulting tree
  fragment is given in Figure~\ref{fig-element-fusion}(b).

\noindent
In XPathLog, this is done by the rules
\begin{expl}
  result[country\fd C] :- cia[cia:country\fd C]. \\
  C[@capital\fd City] :- \\
  \hspace*{0.5cm}
      result/country\fd C[@cia:capital\fd Name and 
                          city\fd City[@name=Name]].   
\end{expl}      
\begin{figure}[htbp]
  \begin{center}
   \begin{tabular}{cccc}
     \rnode{cia}{\fbox{\textsl{\strut cia}}}&
     &\rnode{gs}{\fbox{\textsl{\strut gs}}} \\[0.8cm]
       \rnode{cia-germany}{\fbox{\textsl{\strut cia-germany}}}%
       \rput[l]{0}(0.2,.4){\small\sf @cia:name=``Germany''}%
       \rput[l]{0}(0.2,.1){\small\sf @cia:area=356910}
       \rput[l]{0}(0.2,-.3){\small\sf @cia:population=83536115}
       \rput[l]{0}(0.2,-.7){\small\sf @cia:capital=``Berlin''}
     &\hspace*{4cm}&
       \rnode{gs-germany}{\fbox{\textsl{\strut gs-germany}}}%
           \rput[l]{0}(0.2,.2){\small\sf @gs:name=``Germany''}
     &\hspace*{5cm}\\
     \\[0.9cm]
     \multicolumn{2}{c}{\hspace*{3cm}
         \rnode{berlin}{\fbox{\textsl{\strut berlin}}}%
           \rput[l]{0}(0.1,.2){\small\sf @gs:name=``Berlin''}%
           \rput[l]{0}(0.1,-.2){\small\sf @gs:pop=3472009}}
     &\multicolumn{2}{l}{\hspace*{2cm}%
         \rnode{hamburg}{\fbox{\textsl{\strut hamburg}}}%
           \rput[l]{0}(0.1,.2){\small\sf @gs:name=``Hamburg''}%
           \rput[l]{0}(0.1,-.2){\small\sf @gs:pop=1705872}}
   \end{tabular}
  \nodeconnections{%
    \ncline{->}{cia}{cia-germany}\naput[npos=0.3]{\small\sf cia:country}
    \ncline{->}{gs}{gs-germany}\naput[npos=0.3]{\small\sf gs:country}
    \ncline{->}{gs-germany}{berlin}\naput[npos=0.3]{\small\sf gs:city}
    \ncline{->}{gs-germany}{hamburg}\naput[npos=0.6]{\small\sf gs:city}
    \ncarc[linestyle=dashed,arcangle=-30]{->}{D}{germany} }
\bigskip

    (a) Element fusion -- before
\vskip0.6cm

   \begin{tabular}{c@{\hspace*{2cm}}c@{\hspace*{2cm}}c}
     \rnode{cia2}{\fbox{\textsl{\strut cia}}} 
     & \rnode{result}{\fbox{\textsl{\strut result}}} 
     & \rnode{gs2}{\fbox{\textsl{\strut gs}}} 
     \\[1.1cm]
     & \rnode{germany2}{\fbox{\textsl{\strut cia-germany}}}%
           \rput[l]{0}(0.2,.5){\small\sf @name=``Germany''}%
           \rput[l]{0}(0.2,.1){\small\sf @area=356910}
           \rput[l]{0}(0.2,-.3){\small\sf @population=83536115}
           \rput[l]{0}(0.2,-.7){\small\sf @cia:capital=``Berlin''}
           \rput[l]{0}(0.2,-1.1){\small\sf @capital=\rnode{toberlin}{$\bullet$}}
    \end{tabular}\\[1.5cm]
     \rnode{berlin2}{\fbox{\textsl{\strut berlin}}}
     \rput[l]{0}(0.2,.2){\small\sf @name=``Berlin''}%
        \rput[l]{0}(0.2,-.2){\small\sf @population=3472009}
     \hspace*{5cm}
     \rnode{hamburg2}{\fbox{\textsl{\strut hamburg}}}
           \rput[l]{0}(0.2,.2){\small\sf @name=``Hamburg''}%
           \rput[l]{0}(0.2,-.2){\small\sf @population=1705872}
          \hspace*{3cm}
  \vskip.1cm
  \nodeconnections{%
    \ncline{->}{cia2}{germany2}\ncput*[npos=0.5]{\small\sf cia:country}
    \ncline{->}{result}{germany2}\ncput*[npos=0.5]{\small\sf country}
    \ncline{->}{gs2}{germany2}\ncput*[npos=0.5]{\small\sf gs:country}
    \ncline{->}{germany2}{berlin2}\naput[npos=0.6]{\small\sf city}
    \ncline{->}{germany2}{hamburg2}\nbput[npos=0.6]{\small\sf city}
    \ncarc[linestyle=dashed,arcangle=-30]{->}{D}{germany}
    \ncarc[linestyle=dashed,angleA=-90,arcangleA=20,arcangleB=-50]{->}{toberlin}{berlin2} }
 
    \caption{(b) Element fusion -- after}
    \label{fig-element-fusion}
  \end{center}
\end{figure}%
\end{Example}%
\paragraph{Projection.}
When the integration and restructuring process is completed,
projections are used to define \emph{result views} of the internal
database. A result view is an XML tree, e.g., specified by a root node
and a DTD.

The complete case study in \cite{lopix-mondial-01} describes the
process of data integration, data cleaning, data restructuring and
distinguishing result tree views.  The program is easily extendible by
additional rules for adding another data source.

\section{Analysis, Related Work, and Conclusion}%
\label{sec-related-and-conclusion}

\subsection{Comparison with other XML Languages}\label{sec-comparison-xpathlog}

\paragraph{XPathLog vs.\ Requirements.}

In \cite{fernandez-simeon-wadler-MISC-99}, XQL, XML-QL, and the
languages YATL \cite{cluet-delobel-simeon-smaga-SIGMOD-98} and
\textsl{Lorel}
\cite{abiteboul-quass-mchugh-widom-wiener-jodl-97,goldman-mchugh-widom-99}
are compared and essential features of an XML query language have
been identified. XPathLog relates to their requirements as follows:

\begin{itemize}
\item existence of some kind of pattern clause, step qualifier clause, and
  constructor clause: pattern and step qualifier clause are the same as in
  XPath, extended with variable bindings. The path patterns are
  superior to XML patterns (as e.g.\ used in XML-QL) since they allow
  for dereferencing and navigation along different axes.  The
  constructor clause uses the same XPath-based syntax.
\item constructs for imposing nesting and order: nested elements in
  the result tree are generated by subsequent rules which stepwise
  generate the result.  Grouping (via stepwise generation) and order
  (via \textsf{child($i$)::name}) is supported.
\item combining data from different sources is supported.
\item tag variables or regular path expressions: tag variables are
  supported, regular path expressions are not included in the basic
  XPathLog language (also not in XPath). They are definable as derived
  relations.
\item alternatives are expressible using a separate rule for each
  alternative.
\item checking for absence of information: existence or non-existence
  of properties can be tested using negation, e.g.\ 
  \textsf{//country[not @indep\_date]}.
\item external functions: aggregation, string functions and some
  data conversion is built-in; the set of functions is extensible.
\item navigation along references: implicit dereferencing is supported.  
\end{itemize}

\paragraph{Semistructured Data Languages.} 

We have already mentioned the use of logic programming style languages
in pre-XML projects on semistructured data in
Section~\ref{sec-introduction}.

\textsl{GraphLog} \cite{consens-mendelzon-pods-90} and
F-Logic/\Florid\ \cite{kifer-lausen-SIGMOD-89,kifer-lausen-wu-JACM-95}
presented logic-programming languages over graph data models that
cover the semistructured data model, but did not yet use that notion.

In GraphLog, graphical queries are defined as patterns that are
matched with an underlying graph database. The matched vertices are
bound to variables that are then used for generating an output
instance or for adding edges to the input graph in the rule head. In
the graphical representation, the ``rule head'' is represented as a
distinguished edge in the graphical pattern (to be added to the input
graph).  The language can be seen as a graphical representation of
Datalog over binary relations.  Thus, according to our criteria stated
in Section~\ref{sec-design-concepts}, GraphLog qualifies as a
logic-programming language. GraphLog excludes recursive rules, but
allows for \emph{closure literals} that represent the closure of a
binary predicate; thus the expressiveness of the language is the same
as for stratified linear Datalog.

F-Logic \cite{kifer-lausen-SIGMOD-89,kifer-lausen-wu-JACM-95} is a
\emph{deductive object-oriented database language} that can be seen as
an early concept of a \emph{semistructured}, \emph{self-describing}
data model.  F-Logic defines a data model, a logic, and a database
query and programming language (similar to the relationship between
the X-Structures, XPath-Logic and XPathLog).  The experiences with
F-Logic as a formal framework and as a language for data extraction
and integration from the Web
\cite{ludaescher-himmeroeder-IS-98,may-MONDIAL-report-99} provided the
background for the design of XPath-Logic and XPathLog as a crossbreed
between XPath and F-Logic, combining the experiences with F-Logic as a
successful (but ``proprietary'') language for data integration with
the standards of XML and XPath was a well-grounded evolution step.
Especially, the power of the graph-based F-Logic data model compared
with the restricted tree model of XML made up a central requirement in
the design of XPathLog, leading to the XTreeGraph data model for
\emph{virtual trees in a graph database}. Another important aspect
taken from F-Logic is to have names as first-order citizens of the
language for a seamless incorporation of metadata information.  Due to
these similarities, it was possible to base the implementation of
XPathLog in the \LoPiX\ system on the F-Logic system \Florid.

The \textsl{OEM (Object Exchange Model)} of the \textsc{Tsimmis}
project
\cite{garcia-molina-papakonstantinou-quass-JIIS-97,abiteboul-quass-mchugh-widom-wiener-jodl-97}
was the first data model that was dedicated explicitly to the notion
of \emph{semistructured data}. OEM is a graph based model, for which
node-labeled and edge-labeled presentations have been given.  With
\emph{WSL} and \emph{MSL (Wrapper/Mediator Specification Language)},
Datalog-style programming languages have been presented. The
\emph{Lorel} language
\cite{mchugh-abiteboul-goldman-quass-widom-SIGMODREC-97} is similar to
OQL, combining navigational access (extended with regular path
expressions) with clauses. Lorel supports SQL-like, procedural update
constructs.  \textsl{Lorel} has been migrated to XML in
\cite{goldman-mchugh-widom-99}. In contrast to the XPathLog/\LoPiX\ 
migration, \textsc{Lorel} does not support the XML axes.

\textsl{UnQL}
\cite{buneman-davidson-hillebrandt-suciu-SIGMOD-96,buneman-fernandez-suciu-VLDBJ-2000}
operates on rooted, edge-labeled graphs.  It embeds \emph{graph
  schemata} that are matched as patterns with the underlying database,
combined with navigational access into SQL-like clauses.  UnQL's
semantics is based on \emph{structural recursion} -- similar to the
later XSL.

\textsl{Strudel/StruQL}
\cite{fernandez-florescu-levy-suciu-SIGMODREC-97,fernandez-florescu-kang-SIGMOD-98}
also uses an edge-labeled graph model.  Its syntax embeds query
patterns that are matched with the underlying database into SQL-like
clauses.  StruQL rules specify what new elementary structures are
created, and what links between them are created. The Strudel
project has been continued for XML with XML-QL.

The \textsl{YATL} language of the \textsl{YAT} system
\cite{cluet-delobel-simeon-smaga-SIGMOD-98} is a pre-XML proposal,
already using SGML and DTDs.  Its trees provide a unified model for
relational, object-oriented (ODMG), and semistructured/document data
(SGML). The YATL language follows a rule-based design for complex
objects in the style of MSL or F-Logic; it supports regular path
expressions and tree algebraic operations.  In
\cite{christophides-cluet-simeon-SIGMOD-2000}, the YAT system is
turned into an XML system for data integration, which still does not
use any XML/XPath language constructs.  After mapping an XML instance
to a YAT tree, there is no notion of attributes.  Dereferencing is not
explicitly supported, and it has no notion of the XML axes (similar to
the same issue for XML-QL).

\paragraph{XML Languages.} 

XML-QL and XQuery embed XML patterns and XPath expressions,
respectively, into SQL-style clauses. Expressions can be nested.

XML-QL \cite{deutsch-fernandez-florescu-etal-WWW-99} uses XML patterns
in the head (\texttt{CONSTRUCT}) and body (\texttt{WHERE}) clause.  In
that aspect, it is the XML-pattern-counterpart to the XPath-based
XPathLog.  The XML-QL patterns for selecting elements do not support
the XML \emph{axes} except the \textsf{child} axis, and indirectly the
\textsf{descendant} by regular path expressions.  XML-QL does not
support updates; a potential combination of XML patterns and updates
is not obvious.

XQuery \cite{XQuery-W3C} embeds XPath expressions in SQL-style
\texttt{FOR - LET - WHERE - RETURN} clauses, where the \texttt{RETURN}
clause specifies the result as an XML pattern.  A proposal for
specifying \emph{updates} in XQuery has been published in
\cite{tatarinov-ives-halevy-weld-SIGMOD-2001}.  A more detailed proposal
is described in \cite{lehti-DA-01} and implemented in \cite{quip}.

XML-GL \cite{xml-gl-WWW8-1999,comai-damiani-fraternali-tois-01}
continued the idea of GraphLog for XML.  In contrast to GraphLog, the
rule body and the rule head are represented by separate graphs, called
\emph{extract-match-construct-clip-queries}.  The rule heads
\emph{generate} separate XML structures. Recursion is excluded.
The \textsc{MIX} \emph{(Mediation in XML)} system
\cite{baru-gupta-ludaescher-etal-SIGMOD-99} uses the \textsc{Xmas}
\emph{(XML Matching and Structuring)} language, derived from XML-QL
for data integration;a graphical user interface similar to XML-GL is
provided.
\textsl{XDuce} \cite{hosoya-pierce-WebDB-00} is a functional-style
tree transformation language which uses \emph{regular expression
  pattern matching} of (originally, SGML) DTDs for formulating queries
against XML instances.

\textsl{Xcerpt} \cite{bry-schaffert-iclp-02} is a \emph{pattern-based}
language for querying and transforming XML data.  It follows a clean,
rule-based design where the query (matching) part in the body is
separated from the generation part in the rule head.  XML instances
are regarded as terms that are matched by a term pattern in the rule
body, generating variable bindings. The semantics and the
implementation is given by \emph{simulation unification} that computes
answer substitutions for the variables in the match pattern
against the underlying XML term (similar to UnQL).  Then, the term in
the rule head is instantiated with these variable bindings.  Since
rule heads have only a \emph{generating} semantics, but not an update
semantics, Xcerpt can only be used for querying and transforming XML
data, but not for updating/extending an existing internal XML
database.  It has a rule-based semantics, but there is no global logic
programming semantics for the evaluation of programs.

\textsl{Elog} \cite{baumgartner-flesca-gottlob-LPAR-01} is a logic
programming language for XML which is used as \emph{internal} language
for XML data extraction in the Lixto project
\cite{baumgartner-flesca-gottlob-VLDB-01}. It is based on flattening
XML data into Datalog with specialized Web Access predicates.

\begin{table}[htbp]
\small
\fbox{\begin{tabular}{l|cccccc@{}}
{\bf SSD} &  GraphLog & WSL/MSL & Lorel & UnQL   & StruQL & F-Logic \\\hline
DataModel &   graph & graph/atoms & graph & graph  & graph & graph  \\
Access    &  patterns &patterns&pat/nav&term unif.&patterns & navigation \\
Views     &       y   &    y    &  y    &   y    &   y    &  y \\
Interfering\\
\hspace*{0.3cm} additions 
          &     y   &      n      &  y    &   n    &   n  & y  \\
Paradigm  &    LP     & rules   & SQL   &  SQL   & SQL    & LP  
\end{tabular}}

\fbox{\begin{tabular}{l|cccccc@{}}
{\bf XML} & XML-QL & XQuery & XML-GL & Xcerpt    & Elog  & XPathLog \\\hline
DataModel & XML    & XML    &XML tree& XML tree & atoms & XTreeGraph \\
Access    &patterns&navigation&patterns&term unif.&(atoms)& navigation \\
Views     &   y    &   y    &   y    &  y      & y     & y         \\
Interf. add.&  n    &(XUpdate) & n   &  n      & (+)   & y         \\
standard-\\
\hspace*{0.3cm} based 
          &   no    & is standard & no & no     & (no)  & yes: Xpath \\ 
Paradigm  &  SQL    & SQL    &  rules & rules    & LP    & LP     
\end{tabular}}
    \caption{Comparison}\label{tab-comparison}
\end{table}

Table~\ref{tab-comparison} gives a comparison of some of the
above-mentioned languages.  The ``paradigm'' column indicates the
underlying semantics of the languages: the semantics of SQL-like
languages is best given as an algebraic semantics that specifies the
type and value of expression, allowing for nested expressions. For
rule-based languages, a denotational specification of the outcome of
the right-hand side (query) and of the result of the left-hand side is
required. Logic programming languages require both a model-theoretic
semantics (to specify the outcome of rule heads, and for the
\emph{global} semantics), and an answer semantics for the querying
part. 

\subsection{Contributions}

We have described XPath-Logic as a logic-based framework for handling
XML data, together with an extended XML data model that is suitable
for XML querying, manipulation, and integration.  XPathLog combines
the intuitive ``local'' semantics of addressing XML data by XPath with
the appeal of the ``global'' logic programming semantics: it is
completely XPath-based, i.e., both the rule bodies and the rule heads
use an extended XPath syntax, thereby defining a \emph{constructive}
semantics for XPath expressions.  Although the syntactic difference
between XPath and XPathLog is small, the extension adds much to the
language by turning it into a data manipulation language.  The close
relationship with XPath ensures that its declarative semantics is well
understood from the XML perspective.  Since both XPath and rule-based
programming by using variable bindings are well-known, intuitive
concepts, the ``effect'' of the language is easy to understand on an
intuitive basis, making programming easy.  The logic programming
background provides a strong theoretical foundation of the language
concept.

The data model and the language are implemented in the \LoPiX\ system.
Its practicability has been demonstrated by the \Mondial\ case study.

\paragraph{Acknowledgments.}  Most of this work has been done when I 
was a member of the database group at Freiburg University.  I want to
thank my former colleagues during that time: Lule Ahmedi, Matthias
Ihle, Georg Lausen, Pedro Marr\'on, Martin Weber, and Fang Wei.


\begin{thebibliography}{}

\bibitem[\protect\citeauthoryear{Abiteboul, Quass, McHugh, Widom, and
  Wiener}{Abiteboul
  et~al\mbox{.}}{1997}]{abiteboul-quass-mchugh-widom-wiener-jodl-97}
{\sc Abiteboul, S.}, {\sc Quass, D.}, {\sc McHugh, J.}, {\sc Widom, J.}, {\sc
  and} {\sc Wiener, J.} 1997.
\newblock The {L}orel {Q}uery {L}anguage for {S}emistructured {D}ata.
\newblock {\em Intl. Journal on Digital Libraries (JODL)\/}~{\em 1,\/}~1,
  68--88.

\bibitem[\protect\citeauthoryear{Baru, Gupta, Lud\"ascher, Marciano,
  Papakonstantinou, Velikhov, and Chu}{Baru
  et~al\mbox{.}}{1999}]{baru-gupta-ludaescher-etal-SIGMOD-99}
{\sc Baru, C.}, {\sc Gupta, A.}, {\sc Lud\"ascher, B.}, {\sc Marciano, R.},
  {\sc Papakonstantinou, Y.}, {\sc Velikhov, P.}, {\sc and} {\sc Chu, V.} 1999.
\newblock {XML}-based information mediation with {MIX}.
\newblock In {\em ACM Intl. Conference on Management of Data (SIGMOD)}.
  597--599.

\bibitem[\protect\citeauthoryear{Baumgartner, Flesca, and Gottlob}{Baumgartner
  et~al\mbox{.}}{2001a}]{baumgartner-flesca-gottlob-LPAR-01}
{\sc Baumgartner, R.}, {\sc Flesca, S.}, {\sc and} {\sc Gottlob, G.} 2001a.
\newblock The {Elog} web extraction language.
\newblock In {\em Intl. Conference on Logic Programming and Automated Reasoning
  (LPNMR)}. Number 2250 in LNCS. Springer, 548--560.

\bibitem[\protect\citeauthoryear{Baumgartner, Flesca, and Gottlob}{Baumgartner
  et~al\mbox{.}}{2001b}]{baumgartner-flesca-gottlob-VLDB-01}
{\sc Baumgartner, R.}, {\sc Flesca, S.}, {\sc and} {\sc Gottlob, G.} 2001b.
\newblock Visual web information extraction with {Lixto}.
\newblock In {\em Intl. Conference on Very Large Data Bases (VLDB)}. 119--128.

\bibitem[\protect\citeauthoryear{Bry and Schaffert}{Bry and
  Schaffert}{2002}]{bry-schaffert-iclp-02}
{\sc Bry, F.} {\sc and} {\sc Schaffert, S.} 2002.
\newblock Towards a declarative query and transformation language for {XML} and
  semistructured data: Simulation unification.
\newblock In {\em Intl.\ Conf.\ on Logic Programming (ICLP)}. 255--270.

\bibitem[\protect\citeauthoryear{Buneman, Davidson, Hillebrandt, and
  Suciu}{Buneman
  et~al\mbox{.}}{1996}]{buneman-davidson-hillebrandt-suciu-SIGMOD-96}
{\sc Buneman, P.}, {\sc Davidson, S.}, {\sc Hillebrandt, G.}, {\sc and} {\sc
  Suciu, D.} 1996.
\newblock A query language and optimization techniques for unstructured data.
\newblock In {\em ACM Intl. Conference on Management of Data (SIGMOD)}.
  Montreal, Canada, 505--516.

\bibitem[\protect\citeauthoryear{Buneman, Fernandez, and Suciu}{Buneman
  et~al\mbox{.}}{2000}]{buneman-fernandez-suciu-VLDBJ-2000}
{\sc Buneman, P.}, {\sc Fernandez, M.}, {\sc and} {\sc Suciu, D.} 2000.
\newblock {UnQL}: A query language and algebra for semistructured data based on
  structural recursion.
\newblock {\em VLDB Journal\/}~{\em 9}, 76--110.

\bibitem[\protect\citeauthoryear{Ceri, Comai, Damiani, Fraternali, Paraboschi,
  and Tanca}{Ceri et~al\mbox{.}}{1999}]{xml-gl-WWW8-1999}
{\sc Ceri, S.}, {\sc Comai, S.}, {\sc Damiani, E.}, {\sc Fraternali, P.}, {\sc
  Paraboschi, S.}, {\sc and} {\sc Tanca, L.} 1999.
\newblock {XML-GL}: a graphical language for querying and restructuring {XML}
  documents.
\newblock In {\em Proc. 8th International World Wide Web Conference (WWW 8)}.
  1171--1187.

\bibitem[\protect\citeauthoryear{Christophides, Cluet, and
  Sim\'eon}{Christophides
  et~al\mbox{.}}{2000}]{christophides-cluet-simeon-SIGMOD-2000}
{\sc Christophides, V.}, {\sc Cluet, S.}, {\sc and} {\sc Sim\'eon, J.} 2000.
\newblock On wrapping query languages and efficient {XML} integration.
\newblock In {\em ACM Intl. Conference on Management of Data (SIGMOD)}.
  141--152.

\bibitem[\protect\citeauthoryear{Clark}{Clark}{1998}]{XT}
{\sc Clark, J.} 1998.
\newblock {XT}: an implementation of {XSL T}ransformations.
\newblock \url{http://www.jclark.com/xml/xt.html}.

\bibitem[\protect\citeauthoryear{Cluet, Delobel, Sim\'eon, and Smaga}{Cluet
  et~al\mbox{.}}{1999}]{cluet-delobel-simeon-smaga-SIGMOD-98}
{\sc Cluet, S.}, {\sc Delobel, C.}, {\sc Sim\'eon, J.}, {\sc and} {\sc Smaga,
  K.} 1999.
\newblock Your mediators need data conversion.
\newblock In {\em ACM Intl. Conference on Management of Data (SIGMOD)}.
  177--188.

\bibitem[\protect\citeauthoryear{Comai, Damiani, and Fraternali}{Comai
  et~al\mbox{.}}{2001}]{comai-damiani-fraternali-tois-01}
{\sc Comai, S.}, {\sc Damiani, E.}, {\sc and} {\sc Fraternali, P.} 2001.
\newblock Computing graphical queries over {XML} data.
\newblock {\em ACM Transactions on Information Systems (TOIS)\/}~{\em 19,\/}~4,
  371--430.

\bibitem[\protect\citeauthoryear{Consens and Mendelzon}{Consens and
  Mendelzon}{1990}]{consens-mendelzon-pods-90}
{\sc Consens, M.} {\sc and} {\sc Mendelzon, A.} 1990.
\newblock {GraphLog}: a visual formalism for real life recursion.
\newblock In {\em ACM Symposium on Principles of Database Systems (PODS)}.
  404--416.

\bibitem[\protect\citeauthoryear{Deutsch, Fernandez, Florescu, Levy, and
  Suciu}{Deutsch et~al\mbox{.}}{1999}]{deutsch-fernandez-florescu-etal-WWW-99}
{\sc Deutsch, A.}, {\sc Fernandez, M.}, {\sc Florescu, D.}, {\sc Levy, A.},
  {\sc and} {\sc Suciu, D.} 1999.
\newblock {XML-QL: A Query Language for XML}.
\newblock In {\em 8th. WWW Conference}. W3C.
\newblock World Wide Web Consortium Technical Report, NOTE-xml-ql-19980819,
  \url{www.w3.org/TR/NOTE-xml-ql}.

\bibitem[\protect\citeauthoryear{Deutsch, Fernandez, and Suciu}{Deutsch
  et~al\mbox{.}}{2000}]{deutsch-fernandez-suciu-SIGMOD-2000}
{\sc Deutsch, A.}, {\sc Fernandez, M.}, {\sc and} {\sc Suciu, D.} 2000.
\newblock Storing semistructured data with {STORED}.
\newblock In {\em ACM Intl. Conference on Management of Data (SIGMOD)}.
  431--442.

\bibitem[\protect\citeauthoryear{??}{DOM-W3C}{1998}]{DOM-W3C}
DOM-W3C 1998.
\newblock Document object model ({DOM}).
\newblock \url{http://www.w3.org/DOM/}.

\bibitem[\protect\citeauthoryear{Fernandez, Florescu, Levy, and
  Suciu}{Fernandez
  et~al\mbox{.}}{1997}]{fernandez-florescu-levy-suciu-SIGMODREC-97}
{\sc Fernandez, M.}, {\sc Florescu, D.}, {\sc Levy, A.}, {\sc and} {\sc Suciu,
  D.} 1997.
\newblock A query language for a web-site management system.
\newblock {\em SIGMOD Record\/}~{\em 26,\/}~3, 4--11.

\bibitem[\protect\citeauthoryear{Fernandez, Sim\'eon, and Wadler}{Fernandez
  et~al\mbox{.}}{1999}]{fernandez-simeon-wadler-MISC-99}
{\sc Fernandez, M.}, {\sc Sim\'eon, J.}, {\sc and} {\sc Wadler, P.} 1999.
\newblock {XML} query languages: Experiences and exemplars.
\newblock draft manuscript, communication to the XML Query W3C Working Group.
\newblock \url{http://www-db.research.bell-labs.com/user/simeon/xquery.ps}.

\bibitem[\protect\citeauthoryear{Fernandez, Florescu, Kang, Levy, and
  Suciu}{Fernandez et~al\mbox{.}}{1998}]{fernandez-florescu-kang-SIGMOD-98}
{\sc Fernandez, M.~F.}, {\sc Florescu, D.}, {\sc Kang, J.}, {\sc Levy, A.~Y.},
  {\sc and} {\sc Suciu, D.} 1998.
\newblock Catching the boat with {Strudel}: Experiences with a web-site
  management system.
\newblock In {\em ACM Intl. Conference on Management of Data (SIGMOD)}.
  414--425.

\bibitem[\protect\citeauthoryear{Florescu and Kossmann}{Florescu and
  Kossmann}{1999}]{florescu-kossman-TR-INRIA-99}
{\sc Florescu, D.} {\sc and} {\sc Kossmann, D.} 1999.
\newblock A performance evaluation of alternative mapping schemes for storing
  {XML} data in a relational database.
\newblock Tech. Rep. 3684, INRIA.

\bibitem[\protect\citeauthoryear{??}{FLORID}{1998}]{FLORID}
FLORID 1998.
\newblock \textsc{Florid} homepage.
\newblock \url{http://www.informatik.uni-freiburg.de/~dbis/florid/}.

\bibitem[\protect\citeauthoryear{Frohn}{Frohn}{1998}]{frohn-PHD-98}
{\sc Frohn, J.} 1998.
\newblock {M}agic-{S}et {T}ransformation in deduktiven, objektorientierten
  {D}atenbanksprachen.
\newblock Ph.D. thesis, Institut f\"ur Informatik, Universit\"at Freiburg.

\bibitem[\protect\citeauthoryear{Frohn, Lausen, and Uphoff}{Frohn
  et~al\mbox{.}}{1994}]{frohn-lausen-uphoff-VLDB-94}
{\sc Frohn, J.}, {\sc Lausen, G.}, {\sc and} {\sc Uphoff, H.} 1994.
\newblock Access to objects by path expressions and rules.
\newblock In {\em Intl.\ Conf.\ on Very Large Data Bases (VLDB)}. 273--284.

\bibitem[\protect\citeauthoryear{Garcia-Molina, Papakonstantinou, Quass,
  Rajaraman, Sagiv, Ullman, Vassalos, and Widom}{Garcia-Molina
  et~al\mbox{.}}{1997}]{garcia-molina-papakonstantinou-quass-JIIS-97}
{\sc Garcia-Molina, H.}, {\sc Papakonstantinou, Y.}, {\sc Quass, D.}, {\sc
  Rajaraman, A.}, {\sc Sagiv, Y.}, {\sc Ullman, J.}, {\sc Vassalos, V.}, {\sc
  and} {\sc Widom, J.} 1997.
\newblock The {TSIMMIS} approach to mediation: Data models and languages.
\newblock {\em Journal of Intelligent Information Systems\/}~{\em 8,\/}~2,
  117--132.

\bibitem[\protect\citeauthoryear{Goldman, McHugh, and Widom}{Goldman
  et~al\mbox{.}}{1999}]{goldman-mchugh-widom-99}
{\sc Goldman, R.}, {\sc McHugh, J.}, {\sc and} {\sc Widom, J.} 1999.
\newblock From semistructured data to {XML}: Migrating the {Lore} data model
  and query language.
\newblock In {\em WebDB 1999}. 25--30.

\bibitem[\protect\citeauthoryear{Hosoya and Pierce}{Hosoya and
  Pierce}{2000}]{hosoya-pierce-WebDB-00}
{\sc Hosoya, H.} {\sc and} {\sc Pierce, B.~C.} 2000.
\newblock Xduce: A typed {XML} processing language.
\newblock In {\em WebDB 2000}. 111--116.

\bibitem[\protect\citeauthoryear{Kifer and Lausen}{Kifer and
  Lausen}{1989}]{kifer-lausen-SIGMOD-89}
{\sc Kifer, M.} {\sc and} {\sc Lausen, G.} 1989.
\newblock F-{L}ogic: A higher-order language for reasoning about objects,
  inheritance and scheme.
\newblock In {\em ACM Intl. Conference on Management of Data (SIGMOD)}.
  134--146.

\bibitem[\protect\citeauthoryear{Kifer, Lausen, and Wu}{Kifer
  et~al\mbox{.}}{1995}]{kifer-lausen-wu-JACM-95}
{\sc Kifer, M.}, {\sc Lausen, G.}, {\sc and} {\sc Wu, J.} 1995.
\newblock Logical foundations of object-oriented and frame-based languages.
\newblock {\em Journal of the ACM\/}~{\em 42,\/}~4 (July), 741--843.

\bibitem[\protect\citeauthoryear{Lakshmanan, Sadri, and Subramanian}{Lakshmanan
  et~al\mbox{.}}{1996}]{lakshmanan-sadri-subramanian-VLDB-96}
{\sc Lakshmanan, L. V.~S.}, {\sc Sadri, F.}, {\sc and} {\sc Subramanian, I.~N.}
  1996.
\newblock {SchemaSQL} -- a language for interoperability in relational
  multi-database systems.
\newblock In {\em Intl. Conference on Very Large Data Bases (VLDB)}. 239--250.

\bibitem[\protect\citeauthoryear{Lehti}{Lehti}{2001}]{lehti-DA-01}
{\sc Lehti, P.} 2001.
\newblock Design and implementation of a data manipulation processor for an
  {XML} query language.
\newblock M.S.\ thesis, Technische Universit\"at Darmstadt.

\bibitem[\protect\citeauthoryear{Lud\"ascher, Himmer\"oder, Lausen, May, and
  Schlepp\-horst}{Lud\"ascher
  et~al\mbox{.}}{1998}]{ludaescher-himmeroeder-IS-98}
{\sc Lud\"ascher, B.}, {\sc Himmer\"oder, R.}, {\sc Lausen, G.}, {\sc May, W.},
  {\sc and} {\sc Schlepp\-horst, C.} 1998.
\newblock Managing semistructured data with \textsc{Florid}: A deductive
  object-oriented perspective.
\newblock {\em Information Systems\/}~{\em 23,\/}~8, 589--612.

\bibitem[\protect\citeauthoryear{May}{May}{1999}]{may-MONDIAL-report-99}
{\sc May, W.} 1999.
\newblock Information extraction and integration with \textsc{Florid}: The
  \textsc{Mondial} case study.
\newblock Tech. Rep. 131, Universit\"at Freiburg, Institut f\"ur Informatik.
\newblock Available from
  \url{http://www.informatik.uni-freiburg.de/~may/Mondial/}.

\bibitem[\protect\citeauthoryear{May}{May}{2001a}]{may-habil-01}
{\sc May, W.} 2001a.
\newblock Habilitation thesis.
\newblock Ph.D. thesis, Universit\"at Freiburg.
\newblock Available from
  \url{http://www.informatik.uni-freiburg.de/~may/lopix/}.

\bibitem[\protect\citeauthoryear{May}{May}{2001b}]{lopix-mondial-01}
{\sc May, W.} 2001b.
\newblock Information integration in {XML}: The \textsc{Mondial} case study.
\newblock Tech. rep.
\newblock Available from
  \url{http://www.informatik.uni-freiburg.de/~may/lopix/lopix-mondial.html}.

\bibitem[\protect\citeauthoryear{May}{May}{2001c}]{may-vldb-demo-01}
{\sc May, W.} 2001c.
\newblock {LoPiX}: A system for {XML} data integration and manipulation.
\newblock In {\em Intl.\ Conf.\ on Very Large Data Bases (VLDB), Demonstration
  Track}. 707--708.

\bibitem[\protect\citeauthoryear{May}{May}{2001d}]{LoPiX-www}
{\sc May, W.} 2001d.
\newblock The {\textsc{lop}\textup{ix}} system.
\newblock \url{http://www.informatik.uni-freiburg.de/~may/lopix/}.

\bibitem[\protect\citeauthoryear{May}{May}{2001e}]{Mondial-www}
{\sc May, W.} 2001e.
\newblock The \textsc{Mondial} database.
\newblock \url{http://www.informatik.uni-freiburg.de/~may/Mondial/}.

\bibitem[\protect\citeauthoryear{May}{May}{2002}]{may-DBPL-01}
{\sc May, W.} 2002.
\newblock A rule-based querying and updating language for {XML}.
\newblock In {\em Workshop on Databases and Programming Languages (DBPL 2001)}.
  Number 2397 in LNCS. 165--181.

\bibitem[\protect\citeauthoryear{May and Behrends}{May and
  Behrends}{2001}]{may-behrends-FMLDO-01}
{\sc May, W.} {\sc and} {\sc Behrends, E.} 2001.
\newblock On an {XML} data model for data integration.
\newblock In {\em Intl.\ Workshop on Foundations of Models and Languages for
  Data and Objects (FMLDO 2001)}.
\newblock Post-conference proceedings to appear with Springer LNCS.

\bibitem[\protect\citeauthoryear{McHugh, Abiteboul, Goldman, Quass, and
  Widom}{McHugh
  et~al\mbox{.}}{1997}]{mchugh-abiteboul-goldman-quass-widom-SIGMODREC-97}
{\sc McHugh, J.}, {\sc Abiteboul, S.}, {\sc Goldman, R.}, {\sc Quass, D.}, {\sc
  and} {\sc Widom, J.} 1997.
\newblock Lore: A database management system for semistructured data.
\newblock {\em SIGMOD Record\/}~{\em 26,\/}~3, 54--66.

\bibitem[\protect\citeauthoryear{Przymusinski}{Przymusinski}{1988}]{przymusins%
ki-INCOLL-minker-88}
{\sc Przymusinski, T.~C.} 1988.
\newblock On the declarative semantics of deductive databases and logic
  programs.
\newblock In {\em Foundations of Deductive Databases and Logic Programming},
  {J.~Minker}, Ed. Morgan Kaufmann, 191--216.

\bibitem[\protect\citeauthoryear{Robie}{Robie}{1999}]{XQL-W3C-99}
{\sc Robie, J.} 1999.
\newblock {XQL (XML Query Language)}.
\newblock \url{http://www.metalab.unc.edu/xql/xql-proposal.html}.

\bibitem[\protect\citeauthoryear{Shanmugasundaram, Gang, Tufte, Zhang, Witt,
  and Naughton}{Shanmugasundaram
  et~al\mbox{.}}{}]{shanmugasundaram-gang-tufte-etal-VLDB-99}
{\sc Shanmugasundaram, J.}, {\sc Gang, H.}, {\sc Tufte, K.}, {\sc Zhang, C.},
  {\sc Witt, D. J.~D.}, {\sc and} {\sc Naughton, J.}
\newblock Relational databases for querying {XML} documents: Limitations and
  opportunities.
\newblock In {\em Intl. Conference on Very Large Data Bases (VLDB)}. 302--314.

\bibitem[\protect\citeauthoryear{{Software AG}}{{Software AG}}{2001}]{quip}
{\sc {Software AG}}. 2001.
\newblock Quip: An xquery implementation.
\newblock \url{http://www.softwareag.com/developer/quip/}.

\bibitem[\protect\citeauthoryear{Tatarinov, Ives, Halevy, and Weld}{Tatarinov
  et~al\mbox{.}}{2001}]{tatarinov-ives-halevy-weld-SIGMOD-2001}
{\sc Tatarinov, I.}, {\sc Ives, Z.~G.}, {\sc Halevy, A.}, {\sc and} {\sc Weld,
  D.} 2001.
\newblock Updating xml.
\newblock In {\em ACM Intl. Conference on Management of Data (SIGMOD)}.
  133--154.

\bibitem[\protect\citeauthoryear{Wadler}{Wadler}{1999}]{wadler-misc-99}
{\sc Wadler, P.} 1999.
\newblock Two semantics for {XPath}.
\newblock \url{http://www.cs.bell-labs.com/who/wadler/topics/xml.html}.

\bibitem[\protect\citeauthoryear{??}{XMLInf}{1999}]{XML-Infoset-W3C}
XMLInf 1999.
\newblock {XML} information set.
\newblock \url{http://www.w3.org/TR/XML-infoset}.

\bibitem[\protect\citeauthoryear{??}{XMQ-A}{2001}]{XML-Query-Algebra-W3C}
XMQ-A 2001.
\newblock {XML Query Algebra}.
\newblock \url{http://www.w3.org/TR/query-algebra}.

\bibitem[\protect\citeauthoryear{??}{XMQ-D}{2001}]{XML-Query-Data-Model-W3C}
XMQ-D 2001.
\newblock {XML Query Data Model}.
\newblock \url{http://www.w3.org/TR/query-datamodel}.

\bibitem[\protect\citeauthoryear{??}{XPath}{1999}]{XPath-W3C}
XPath 1999.
\newblock {XML Path Language (XPath)} version 1.0: 1999.
\newblock \url{http://www.w3.org/TR/xpath}.

\bibitem[\protect\citeauthoryear{??}{XPQOF}{2001}]{XPath-XQuery-Functions-Oper%
ators-W3C}
XPQOF 2001.
\newblock {XQuery 1.0 and XPath 2.0 Functions and Operators}.
\newblock \url{http://www.w3.org/TR/xquery-operators}.

\bibitem[\protect\citeauthoryear{??}{XQFS}{2001}]{XQuery-Formal-Semantics-W3C}
XQFS 2001.
\newblock {XQuery 1.0 Formal Semantics}.
\newblock \url{http://www.w3.org/TR/query-semantics}.

\bibitem[\protect\citeauthoryear{??}{XQuery}{2001}]{XQuery-W3C}
XQuery 2001.
\newblock {XQuery: A Query Language for XML}.
\newblock \url{http://www.w3.org/TR/xquery}.

\bibitem[\protect\citeauthoryear{??}{XSLT}{1999}]{XSLT-W3C}
XSLT 1999.
\newblock {XSL T}ransformations ({XSLT}).
\newblock \url{http://www.w3.org/TR/xslt}.

\end{thebibliography}

\begin{appendix}
\section{Proofs}\label{sec-proofs}

\begin{Proof} of Theorem~\ref{theo-semantics-S-wadler} and
  Lemma~\ref{lemma-semantics-S-wadler}: \ The proof is done by
  structural induction.  The enumeration is the same as in
  Definition~\ref{def-semantics-xpathlogic}.  Below, $\bel$ is an
  assignment of the pseudo variables $Size$ and $Pos$ (often even
  empty).  We write $\stackrel{**}{=}$ for ``equals by definition in
  \cite{wadler-misc-99}''. The individual items of the theorem 
  are referred to below by $\textrm{IH}1,\ldots,\textrm{IH}4$
  (induction hypotheses).

\begin{enumerate}
\item For closed, absolute expressions (i.e., without free variables), \\
   \hspace*{0.3cm}
  $\pl S_{\pl X}(/expr) 
        \stackrel{\rm Def}{=}
     \pl S^{any}_{\pl X}(/expr,any,\emptyset)
        \stackrel{\rm IH1}{=}
      \pl S [[/expr]](x)$ \
   for arbitrary $x$.
\item Reference expressions (\cite{wadler-misc-99}: only absolute 
   expressions): \\
   \hspace*{0.3cm} $\pl S^{any}_{\pl X}(/p,any,\bel) \stackrel{\rm
     Def}{=} \pl S^{any}_{\pl X}(p,root,\bel) \stackrel{\rm IH2}{=}
   \pl S^{any} [[/expr]](root)$~.
\item Axis step: \\
   \hspace*{0.3cm}
   $\pl S^{any}_{\pl X}(axis::pattern,x,\bel) 
        \stackrel{\rm Def}{=}
    \pl S^{axis}_{\pl X}(pattern,x,\bel) 
        \stackrel{\rm IH2}{=}
      \pl S^{axis} [[/pattern]](x)$~.
\item The node test is the base case which is directly mapped to 
   the axes: \\
   \hspace*{0.3cm}
   $\pl S^a_{\pl X}(name,x,\bel)  \stackrel{\rm Def}{=}
      \textsf{list}_{(v,n) \in \pl A_{\pl X}(a,x)} (v \st n = name)$\\
   which is characterized in \cite{wadler-misc-99} ($\pl A[[a]]$ 
   enumerates the axes, $\pl P(a)$ gives the axes' principal nodetype)  
   by \\
   \hspace*{0.3cm}
   $\{ x_1 \st x_1 \in \pl A[[a]]x,\ nodetype(x_1) = \pl P(a),\
        \textsf{name}(x_1) = name \}$ \\
   which is the definition of \ $\pl S^a [[name]](x)$. 
  Note that dereferencing \texttt{IDREF(S)} and
  splitting \texttt{NMTOKENS} has been excluded, thus, the result list
  is still in document order.  Similar (note that \textsf{node}() is
  not defined in \cite{wadler-misc-99}, we extend the definition
  according to the XPath specification)
  \[\begin{array}{l}
    \pl S^a_{\pl X}(\textsf{node()},x,\bel) 
     \ \stackrel{\rm Def}{=}  \
        \textsf{list}_{(v,n) \in \pl A_{\pl X}(a,x)} (v \st v \in \pl V)  \\
    \hspace*{1cm}     
      = \
      \{ x_1 \st x_1 \in \pl A[[a]]x,\ nodetype(x_1) = element\} 
     \ = \
      \pl S^a [[\textsf{node}()]](x) \\
    \pl S^a_{\pl X}(\textsf{text()},x,\bel) 
     \ \stackrel{\rm Def}{=} \
        \textsf{list}_{(v,n) \in \pl A_{\pl X}(a,x)} (v \st v \in \pl V)  \\
    \hspace*{1cm}     
       = \
      \{ x_1 \st x_1 \in \pl A[[a]]x,\ nodetype(x_1) = {\it Text}\}
     \ = \
      \pl S^a [[\textsf{text}()]](x)~.
   \end{array}
   \]
\item Step with variable binding: obvious
\item Step qualifiers:
 \[ \pl S^a_{\pl X}(pattern[\textsl{stepQ}],x,\bel)
     \stackrel{\rm Def}{=}
     \textsf{list}_{y\in \pl S^a_{\pl X}(pattern,x,\bel)} 
     (y \st \pl Q_{\pl X}(\textsl{stepQ},y,
                          \bel^{k,n}_{\small\textsl{Pos},\textsl{Size}})) 
     \]
   where
   $L_1 \stackrel{\rm Def}{=} \pl S^a_{\pl X}(pattern,x,\bel)$ which 
   equals $\pl S^a [[pattern]](x,k,n)$ by induction hypothesis 
   $\textrm{IH}3$ and $n := size(L_1)$, 
   and for every $y$, let $j$ the index of $y$ in $L_1$
   (which equals $size(\{x_1\st x_1\in L_1, x_1 \le_{doc} y\}$)),
   $k := j$ if $a$ is a forward axis, and $k := n+1-j$ if $a$ is
   a backward axis.  This is the same as defined for \
   $\pl S^a [[pattern[\textsl{stepQualifier}]]](x)$ and, 
   by induction hypothesis IH3, the same as
   \[ \stackrel{\rm IH3}{=} \ 
     \textsf{list}_{y\in \pl S^a_{\pl X}(pattern,x,\bel)} 
     (y \st \pl Q[[\textsl{stepQ}]](,y,k,n))
     \ \stackrel{**}{=} \ 
     \pl S^a [[pattern[\textsl{stepQ}]]](x)~.
   \]

\item Path: \
 $\begin{array}[t]{lcl} 
    \pl S^a_{\pl X}(p_1/p_2,x,\bel) 
    & \stackrel{\rm Def}{=} &
      \textsf{concat}_{y\in \pl S^a_{\pl X}(p_1,x,\bel)}
         (\pl S^{any}_{\pl X}(p_2,y,\bel)) \\
    & \stackrel{\rm IH2}{=} &
      \textsf{concat}_{y\in \pl S^a [[p_1]](x)} (\pl S^a [[p_2]](y))
       \ \stackrel{**}{=} \
       \pl S^a [[p_1/p_2]](x)~.
     \end{array}$
\item Reference expressions (existential semantics) in step qualifiers:
\[ \begin{array}{lcl} 
    \pl Q_{\pl X}({\it refExpr},x,\bel)
    & \stackrel{\rm Def}{\Iff} &
      \pl S^{any}_{\pl X}({\it refExpr},x,\bel) \neq \emptyset \\
    & \stackrel{\rm IH3}{\Iff} &
      \pl S^{child} [[{\it refExpr}]](x) \neq \emptyset
      \stackrel{**}{\Iff}
      \pl Q [[{\it refExpr}]](x, k, n)~.
    \end{array}
    \]
   (for all $k,n$ since these are not used in ${\it refExpr}$).
 \item Predicates: \cite{wadler-misc-99} knows only the ``=''
   comparison. The definition is although not complete: e.g.\ for step
   qualifiers of the form \textsf{[a/b/c = ``foo'']} which are allowed
   in XPath, there is no semantics defined.  We extend the semantics
   according to the XPath specification, applying either $\pl S$ or
   $\pl E$.
 \[ \begin{array}{@{}l@{}c@{}l}
    \multicolumn{3}{l}%
    {\pl Q_{\pl X}(pred(expr_1,\ldots,expr_n), x, \bel)} \\
    \hspace*{.1cm}
     & \stackrel{\rm Def}{\Iff} &
      \begin{array}[t]{@{}l}
      \mbox{ there are }
      x_1 \in \pl S^{any}_{\pl X}(expr_1,x,\bel),\ldots,
      x_n \in \pl S^{any}_{\pl X}(expr_n,x,\bel) \\
      \mbox{ such that }
       (x_1, \ldots, x_n) \in \pl I_P(pred)
    \end{array} \smallskip\\
    & \stackrel{\rm IH2/4}{\Iff} &
      \mbox{ there are }
      \begin{array}[t]{l@{}}
      x_1 \in \pl S^{child} [[expr_1]](x) \mbox{ or } 
      x_1 \in \pl E[[expr_1]](x, \bel(Pos), \bel(Size)), \ldots, \\
      x_n \in \pl S^{child} [[expr_n]](x) \mbox{ or } 
      x_n \in \pl E[[expr_n]](x, \bel(Pos), \bel(Size))
      \end{array} \smallskip\\
    && \mbox{ such that $(x_1, \ldots, x_n) \in \pl I(pred)$.}
  \end{array}
    \] 
\item -- \ref{def-semantics-xpathlogic-item-functions}. Boolean
    connectives and quantification, constants, and variables: obvious.
    Functions are not defined in \cite{wadler-misc-99}, but the
    extension is obvious.
\item[\ref{def-semantics-xpathlogic-item-context-functions}.]
  Context-related functions use the extension of variable bindings
  by pseudo-variables $Size$ and $Pos$ in rule
  (\ref{def-semantics-xpathlogic-item-filter}):
 \[ \begin{array}{lclcl}
    \pl S^{any}_{\pl X}(\textsf{position()}, x, \bel) 
     &\stackrel{\rm Def}{\Iff}&
        \bel(\textsl{Pos})
     &\stackrel{**}{=}&
        \pl E[[\textsf{position}()]](x, \bel(Pos), \bel(Size)) 
    \\
    \pl S^{any}_{\pl X}(\textsf{last()}, x, \bel) 
     &\stackrel{\rm Def}{\Iff}&
    \bel(\textsl{Size})
     &\stackrel{**}{=}&
        \pl E[[\textsf{last}()]](x, \bel(Pos), \bel(Size)) ~.
  \end{array}
  \]
\end{enumerate}
\end{Proof}

\begin{Proof} of Lemma~\ref{lemma-correctness-SB-QB}: \
\begin{quote}
  \psframebox{\begin{tabular}{p{.85\textwidth}}
  {\bf Note:} A bit sloppy, we write 
  $(x,\bel) \in \pl{SB}_{\pl{X}}(expr)$ for
  ``$x \in \Res{\pl{SB}_{\pl{X}}(expr)}$ and 
    $\bel \in \Bdgs{\pl{SB}_{\pl{X}}(expr),x}$''. 
  \end{tabular}}
\end{quote}
\begin{enumerate}
\item For closed expressions, \
     $x \in \Res{\pl{SB}_{\pl{X}}({\it refExpr})} \
      \stackrel{\rm Def}{\Iff} $
  \[   x\in \Res{\pl{SB}_{\pl{X}}({\it refExpr},\emptyset)} 
     \ \stackrel{\rm IH}{\Iff} \
       x \in \pl{S}_{\pl{X}}({\it refExpr},\emptyset)
     \ \stackrel{\rm Def.\ref{def-semantics-xpathlogic}}{\Iff}
     x \in \pl{S}_{\pl{X}}({\it refExpr})~.
   \] 
\item Reference expressions are translated into path expressions wrt.\
  a start node:
  \begin{itemize}
  \item  entry points: rooted path
   \[\begin{array}{@{}l@{}c@{}l}
      \multicolumn{3}{@{}l}{(x,\bel) \in \pl{SB}_{\pl{X}}(/p,Bdgs) \
      \stackrel{\rm Def}{\Iff} \
      (x,\bel) \in \pl{SB}^{any}_{\pl{X}}(p,root,Bdgs)} \\
      \hspace*{.5cm}
      &\stackrel{\rm IH}{\Iff}&
      x \in \pl{S}^{any}_{\pl{X}}(p,root,\bel)
       \mbox{ and $\bel$ completes some $\bel'\in Bdgs$ with $\free(/p)$}) \\
      &\stackrel{\rm Def.\ref{def-semantics-xpathlogic}}{\Iff}&
      x \in \pl{S}^{any}_{\pl{X}}(/p,\bel)
       \mbox{ and $\bel$ completes some $\bel'\in Bdgs$ with $\free(/p)$})~.
     \end{array}
     \]
  \item  entry points: constants $c \in \pl V$ analogously (set $c$   
     instead of $root$ above).
  \item  entry points: variables $V\in\Var$. By definition,
   \[\begin{array}{lcl}
      (x,\bel) \in \pl{SB}_{\pl{X}}(V/p,Bdgs) 
      \stackrel{\rm Def}{\Iff}\\
     \hspace*{1cm}
      (x,\bel) \in \textsf{concat}_{x \in \pl A_{\pl{X}}({\sf descendants},root)\!\downarrow_1} 
             (\pl{SB}^{any}_{\pl{X}}(p,x,Bdgs \join \{V/x\}))
     \end{array}
    \]
    which is exactly the case if there is an 
    $x \in \pl A_{\pl{X}}({\sf descendants},root)\!\downarrow_1$
    such that \
    $(x,\bel) \in (\pl{SB}^{any}_{\pl{X}}(p,x,Bdgs \join \{V/x\}))$~. \\
    By induction hypothesis, this is equivalent with
    \[  x \in \pl{S}^{any}_{\pl{X}}(p,x,\bel)
       \mbox{ and $\bel$ completes some $\bel'\in Bdgs\join \{V/x\}$ with $\free(p)$} 
    \]
    which is exactly the case if $x = \beta(V)$ and 
    $\bel$ completes some $\bel'\in Bdgs$ with $\free(V/p)$. By
    Def.~\ref{def-semantics-xpathlogic}, this again is equivalent with
    \[ x \in \pl{S}^{any}_{\pl{X}}(V/p,\bel)
       \mbox{ and $\bel$ completes some $\bel'\in Bdgs$ with $\free(V/p)$}~.\]
   \end{itemize}
\item Axis step: \
  $(x,\bel) \in \pl{SB}^{any}_{\pl{X}}(axis::pattern,z,Bdgs)$
 \[\begin{array}{@{}lcl}
    &
     \stackrel{\rm Def}{\Iff} &
      (x,\bel) \in \pl{SB}^{axis}_{\pl{X}}(pattern,z,Bdgs) \\
    &\stackrel{\rm IH}{\Iff}&
    x \in \pl{S}^{axis}_{\pl{X}}(pattern,z,\bel) \\
    &&
     \mbox{ and $\bel$ completes some $\bel'\in Bdgs$ with 
      $\free(pattern)$} \\
    &\stackrel{\rm Def.\ref{def-semantics-xpathlogic}}{\Iff}&
    x \in \pl{S}^{any}_{\pl{X}}(axis::pattern,z,\bel) \\
    && \mbox{ and $\bel$ completes some $\bel'\in Bdgs$ with 
              $\free(axis::pattern)$}~.
     \end{array}
  \]
\item Node test: \ 
 $\begin{array}[t]{l}
    (x,\bel) \in \pl{SB}^a_{\pl{X}}(name,z,Bdgs)
    \stackrel{\rm Def}{\Iff} \\
    \hspace*{2cm}
    (x,\bel) \in  
      \textsf{list}_{(v,n) \in \pl A_{\pl{X}}(a,z),\ n = name} 
            (v,\{true\} \join Bdgs) 
   \end{array}$
   
   \noindent
   which is exactly the case if 
   $x \in \textsf{list}_{(v,n) \in \pl A_{\pl{X}}(a,z),\ n = name} (v)$
   and $\bel \in Bdgs$ which, by
   Def.~\ref{def-semantics-xpathlogic} is equivalent with
   $x \in \pl{S}^a_{\pl{X}}(name,z,\bel)$
   and $\bel$ completes some $\bel'\in Bdgs$ with 
   $\free(name) =\emptyset$.
   Analogously for \textsf{node()} and \textsf{text()}. \\
   Variables at nodetest position:
   \[(x,\bel) \in \pl{SB}^a_{\pl{X}}(N,z,Bdgs) 
    \ \stackrel{\rm Def}{\Iff} \
    (x,\bel) \in  
    \textsf{list}_{(v,n) \in \pl A_{\pl{X}}(a,z)} (v,\{N/n\} \join Bdgs) 
   \]  
   which is exactly the case if 
   $x \in \textsf{list}_{(v,n) \in \pl A_{\pl{X}}(a,z)} (v)$
   and $\bel \in \{N/n\} \join Bdgs$ which, by
   Def.~\ref{def-semantics-xpathlogic} is equivalent with
   $x \in \pl{S}^a_{\pl{X}}(N,z,\bel)$
   and $\bel$ completes some $\bel'\in Bdgs$ with 
   $\free(N) =\{N\}$.
\item Step with variable binding: 
 \[\begin{array}{@{}l@{}c@{}l}
   \multicolumn{3}{@{}l}
    {(x,\bel) \in \pl{SB}^a_{\pl{X}}(pattern\fd V,z,Bdgs)} \\
   \hspace*{0.2cm}
    &\stackrel{\rm Def}{\Iff}&
    (x,\bel) \in 
      \textsf{list}_{(y,\xi) \in \pl{SB}^a_{\pl{X}}(pattern,z,Bdgs)}
            (y, \xi \join \{V/y\}) \\
    &\Iff&
    \mbox{there is a $\bel''$ s.t. }
      (x,\bel'') \in \pl{SB}^a_{\pl{X}}(pattern,z,Bdgs)
    \mbox{ and } \bel = \bel'' \join \{V/x\}~.
   \end{array}
   \]
By induction hypothesis, this is exactly the case if
   there is a $\bel''$ such that
   $x \in \pl S^a_{\pl{X}}(pattern,z,\bel'')$ and
   $\bel''$ completes some $\bel'\in Bdgs$ with $\free(pattern)$,
   and $\bel = \bel'' \join \{V/x\}$. Exactly then, since $x=\bel(V)$, by 
   Definition~\ref{def-semantics-xpathlogic},
   $x \in \pl S^a_{\pl{X}}(pattern\fd V,z,\bel)$ and $\bel$
   completes $\bel'$ with 
   $\free(pattern\fd V) = \free(pattern) \cup \{V\}$.
\item Step Qualifier(s): \
   $(x,\bel) \in 
   \pl{SB}^a_{\pl{X}}(pattern[\textsl{stepQualifier}],z,Bdgs)$
 \[\begin{array}{@{}l@{}c@{}l}
   \hspace*{0.2cm}
    &\stackrel{\rm Def}{\Iff}&
     (x,\bel) \in
      \textsf{list}_{\begin{array}[t]{@{}l}
         \scriptstyle
          (y,\xi)\in \pl{SB}^a_{\pl{X}}(pattern,z,Bdgs),\\[-0.2ex]
         \scriptstyle
          \pl{QB}_{\pl{X}}(\textsl{\small stepQualifier},y,\xi') \neq\emptyset
         \end{array}}
       \hspace*{-1.2cm}
         (y,\ \pl{QB}_{\pl{X}}(\textsl{\small stepQualifier},y,\xi')
                                    \setminus\{Pos,Size\})
   \end{array}
   \]
   for $\xi$ as defined in Definition~\ref{def-semantics-xpathlog}%
     (\ref{def-semantics-xpathlog-item-filter}).
   This is exactly the case if
   (i) there is a $\bel''$ s.t. 
      $\bel''\in \pl{QB}_{\pl{X}}(\textsl{stepQualifier},x,\xi')$
    \mbox{ and } $\bel = \bel''\setminus\{Pos,Size\}$, and
   (ii) $(x,\xi) \in \pl{SB}^a_{\pl{X}}(pattern,z,Bdgs)$   
     i.e., $\xi$ is the corresponding set of variable bindings, and 
   (iii) $\pl{QB}_{\pl{X}}(\textsl{\small stepQualifier},x,\xi') 
      \neq\emptyset$.

   The first item is by induction hypothesis equivalent to
   $\pl Q_{\pl{X}}({\it stepQualifier},x,\bel'')$ 
       and $\bel''$ completes some $\bel'\in \xi'$ with 
       $\free({\it stepQualifier})$~
        (*).\\
   The third item is redundant here (it avoids the addition
   of elements with empty bindings list to the result).
   Since $\bel''$ completes some $\bel'\in \xi'$ with 
   $\free({\it stepQualifier})$, we know that
   $\gamma :=\bel'\setminus\{Pos,Size\}$ is an element of $\xi$.
   Specializing the second item to $\gamma$ yields 
   \
   $(x,\gamma) \in \pl{SB}^a_{\pl{X}}(pattern,z,Bdgs)$~.
   \\
   By induction hypothesis, \ \
   $x \in \pl S^a_{\pl{X}}(pattern,z,\gamma)$ \ (**) 
   \ \
   and $\gamma$ completes some $\gamma'\in Bdgs$ with $\free(pattern)$.
   Above, we derived $\gamma =\bel'\setminus\{Pos,Size\}$.
   Using $(*)$, since $\bel''$ is a completion of $\bel'$ 
   with $\free({\it stepQualifier})$, completing $\gamma'\in Bdgs$ first 
   to $\gamma$ (binding $\free(pattern)$), then to $\bel'$
   (binding $Size$ and $Pos$), then to $\bel''$ 
   (binding $\free({\it stepQualifier})$), we have \
   $ \pl Q_{\pl{X}}({\it stepQualifier},y,\bel'')$~. \\
   From $(**)$, since $\bel''$ completes $\gamma$, 
   \
   $x \in \pl S^a_{\pl{X}}(pattern,z,\bel'')$ 
   \
   thus by Def.~\ref{def-semantics-xpathlogic}, \
   the desired result
   \
   $x \in \pl{S}^a_{\pl{X}}(pattern[\textsl{stepQualifier}],z,Bdgs)$ \
   for $\bel''$ which completes $\gamma'\in Bdgs$
   with $\free(pattern[\textsl{stepQualifier}])$.

   \noindent
   The argumentation showed the ``$\Rightarrow$'' direction (which 
   is the more difficult direction since  $\gamma$ must be guessed).
   ``$\Leftarrow$'' uses the same relationships and variable bindings.
\item\label{def-semantics-xpathlog-item-concat} Path: \
  $(x,\bel)\in \pl{SB}^a_{\pl{X}}(p_1/p_2,z,Bdgs)$
 \[\begin{array}{lcl}
   \hspace*{0.2cm} 
    &\stackrel{\rm Def}{\Iff}&
    (x,\bel)\in
       \textsf{concat}_{(y,\xi)\in \pl{SB}^{any}_{\pl{X}}(p_1,z,Bdgs)}
         \pl{SB}^a_{\pl{X}}(p_2,y,\xi)\smallskip\\
    &\Iff&
    \mbox{there is an $(y,\xi)\in \pl{SB}^{any}_{\pl{X}}(p_1,z,Bdgs)$
          s.t. } 
     (x,\bel)\in\pl{SB}^a_{\pl{X}}(p_2,y,\xi)  \\
    & \stackrel{\rm IH}{\Iff} &
    \mbox{there is a $\gamma \in \xi$ s.t. there is a $\gamma'$ s.t.
        $x \in\pl S^a_{\pl{X}}(p_2,y,\gamma')$ and}\\
    && \mbox{$\gamma'$ completes $\gamma$ with $\free(p_2)$}~.
    \end{array}  
  \]
  For this $\gamma$,
  $(y,\gamma)\in \pl{SB}^{any}_{\pl{X}}(p_1,z,Bdgs)$
  and by induction hypothesis again
  $y \in\pl S^a_{\pl{X}}(p_1,z,\gamma)$
  and $\gamma$ completes some $\bel'\in Bdgs$ with $\free(p_1)$. 
  Thus, also $x \in\pl S^a_{\pl{X}}(p_2,y,\gamma')$ and 
  $y \in\pl S^a_{\pl{X}}(p_1,z,\gamma')$ and by 
  Def.~\ref{def-semantics-xpathlogic},
  $x \in\pl S^a_{\pl{X}}(p_1/p_2,z,\gamma')$.
  $\gamma'$ completes some $\bel'\in Bdgs$ with 
  $\free(p_1)\cup\free(p_2)$.

\item Reference expressions (existential semantics) in step qualifiers:
\[\begin{array}{@{}lcl}
   \multicolumn{3}{@{}l}
    {\bel \in \pl{QB}_{\pl{X}}({\it refExpr}, z, Bdgs)
    \ \stackrel{\rm Def}{\Iff} \
     \bel \in
     \bigcup_{(y,\xi)\in \pl{SB}^{any}_{\pl{X}}({\it refExpr},z,Bdgs)} \xi} \\
    &\Iff&
     \mbox{there is a $y$ s.t.
     $(y,\bel)\in \pl{SB}^{any}_{\pl{X}}({\it refExpr},z,Bdgs)$} \\
    & \stackrel{\rm IH}{\Iff} &
      y \in \pl{S}^{any}_{\pl{X}}({\it refExpr},z,\bel) \\
    &&  \mbox{ and $\bel$ completes some $\bel'\in Bdgs$ with 
             $\free({\it refExpr})$} \\
    &\stackrel{\rm Def.\ref{def-semantics-xpathlogic}}{\Iff}&
      \pl Q_{\pl{X}}({\it refExpr},z,\bel)
      \mbox{ and $\bel$ completes some $\bel'\in Bdgs$ with 
             $\free({\it refExpr})$}~.
   \end{array}
\]

\item Built-in equality predicate ``$=$'': similar to predicates and
  variable assigments by $\fd V$.  All other predicates: \
  $\bel \in \pl{QB}_{\pl{X}}(pred(arg_1,\ldots,arg_n), z, Bdgs)$
  \[\begin{array}{@{}lcl}
    \hspace*{0.5cm}
    &\stackrel{\rm Def}{\Iff}&
     \bel \in
       \bigcup\nolimits_{\begin{array}{c}
          \scriptstyle
          (x_i,\xi_i) \in \pl{SB}^{any}_{\pl{X}}(arg_1,z,Bdgs), \
          (x_1, \ldots, x_n) \in \pl I(pred)
         \end{array}
         }
        \xi_1 \join \ldots \join \xi_n \\
    &\Iff&
      \mbox{there are $(x_1,\xi_1),\ldots, (x_n,\xi_n)$
     s.t. $(x_i,\xi_i) \in \pl{SB}^{any}_{\pl{X}}(arg_i,z,Bdgs)$} \\
    && \mbox{and } (x_1, \ldots, x_n) \in \pl I(pred)
       \mbox{ and }  \bel\in \xi_1 \join \ldots \join \xi_n \\
    &\Iff& \mbox{(take the right $\bel_i\in\xi_i$)} \\
    && \mbox{there are $(x_1,\bel_1),\ldots, (x_n,\bel_n)$ s.t. } 
       \mbox{$(x_i,\bel_i) \in \pl{SB}^{any}_{\pl{X}}(arg_i,z,Bdgs)$} \\
    && \mbox{ and } (x_1, \ldots, x_n) \in \pl I(pred)
       \mbox{ and } \bel = \bel_1 \join \ldots \join \bel_n \\
    & \stackrel{\rm IH}{\Iff} &
      \mbox{there are $(x_1,\bel_1),\ldots, (x_n,\bel_n)$
     s.t. $x_i \in \pl S^{any}_{\pl{X}}(arg_i,z,\bel_i)$} \\
    && \mbox{and $\bel_i$ extends some 
              $\bel'_i\in Bdgs$ with $\free(arg_i)$} \\
    && \mbox{and } (x_1, \ldots, x_n) \in \pl I(pred)
       \mbox{ and } \bel = \bel_1 \join \ldots \join \bel_n \\
    &\Iff& \mbox{(the join guarantees that 
        $\bel':=\bel'_1=\ldots=\bel'_n$ holds)} \\
    &&  \mbox{there are $x_1,\ldots,x_n$
        s.t. $x_i \in \pl S^{any}_{\pl{X}}(arg_i,z,\bel_i)$} \\
    && \mbox{and $\bel$ extends some $\bel'\in Bdgs$ with 
       $\free(arg_1)\cup\ldots\cup\free(arg_n)$} \\
    &\stackrel{\rm Def.\ref{def-semantics-xpathlogic}}{\Iff}&
      \pl Q_{\pl{X}}(pred(arg_1,\ldots,arg_n),z,\bel) \\
    &&
      \mbox{ and $\bel$ completes some $\bel'\in Bdgs$ with 
             $\free(pred(arg_1,\ldots,arg_n))$}~.
         \end{array}
         \]

\item Negated expressions which do \emph{not contain any free variable}:
  trivial. \\
  For negated expressions which contain free variables: Note that
  all variables in $\free(\mbox{\sf not } expr)$ are required to be
  bound by $Bdgs$ (safety).
  
 \[\begin{array}{lcl}
   \multicolumn{3}{l}
    {\bel \in \pl{QB}_{\pl{X}}(\mbox{\sf not } expr, z,Bdgs)}\\ 
   &\stackrel{\rm Def}{\Iff}&
    \bel \in Bdgs \mbox{ and there is no 
      $\bel' \in \pl{QB}_{\pl{X}}(expr,z,Bdgs)$
                      s.t.\ $\bel\le\bel'$} \\
    & \stackrel{\rm IH}{\Iff} &
    \bel \in Bdgs \mbox{ and there is no 
      $\bel''$ such that}\\
    && \mbox{$\pl Q_{\pl{X}}(expr,z,\bel'')$ and $\bel''$ extends $\bel'$ 
      with $\free(expr)$ and $\bel\le\bel'$} \\
    &\stackrel{\rm Safety}{\Iff}&
      \bel \in Bdgs \mbox{ and not $\pl Q_{\pl{X}}(expr,z,\bel)$} \\
    &\stackrel{\rm Def.\ref{def-semantics-xpathlogic}}{\Iff}&
      \bel \in Bdgs 
      \mbox{ and $\pl Q_{\pl{X}}(\mbox{\sf\ not } expr,z,\bel)$}~.
    \end{array}
   \]
 Conjunction: \
 $\bel \in \pl{QB}_{\pl{X}}(expr_1 \mbox{\sf\ and } expr_2, z, Bdgs)$
 \[\begin{array}{@{}l@{}c@{}l}
    &\stackrel{\rm Def}{\Iff}&
    \bel \in 
       \pl{QB}_{\pl{X}}(expr_1,z,Bdgs) \join 
       \pl{QB}_{\pl{X}}(expr_2,z,\pl{QB}_{\pl{X}}(expr_1,z,Bdgs)) \\
    & \stackrel{\rm IH}{\Iff} &
       \mbox{there are $\gamma_1\in\pl{QB}_{\pl{X}}(expr_1,z,Bdgs)$}\\
    &&\mbox{and $\gamma_2\in\pl{QB}_{\pl{X}}(expr_1,z,
                   \pl{QB}_{\pl{X}}(expr_1,z,Bdgs))$ s.t.
            $\pl Q_{\pl{X}}(expr_1,z,\gamma_1)$}\\
    && \mbox{and $\gamma_1$ completes some $\bel'\in Bdgs$ with 
             $\free({expr_1})$ and} \\
    && \pl Q_{\pl{X}}(expr_2,z,\gamma_2)
       \mbox{ and $\gamma_1$ completes some 
            $\gamma''\in \pl{QB}_{\pl{X}}(expr_1,z,Bdgs)$} \\
    && \mbox{with $\free({expr_2})$ and $\bel = \gamma_1\join\gamma_2$}.
    \end{array}
  \]
  \[\begin{array}{@{}lcl}
    &\Iff& \mbox{(join condition: $\gamma_1=\gamma''\le\gamma_2$)}
      \ \ \pl Q_{\pl{X}}(expr_1,z,\gamma_2) \mbox{ and } 
       \pl Q_{\pl{X}}(expr_2,z,\gamma_2) \\
    &&   \mbox{ and $\gamma_2$ completes some 
            $\bel'\in Bdgs$ with 
             $\free({expr_1})\cup\free({expr_2})$} \\
    &\Iff&
     \pl Q_{\pl{X}}(expr_1 \mbox{\sf and } expr_2,z,\gamma_2) \\
    &&   \mbox{ and $\gamma_2$ completes some 
            $\bel'\in Bdgs$ with 
             $\free({expr_1})\cup\free({expr_2})$}~.
         \end{array}    
   \]         

\item -- \ref{def-semantics-xpathlogic-item-functions}.: trivial.
   (safety for variables; functions similar to predicates).
\item[\ref{def-semantics-xpathlogic-item-context-functions}.] 
  Context-related functions use the extension of variable bindings
  by pseudo-variables $Size$ and $Pos$ in rule
  (\ref{def-semantics-xpathlog-item-filter}):
 \[\begin{array}{@{}lcl}
    \multicolumn{3}{l}{(x,\bel)\in 
        \pl{SB}^{any}_{\pl{X}}(\textsf{position()}, z, Bdgs)} \\
    &\stackrel{\rm Def}{\Iff}&
     (x,\bel)\in \textsf{list}_{\bel \in Bdgs} 
           (\bel(Pos), \{\bel' \in Bdgs\st \bel(Pos)=\bel'(Pos)\}) \\
    &\Iff& \bel(Pos) = x \mbox{ for some $\bel\in Bdgs$} \\
    &\Iff&  x  \in \pl S^{any}_{\pl{X}}(\textsf{position()}, z, \bel)
            \mbox{ for some $\bel\in Bdgs$} \\
    &\Iff&  x  \in \pl S^{any}_{\pl{X}}(\textsf{position()}, z, \bel) \\
    && \mbox{and $\bel$ completes some $\bel' \in Bdgs$ by 
       $\free(\textsf{position()})$ (which is empty)}.
  \end{array}
  \]
  Analogously for \textsf{last()}.
\end{enumerate}
\end{Proof}

\begin{Proof} of Lemma~\ref{lem-correctness-atomize}: \
Structural induction.\
\begin{itemize}
\item entry case (using $\bel = \bel'$): \ \
  $(\pl{X},\bel) \models /p
    \stackrel{\rm Def.\ref{def-semantics-xpathlog-formulas}}{\Iff}
    (\pl S_\pl{X}(/p,\bel)) \neq\emptyset$
  \[\begin{array}{cl}
    \stackrel{\rm Def.\ref{def-semantics-xpathlogic}}{\Iff}
    & (\pl S_\pl{X}(p,root,\bel)) \neq\emptyset
     \stackrel{\rm Def.\ref{def-semantics-xpathlogic}}{\Iff}
     (\pl S_\pl{X}(root/p,\bel)) \neq\emptyset
      \stackrel{\rm Def.\ref{def-semantics-xpathlog-formulas}}{\Iff}
      (\pl{X},\bel) \models root/p  \smallskip\\
    \stackrel{\rm IH}{\Iff}
    & (\pl{X},\bel) \models \atomize(root/p) 
      \stackrel{\rm Def}{\Iff}
     (\pl{X},\bel) \models \atomize(/p)~. 
    \end{array}
    \]
\item Paths are resolved into steps and step qualifiers are isolated
  (the case where a don't care variable is introduced is shown; w.l.o.g.,
    $path$ is an absolute path expression)

  \[\begin{array}{@{}l@{}c@{}l}
    \multicolumn{3}{@{}l}
      {(\pl{X},\bel) \models path/axis::nodetest[\textit{stepQualifier}]~/remainder}
    \\
    & \Iff
    & \pl S_\pl{X}(path/axis::nodetest[\textit{stepQualifier}]~/remainder,\bel) 
       \neq\emptyset \\
    & \Iff
    & \pl S_\pl{X}(path/axis::nodetest[\textit{stepQualifier}]~/remainder,root,\bel) 
       \neq\emptyset \\
    & \Iff
    & 
      \textsf{concat}_{y\in 
           \pl S^a_{\pl{X}}(path/axis::nodetest[\textit{stepQualifier}],root,\bel)}
         (\pl S^{any}_{\pl{X}}(remainder,y,\bel)) 
       \neq\emptyset \\
    & \Iff
    & \mbox{there is a node } 
      v \in \pl S^a_{\pl{X}}(path/axis::nodetest[\textit{stepQualifier}],root,\bel) \\
    && \mbox{ s.t. }     
      \pl S^{any}_{\pl{X}}(remainder,v,\bel) \neq\emptyset \\ 
    &\Iff
    & \mbox{there is a node } 
      v \in \textsf{list}_{y\in \pl S^a_{\pl{X}}(path/axis::nodetest,x,\bel)} 
       (y \st \pl Q_{\pl{X}}(\textit{stepQualifier},y,\bel)) \\
    &&  \mbox{ s.t. }   
      \pl S^{any}_{\pl{X}}(remainder,v,\bel) \neq\emptyset \\ 
    &\Iff&  \mbox{there is a node $v$ s.t. }
      v \in \pl S^a_{\pl{X}}(path/axis::nodetest,x,\bel) \\
    &&\mbox{ and }
      \pl Q_{\pl{X}}(\textit{stepQualifier},v,\bel) \mbox{ and }
      \pl S^{any}_{\pl{X}}(remainder,v,\bel) \neq\emptyset \\ 
    &\Iff&  \mbox{there is a node $v$ s.t. }
      v \in \pl S^a_{\pl{X}}(path/axis::nodetest\fd \_X,x,\bel^v_{\_X}) \\
    && \mbox{ and }
      \pl Q_{\pl{X}}(\textit{V[stepQualifier]},v,\bel^v_{\_X}) \mbox{ and }
      \pl S^{any}_{\pl{X}}(V/remainder,v,\bel^v_{\_X}) \neq\emptyset \\ 
    &\Iff&  \mbox{there is a node $v$ s.t. }
      v \in \pl S^a_{\pl{X}}(path[axis::nodetest\fd \_X],x,\bel^v_{\_X}) \\
    && \mbox{ and } 
      \pl Q_{\pl{X}}(\textit{V[stepQualifier]},v,\bel^v_{\_X}) \mbox{ and }
      \pl S^{any}_{\pl{X}}(V/remainder,v,\bel^v_{\_X}) \neq\emptyset \\ 
    &\Iff&  \mbox{there is a node $v$ s.t. }
      \pl S^a_{\pl{X}}(path[axis::nodetest\fd \_X],x,\bel^v_{\_X})
             \neq\emptyset \\
    && \mbox{ and }
      \pl Q_{\pl{X}}(\textit{V[stepQualifier]},v,\bel^v_{\_X})
      \mbox{ and }
      \pl S^{any}_{\pl{X}}(V/remainder,x,\bel^v_{\_X}) \neq\emptyset \\ 
    &\Iff&  \mbox{there is a node $v$ s.t. }
      \pl Q_{\pl{X}}(path[axis::nodetest\fd \_X],\bel^v_{\_X}) \\
    && \mbox{ and }
      \pl Q_{\pl{X}}(\textit{V[stepQualifier]},\bel^v_{\_X})
      \mbox{ and }
      \pl Q_{\pl{X}}(V/remainder,\bel^v_{\_X})
  \end{array}
  \]
  \[\begin{array}{@{}l@{}c@{}l}
    &  \stackrel{\rm IH}{\Iff}
    & \mbox{there is a node $v$ s.t. }
      \pl Q_{\pl{X}}(\atomize(path[axis::nodetest\fd \_X]),\bel^v_{\_X}) \\
    && \mbox{ and }
      \pl Q_{\pl{X}}(\atomize(\textit{V[stepQualifier]}),\bel^v_{\_X})
      \mbox{ and }
      \pl Q_{\pl{X}}(\atomize(V/remainder),\bel^v_{\_X}) \\
    &\Iff&  \mbox{there is a node $v$ s.t. }
      \pl Q_{\pl{X}}(\atomize(\ldots),\bel^v_{\_X})~.
   \end{array}
   \]
\item Conjunctions in step qualifiers: obvious.
\item Predicates in step qualifiers: W.l.o.g., consider a unary predicate
  with a relative argument expression:
  \[\begin{array}{@{}l@{}c@{}l}
    \multicolumn{3}{@{}l}
      {(\pl{X},\bel) \models V[pred(expr)]} \\
    \hspace*{0.2cm}
    &\Iff
    &\pl S_\pl{X}(V[pred(expr)],\bel(V),\bel)\neq\emptyset \\
    &\Iff
    & \textsf{list}_{y\in \pl S^a_{\pl{X}}(V,\bel(V),\bel)} 
       (y \st \pl Q_{\pl{X}}(pred(expr),y,\bel)) 
       \neq \emptyset \\
    &\Iff
    & \mbox{($\bel(V)$ is the only element in 
              $\pl S^a_{\pl{X}}(V,\bel(V),\bel)$) s.t. }
     \pl Q_{\pl{X}}(pred(expr),\bel(V),\bel) \\
    &\Iff& \mbox{there is an $x \in \pl S_\pl{X}(expr,\bel(V),\bel)$
         such that $pred(x) \in \pl{X}$} \\
    &\Iff& \mbox{there is an $x$ s.t.
         $x \in \pl S_\pl{X}(V/expr\fd \_X,root,\bel^x_{\_X})$
         and  $(\pl{X},\bel^x_{\_X}) \models pred(\_X)$} \\
    &\Iff&  \mbox{there is an $x$ s.t.
         $(\pl{X},\bel^x_{\_X})\models V/expr\fd \_X$ 
         and $(\pl{X},\bel^x_{\_X}) \models pred(\_X)$} \\
    & \stackrel{\rm IH}{\Iff}
    & \mbox{there is an $x$ s.t.
      $(\pl{X},\bel^x_{\_X})\models \atomize(V/expr\fd \_X)$} \\
    && \mbox{and $(\pl{X},\bel^x_{\_X}) \models pred(\_X)$} \\
    & \Iff& \mbox{there is an $x$ s.t.
      $(\pl{X},\bel^x_{\_X})\models \atomize(V[pred(expr)])$}~.
       \end{array}
    \]
\item Predicate atoms: analogous.
\end{itemize}
\end{Proof}

\end{appendix}

\end{document}